\documentclass[aps,prd,twocolumn,nofootinbib,superscriptaddress,preprintnumbers,floatfix]{revtex4-2}

\usepackage{graphicx}
\usepackage{dcolumn}
\usepackage{bm}
\usepackage{hyperref}
\usepackage{physics}
\usepackage{xcolor}
\usepackage[normalem]{ulem}

\newcommand{\me}[3]{\langle #1\vert\ #2\ \vert #3\rangle}
\newcommand{\pvec}{\bm{p}}
\newcommand{\xvec}{\bm{x}}
\newcommand{\beq}{\begin{equation}}
\newcommand{\eeq}{\end{equation}}
\newcommand{\Bp}{B^{(\bm{P})}}
\newcommand{\Ecm}{E_{\rm cm}}
\newcommand{\etaop}{\tilde\eta}
\newcommand{\phiop}{\tilde\phi}

\newcommand{\bern}{
	Institute for Theoretical Physics,
	Albert Einstein Center for Fundamental Physics,
	University of Bern, 3012 Bern, Switzerland}
\newcommand{\bochum}{
	Fakult\"at f\"ur Physik und Astronomie,
	Institut f\"ur Theoretische Physik II,
	Ruhr-Universit\"at Bochum, 44780 Bochum, Germany}
\newcommand{\cmu}{
    Department of Physics, Carnegie Mellon University,
    Pittsburgh, Pennsylvania 15213, USA}
\newcommand{\kentstate}{
	Department of Physics, Kent State University,
	Kent, Ohio 44242, USA}
\newcommand{\lblnsd}{
    Nuclear Science Division,
    Lawrence Berkeley National Laboratory,
	Berkeley, California 94720, USA}

\begin{document}

\title{Investigating the role of tetraquark operators in lattice QCD\\
       studies of the \boldmath $a_0(980)$ and $\kappa$ resonances}

\author{Andrew~D.~Hanlon}
\affiliation{\kentstate}

\author{Daniel Darvish}
\affiliation{\cmu}

\author{Sarah Skinner}
\affiliation{\cmu}

\author{John Meneghini}
\affiliation{\cmu}

\author{Ruair\'i Brett}
\affiliation{\cmu}

\author{John~Bulava}
\affiliation{\bochum}

\author{Jacob Fallica}
\affiliation{\cmu}

\author{Colin Morningstar}
\affiliation{\cmu}

\author{Fernando Romero-L\'{o}pez}
\affiliation{\bern}

\author{Andr\'{e} Walker-Loud}
\affiliation{\lblnsd}

\date{March 27, 2026}

\begin{abstract}
The role of tetraquark operators in studying the isodoublet strange $\kappa$ and isovector nonstrange
$a_0(980)$ scalar mesons in lattice QCD is examined using an ensemble with $m_\pi\approx230$~MeV and
spatial extent $L$ such that $m_\pi L\approx4.4$. Hermitian correlation matrices using both
single-meson, meson-meson, and tetraquark interpolating operators are used to extract the spectrum of
finite-volume stationary states in the appropriate symmetry channels.  Hundreds of local and extended
tetraquark operators are explored. Determinations of the spectrum in each channel are found to be
unreliable without the inclusion of at least one tetraquark operator. For example, the inclusion of
tetraquark operators with isospin 1/2 and strangeness 1 quantum  numbers reveals the existence of
an additional energy level in the $K\eta$ sub-system below the $K\eta$ threshold. The implications
of this on parametrizing the scattering $K$-matrix through a well-known quantization condition to
extract properties of the $\kappa$ and $a_0(980)$ scalar meson resonances are discussed.
\end{abstract}

\keywords{lattice QCD, scattering amplitudes}
\maketitle

\section{Introduction}
\label{sec:intro}

The low-lying scalar mesons of quantum chromodynamics (QCD) present significant challenges to
experimental and theoretical studies compared to other hadrons.  The large decay widths of the scalar
resonances make them difficult to resolve from background processes in scattering experiments, with
multiple decay channels opening up within small energy ranges.  The scalar mesons defy description
by the naive quark model\cite{Godfrey:1985xj,Rupp:2018azz}. These systems are also difficult to study
in lattice QCD, requiring correlation matrices which involve a large variety of different
interpolating operators and whose evaluations are hampered by noisy fermionic disconnected
contributions.

It has been suggested that the $a_0(980)$ and $K_0^*(700)$ (also known as the $\kappa$) scalar
resonances could have tetraquark content~\cite{Jaffe:2004ph, Amsler:2004ps, Close:2002zu,
Maiani:2004uc}. Tetraquark operators have been 
shown\cite{Stump:2025owq,Prelovsek:2025vbr,Alharazin:2026lno} to be
important for the $T_{cc}(3875)^+$ resonance. To date, there has been only a small number of
lattice QCD studies investigating the tetraquark content of light scalar resonances. In 2010,
Prelovsek et al.~\cite{Prelovsek:2010kg} investigated the $\sigma$ and $\kappa$ as possible tetraquark
candidates, but neglected disconnected diagrams in their calculations. Using tetraquark interpolators,
they found an additional light state in both the $\sigma$ and $\kappa$ channels. In 2013, the ETM
collaboration examined the $a_0(980)$ and $\kappa$ using four-quark operators~\cite{Alexandrou:2012rm},
though they also neglected disconnected diagrams in their calculations. They found no evidence of an
additional state that can be interpreted as a tetraquark. In 2018, Alexandrou
et al.~\cite{Alexandrou:2017itd} conducted a study of the $a_0(980)$ with four-quark operators,
including disconnected contributions. In their study, they found an additional finite-volume state
in the sector containing the $a_0(980)$ meson, which has primarily quark-antiquark content but also
has sizeable diquark-antidiquark content, in the range of 1100 to 1200 MeV. Additionally, they concluded
that disconnected diagrams have drastic effects on their results, and thus, cannot be neglected. The
importance of including disconnected diagrams when studying the $\kappa$ and $a_0(980)$ (as well as
other scalar mesons) is also shown in Ref.~\cite{Guo:2013nja}.

In this work, we examine the role of tetraquark operators in lattice QCD studies of the $a_0(980)$
and $\kappa$ scalar meson resonances. We make two different determinations of the spectrum in each
symmetry sector: one using a basis of only single- and two-meson operators, and one using a basis
that also includes a tetraquark operator selected from hundreds of tetraquark operators which were
tested in preliminary low-statistics Monte Carlo calculations. The stochastic LapH
method~\cite{Morningstar:2011ka} is used to evaluate the diagrams in our calculations, including
\textit{all} disconnected contributions. We find that including a tetraquark operator yields an
additional low-lying finite-volume state in the isodoublet strange $A_{1g}$ sector.  In the
$a_0(980)$ channel, a significant change in the extracted spectrum occurs upon inclusion of the
tetraquark operator.

To obtain hadron scattering amplitudes and resonance information in lattice QCD, the first step is
to compute the energies of the finite-volume (FV) stationary states in the relevant channels of
interest involving the scattered hadrons. Once the FV spectra are obtained, the elements of the
infinite-volume scattering $K$ matrix must be appropriately parametrized, then best fit values of
the parameters are obtained using an appropriate quantization
condition\cite{Luscher:1990ux,Rummukainen:1995vs,Kim:2005gf,Briceno:2014oea}.  In this exploratory
work, we focus mainly on the first step, extracting the FV spectra, highlighting the importance of
using judiciously chosen interpolating operators.  In Sec.~\ref{sec:opinterps}, we describe our
lengthy exploration of interpolating operators and find that at least one tetraquark operator
seems to be necessary for reliably extracting all of the needed energies of the FV spectra relevant
for lattice QCD studies of the $\kappa$ and $a_0(980)$ resonances.  Our results for the FV spectra
are then presented in Sec.~\ref{sec:fvspec}.  In Sec.~\ref{sec:Kamp}, we address the impact of the
tetraquark operators on determining the $K$-matrix parameters, followed by concluding remarks in
Sec.~\ref{sec:conclude}.

\section{Interpolating operators}
\label{sec:opinterps}

The interpolating operators we use to determine the low-lying energies of the stationary states
of QCD in finite volume in the isodoublet strange $A_{1g}$ and isotriplet nonstrange $A_{1g}^-$
channels of total zero momentum are described in this section. First, we describe how we obtain
these energies from matrices of temporal correlations of interpolating operators in
Sec.~\ref{sec:temporal}.  The construction of our single-meson and two-meson operators is then
outlined in Sec.~\ref{sec:mesonops}, and our construction of hundreds of tetraquark operators
is detailed in Sec.~\ref{sec:tetrops}.  Calculation details, such as the parameters of our
ensemble of gauge configuration, our gluon and quark field smearing, and our use of the
stochastic LapH method, are outlined in Sec.~\ref{sec:calcdetails}.  The final sets of
interpolating operators to use are presented in Sec.~\ref{sec:opselect}, along with an outline
of the procedure used to select them.

\subsection{Energies from temporal correlations}
\label{sec:temporal}

In lattice QCD, the energies of the stationary states in finite volume are extracted from
matrices of temporal correlations
\begin{equation}
 {\cal C}_{ij}(t)= \langle 0 \vert O_i(t+t_s) \overline{O}_j(t_s) \vert 0 \rangle,
\end{equation}
where $t$ is the time separation, $t_s$ is some reference source time, and $O_i(t)$ denotes
the $i$-th interpolating operator constructed from quark and gluon fields to annihilate single- and
multi-hadron states having appropriate transformation properties.  The corresponding creation
operators are denoted by $\overline{O}_i(t)$.  The operators $O_i$ and $\overline{O}_i$ are the
analytic continuations to Euclidean space of the Minkowski-space operators $O_i^{(M )}$ and
$O_i^{(M)\dagger}$, respectively. In the imaginary time formalism, estimates of these ${\cal C}_{ij}(t)$
correlators can be obtained using a Markov-chain Monte Carlo method.  The various diagonal elements
of this matrix at early times can differ drastically in magnitude due to different operator
normalizations.  To ameliorate any possible numerical instabilities from these physically
irrelevant normalizations, the following rescaled matrix is formed and subsequently used:
\begin{equation}
  C_{i j}(t) = \frac{\mathcal{C}_{i j}(t)}{\sqrt{\mathcal{C}_{i i}\left(t_N\right)
   \mathcal{C}_{j j}\left(t_N\right)}},
\end{equation}
where the normalization time $t_N=\tau_Na_t$ is taken at some early time, such as
$t_N < 3\,a_t \sim 0.1~$fm, and $a_t$ is the temporal lattice spacing.
In finite volume, the stationary-state energies $E_n$ are discrete,
so by inserting a complete set of states, these correlators can be expressed in terms of the
energies using
\begin{equation}
   C_{ij}(t) = \sum_{n=0} Z_i^{(n)} Z_j^{(n)\ast}\, e^{-E_n t},
   \quad Z_j^{(n)}=  \me{0}{O_j}{n},
\label{eq:corrt}
\end{equation}
ignoring negligible effects from the temporal boundary and taking $E_{n+1}\ge E_n$.

Given the large number of complex-valued overlap factors $Z_i^{(n)}$ in Eq.~(\ref{eq:corrt}),
it is not practical to do simultaneous fits to the entire $C(t)$ matrix to determine the
low-lying energies, so instead, separate fits are carried out to each of the diagonal elements
of the matrix $\widetilde{C}(t)$ obtained using a single-pivot
rotation\cite{Fox:1981xz,Michael:1982gb,Luscher:1990ck}
\begin{equation}
  \widetilde{C}(t) = U^\dagger\ C(t_0)^{-1/2}\ C(t)\ C(t_0)^{-1/2}\ U,
\label{eq:rotatedcorr}
\end{equation}
where the columns of the unitary matrix $U$ are the orthonormal eigenvectors of
$C(t_0)^{-1/2}\,C(t_D)\,C(t_0)^{-1/2}$, determined by solving a generalized eigenvector
problem (GEVP) given by $C(t_D)W=C(t_0)W\Lambda$, where the matrix $W=C(t_0)^{-1/2}U$ and $\Lambda$
is the diagonal matrix of eigenvalues. We choose $t_0=\tau_0\,a_t$ and $t_D=\tau_D\,a_t$ large enough so that
$\widetilde{C}(t)$ remains diagonal within statistical errors for $t>t_D$ and such that the
extracted energies are insensitive to increases in these parameters. Typically,
two-exponential fits of the form $A_\alpha e^{-E_{\alpha}t}(1+B_\alpha e^{-\Delta t})$
to the diagonal elements $\widetilde{C}_{\alpha\alpha}(t)$ yield the
energies $E_\alpha$, and the magnitudes of the overlap factors are determined using
$\vert Z_j^{(n)}\vert =\vert C(t_0)^{1/2}_{jk}U_{kn}A_n^{1/2}\vert$ (no sum over $n$). 
Such fits are done using the standard
correlated-$\chi^2$ method with the full covariance matrix of the correlator estimates
in the $\chi^2$ evaluated by bootstrap resampling and the fit parameter uncertainties
estimated by bootstrap resampling.

Since stochastic estimates of $C_{ij}(t)$ are obtained using the Monte Carlo method and
the signal-to-noise ratios of such estimates fall quickly with $t$, it is very important
to use judiciously-constructed operators $O_j(t)$ to maximize the $\vert Z_j^{(n)}\vert$ factor
magnitudes for the
states of interest and minimize those for unwanted higher-lying states in order to reliably
extract the low-lying energies.
If the $N$ lowest-lying energies in a particular symmetry channel are needed, a correlation
matrix using at least $N$ interpolating operators must be evaluated.  To improve the energy
extractions, more than $N$ operators are usually required to help remove contamination from
levels above the lowest $N$ states.  A crucial point, however, is that in practice, for each
of the $N$ lowest-lying energies, an operator should be present which produces a state having
significant overlap with the eigenstate associated with that energy.
Without such an operator set, the limited temporal range
of the correlations which can be reliably estimated due to the limited statistics
possible with current Monte Carlo methods can lead to missed energy levels.

For each symmetry channel of interest specified by a total momentum and an irreducible
representation (irrep) of its little group, a basic set of operators to use should include
operators that contain the incoming particles as well as operators that contain the outgoing
particles of the scattering process in all of the allowed individual momenta and in
spin/orbital combinations that transform according
to the irrep of the channel.  Determining if \textit{other} operators are also needed and
what those operators should be is an important issue.

\subsection{Meson and meson-meson operator construction}
\label{sec:mesonops}

Our meson and meson-meson operator construction is detailed in
Refs.\cite{Basak:2005aq,Morningstar:2013bda}, but
key points are summarized here. Individual hadron operators are constructed
by combining basic building blocks, which are covariantly-displaced LapH-smeared quark
fields\cite{Morningstar:2011ka} with stout link smearing\cite{Morningstar:2003gk}.
The smeared quark field for flavor $A$ is here denoted by
$\widetilde{\psi}^{(A)}_{a\alpha}(x)$, where $a$ is a color index, $\alpha$ is a Dirac spin
index, and $x$ is a lattice site. The gauge-covariantly-displaced, LapH-smeared quark fields
are formed using
\begin{equation}
  q^A_{a\alpha j}= (D^{(j)}\widetilde{\psi})_{a\alpha}^{(A)},
  \qquad  \overline{q}^A_{a\alpha j} = (\widetilde{\overline{\psi}}
  \gamma_4\, D^{(j)\dagger})_{a\alpha}^{(A)},
\label{eq:quarkdef}
\end{equation}
where $\gamma_4$ is the temporal Dirac $\gamma$-matrix, and $D^{(j)}$ is a gauge-covariant
displacement of type $j$.  The displacement type is a sequence of $p$ spatial directions
on the lattice $j=(j_1,j_2,\cdots,j_p)$.  This displacement can be trivial ($j=0$ meaning
no displacement), a displacement in a given single spatial direction on the lattice by
some number of links (typically two or three), or a combination of two or more spatial
lattice directions.  If we define $d_r = \hat{j}_1+\hat{j}_2+\dots+\hat{j}_{r-1}$,
then the displacement $D^{(j)}$ is defined as a product of smeared link variables:
\begin{align}
  D^{(j)}(x,x^\prime) &= \widetilde{U}_{j_1}(x)\ \widetilde{U}_{j_2}(x\!+\!d_2)
  \ \widetilde{U}_{j_3}(x\!+\!d_3)\dots \nonumber \\
  &\times   \widetilde{U}_{j_p}(x\!+\!d_p) \delta_{x^\prime,x+d_{p+1}}.
\end{align}
The use of $\gamma_4$ in Eq.~(\ref{eq:quarkdef}) is convenient for obtaining baryon
correlation matrices that are Hermitian.

To construct our meson operators, we first form so-called \emph{elemental} mesonic
operators.  Such an operator which destroys a state of three-momentum $\pvec$ is given by
\begin{equation}
  \Phi^{AB}_{\alpha\beta}(t)=
  \sum_{\bm{x}}e^{-i\pvec\cdot(\xvec+\frac{1}{2}(\bm{d}_\alpha+\bm{d}_\beta))}
  \delta_{ab}  \ \overline{q}^A_{a\alpha}(\bm{x},t)\ q^B_{b\beta}(\bm{x},t),
\label{eq:mesonelemental}
\end{equation}
where $q,\overline{q}$ are defined in Eq.~(\ref{eq:quarkdef}),
$\bm{d}_\alpha, \bm{d}_\beta$ are the spatial displacements of the
$\overline{q},q$ fields, respectively, from $\xvec$,
$A,B$ indicate flavor, and $\alpha,\beta$ are compound indices
incorporating both spin and quark-displacement types.  The phase factor
involving the quark-antiquark displacements is needed to ensure proper
transformation properties under $G$-parity for arbitrary displacement types.
The ``barred'' operators  which create a state of momentum $\pvec$ then take the form
\begin{equation}
  \overline{\Phi}_{\alpha\beta}^{AB}(t)=
  \sum_{\bm{x}} e^{i\pvec\cdot(\xvec+\frac{1}{2}(\bm{d}_\alpha+\bm{d}_\beta))}
  \delta_{ab}\ \overline{q}^B_{b\beta}(\bm{x},t)\ q^A_{a\alpha}(\bm{x},t).
\end{equation}

Each meson sink operator is then a superposition of these elemental operators
\begin{equation}
  M_{l}(t)= c^{(l)}_{\alpha\beta}\ \Phi^{AB}_{\alpha\beta}(t),
\label{eq:mesonstart}
\end{equation}
(or is a flavor combination of the above form), where $l$ is a compound index comprised
of a three-momentum $\pvec$, an irreducible representation $\Lambda$ of the little
group of $\pvec$, the row $\lambda$ of the irrep, total isospin $I$, isospin projection
$I_3$, strangeness $S$, and an identifier labeling the different operators in each
symmetry channel.  The superposition coefficients are determined using group-theoretical
projections.  The corresponding source operators are
\begin{equation}
  \overline{M}_{l}(t)= c^{(l)\ast}_{
  \alpha\beta}\ \overline{\Phi}^{AB}_{\alpha\beta}(t).
\end{equation}
The spatial configurations we use for our single meson operators are shown in Table I
of Ref.~\cite{Morningstar:2013bda}.  We use single-site meson operators (SS),
singly-displaced (SD), doubly-displaced in an L-shape (DDL), and triply-displaced in a
U-shape (TDU) and using three orthogonal directions (TDO).  For mesons having
nonzero momenta, quark displacements can be either transverse or longitudinal to
the meson momentum.  The flavor structures of our single meson operators are
presented in Table~\ref{tab:mesonflav}.

\begin{table}
\caption{Flavor content of our meson annihilation operators. Ordering of the quark/antiquark
 fields is the same as the annihilation operators shown in Eq.~(\ref{eq:mesonelemental}).
 The names $\etaop$ and $\phiop$ are used to denote $\overline{u}u+\overline{d}d$ and
 $\overline{s}s$ elemental operators, respectively, to simplify notation; they are not
 meant to indicate physical $\eta$ and $\phi$ meson operators which would be superpositions
 of $\overline{u}u+\overline{d}d$ and $\overline{s}s$ elementals.
\label{tab:mesonflav}}
\begin{ruledtabular}
\begin{tabular}{cc}
    Operator name & Flavor content \\ \hline
    $\pi^+$ & $\overline{d}u$ \\
    $\pi^0$ & $(\overline{d}d-\overline{u}u)/\sqrt{2}$\\
    $\pi^-$ & $-\overline{u}d$ \\
    $\etaop$ &  $\overline{u}u+\overline{d}d$\\
    $\phiop$ &  $\overline{s}s$ \\
    $K^+$ & $\overline{s}u$ \\
    $K^0$ & $\overline{s}d$ \\
    $\overline{K}^0$ & $\overline{d}s$ \\
    $\overline{K}^-$ & $-\overline{u}s$\\
\end{tabular}
\end{ruledtabular}
\end{table}

Our hadron operator construction generalizes to two, three, or more hadrons. To form a
two-meson operator $\mathcal{O}_l(t)$, we follow a similar procedure and project the product
of two final meson operators $M^{a}_{l_a}(t) M^{b}_{l_b}(t)$ onto a final symmetry channel
$l$: $\mathcal{O}_l(t) = c^{(l)}_{l_a l_b} M^{a}_{l_a}(t) M^{b}_{l_b}(t)$.

\subsection{Tetraquark operator construction}
\label{sec:tetrops}

As previously mentioned, some past works have suggested that the $\kappa$ and $a_0(980)$
resonances might require tetraquark operators to be investigated
reliably\cite{Prelovsek:2010kg,Alexandrou:2017itd}.  To test this, we designed and
implemented a large variety of tetraquark operators.

Quark fields at each lattice site $x$ transform under a local gauge transformation (color
rotation) according to the fundamental $3$ irrep of $SU(3)$, whereas antiquark fields
transform according to the complex-conjugate $\overline{3}$ irrep. Color singlet tetraquark
operators can be built from the product of two 3 and two $\overline{3}$ objects:
\begin{align}
  3 \otimes 3 \otimes \overline{3} \otimes \overline{3}&=\nonumber\\
  1\oplus1\oplus8&\oplus8\oplus8\oplus8\oplus10\oplus\overline{10}\oplus27.
\label{eq:cg}
\end{align}
which yields two linearly independent color-singlet objects, as evidenced by the two-fold
occurrence of the $1$ irrep in Eq.~(\ref{eq:cg}). Let $p(x)$, $q(x)$, $r(x)$, $s(x)$ denote
distinct fields which transform as color vectors in the fundamental $3$ irrep, then
$p^{*}(x)$, $q^{*}(x)$ transform according to the complex-conjugate $\overline{3}$ irrep.
For distinct fields, there are $3^4=81$ basis operators formed by the color fields,
$p_{a}^{*}(x) q_{b}^{*}(x) r_{c}(x) s_{d}(x)$. The following combinations are both
linearly independent and gauge-invariant:
\begin{align}
   T_{S} &=\left(\delta_{a c} \delta_{b d}+\delta_{a d} \delta_{b c}\right)
   p_{a}^{*}(x) q_{b}^{*}(x) r_{c}(x) s_{d}(x), \label{eq:TSdef}\\
   T_{A} &=\left(\delta_{a c} \delta_{b d}-\delta_{a d}
   \delta_{b c}\right) p_{a}^{*}(x) q_{b}^{*}(x) r_{c}(x) s_{d}(x),\label{eq:TAdef}
\end{align}
and so they form a basis with which to construct our elemental tetraquark operators, fulfilling
the need for \emph{two} linearly independent color singlet operators from Eq.~(\ref{eq:cg}).
Other color objects can be formed using, for example, link variables in different color
irreps, but as long as only two quark and two antiquark fields are present, Clebsch-Gordan
coefficients can be used to show that such objects are expressible in terms of the above
operators, with possibly gauge-invariant gluon loops.

\begin{figure}
\begin{center}
\includegraphics[width=0.85\columnwidth]{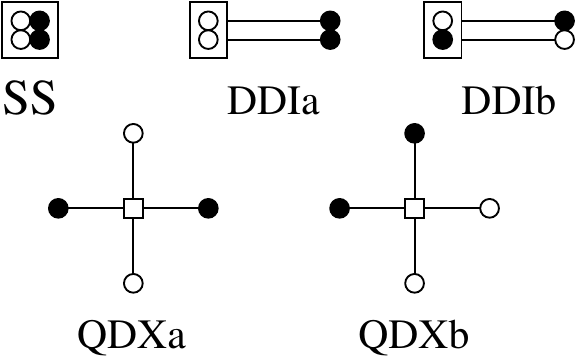}
\end{center}
\caption[tetra]{Our tetraquark operators consist of two gauge-covariantly
 displaced quarks (open circles) and two displaced antiquarks (solid circles)
 arranged in the spatial configurations shown above.  The boxes indicate locations
 where color and spin indices are connected.  The different spatial configurations
 we use are single-site operators (SS), doubly-displaced operators
 in an I configuration (DDIa and DDIb), and quadruply-displaced operators in a cross
 configuration (QDXa and QDXb).  The letters ``a'' and ``b'' in the labels denote
 the different orderings of displacing the quarks and antiquarks.  For each label,
 there are two color structures, given by Eqs.~(\ref{eq:TSdef})
 and (\ref{eq:TAdef}).
\label{fig:tqspatial}}
\end{figure}

Each of the color-vector fields $p(x)$, $q(x)$, $r(x)$, $s(x)$ above can be one spin component of
a local quark field, a smeared quark field, or an appropriately covariantly-displaced and
smeared quark field, but each must transform as a color three-vector at site $x$.  Using a variety
of different combinations of the Dirac indices and covariant displacements of the quark/antiquark
fields allows us to construct a very large number of tetraquark operators having differing spin,
orbital, and radial structures.  We designed and implemented hundreds of tetraquark operators
with different flavor, spin, and spatial constructions.  The spatial configurations we use are
shown in Fig.~\ref{fig:tqspatial}.  We employ single-site operators (SS), doubly-displaced operators
in an I configuration (DDIa and DDIb), and quadruply-displaced operators in a cross
configuration (QDXa and QDXb).  The letters ``a'' and ``b'' in the label denote
different orderings of displacing the quarks and antiquarks.  For each label, there are
two color structures corresponding to Eqs.~(\ref{eq:TSdef}) and (\ref{eq:TAdef}).

Each of our tetraquark operators which annihilates a state of
three-momentum $\pvec$ is a linear superposition of gauge-invariant elemental operators
\begin{align}
  &\Phi^{ABCD\pm}_{\alpha\beta\mu\nu;ijkl}(t)=\sum_{\bm{x}}e^{-i\pvec\cdot\xvec}
  (\delta_{ab}\delta_{cd}\pm \delta_{ad}\delta_{bc}) \nonumber\\
  & \qquad \times\overline{q}^A_{a\alpha i}(\bm{x},t)\ q^B_{b\beta j}(\bm{x},t)
  \ \overline{q}^C_{c\mu k}(\bm{x},t)\ q^D_{d\nu l}(\bm{x},t),
\label{eq:tetraquarkelemental}
\end{align}
where $q,\overline{q}$ are defined in Eq.~(4) of Ref.~\cite{Morningstar:2013bda},
$A,B,C,D$ indicate flavor, and $\alpha,\beta,\mu,\nu$ are spin indices,
$i,j,k,l$ indicate quark-displacement types, and $a,b,c,d$ are color
indices.  In this form, we do not need any additional phase factors to
properly handle $G$-parity, as we do with the moving mesons.
The ``barred'' operators  which create a state of momentum $\pvec$ then take the form
\begin{align}
  &\overline{\Phi}_{\alpha\beta\mu\nu;ijkl}^{ABCD\pm}(t)=
  \sum_{\bm{x}} e^{i\pvec\cdot\xvec}
  (\delta_{ab}\delta_{cd}\pm \delta_{ad}\delta_{bc}) \nonumber\\
  &\qquad\times \overline{q}^D_{d\nu l}(\bm{x},t)\ q^C_{c\mu k}(\bm{x},t)
  \ \overline{q}^B_{b\beta j}(\bm{x},t)\ q^A_{a\alpha i}(\bm{x},t).
\end{align}
Each tetraquark sink operator has the form
\begin{equation}
  T_{l\pm}(t)= c^{(l)}_{\alpha\beta\mu\nu;ijkl}\ \Phi^{ABCD\pm}_{\alpha\beta\mu\nu;ijkl}(t),
\label{eq:tetrastart}
\end{equation}
(or is a flavor combination of the above form),
where $l$ is a compound index comprised of a three-momentum $\pvec$, an
irreducible representation $\Lambda$ of the lattice symmetry group, the
row $\lambda$ of the irrep, total isospin $I$ and isospin projection $I_3$,
strangeness $S$, and an identifier labeling the different operators in each
symmetry channel.  Here, we focus on mesons containing only $u,d,s$ quarks.
The corresponding source operators are
\begin{equation}
  \overline{T}_{l\pm}(t)= c^{(l)\ast}_{
  \alpha\beta\mu\nu;ijkl}\ \overline{\Phi}^{ABCD\pm}_{\alpha\beta\mu\nu;ijkl}(t).
\end{equation}
Group-theoretical projections are carried out to determine the superposition
coefficients $c^{(l)}_{\alpha\beta\mu\nu;ijkl}$ so that the resulting operators
transform irreducibly under the little group of $\pvec$. These superposition
coefficients are available upon request.  The different flavor structures of our tetraquark
annihilation operators are presented in Table~\ref{tab:flavor_structs}.

\begin{table}[t]
\caption{Flavor content of our isodoublet strange and isotriplet nonstrange tetraquark
 annihilation operators for isospin projection $I_3=I$. Ordering of the quark/antiquark
 fields is the same as the annihilation operators shown in Eq.~(\ref{eq:tetraquarkelemental}).
\label{tab:flavor_structs}}
\begin{ruledtabular}
\begin{tabular}{ccc}
    Isospin & Strangeness & Flavor content  \\ \hline
    $\frac{1}{2}$ & 1 & $\overline s u \overline s s$ \\
    $\frac{1}{2}$ & 1 & $\overline{s} u(\overline{u} u+\overline{d} d)$ \\
    $\frac{1}{2}$ & 1 &  $\overline{s} u(\overline{u} u-\overline{d} d)
               -\sqrt{2}\ \overline{s} d \overline{d} u$ \\
    1 & 0 & $(\overline{u} u+\overline{d} d) \overline{d} u$ \\
    1 & 0 & $\overline s s \overline d u$ \\
    1 & 0 & $\overline{d} u(\overline{u} u-\overline{d} d)-(\overline{u} u-\overline{d} d)
    \overline{d} u$ \\
\end{tabular}
\end{ruledtabular}
\end{table}

Although the color structures of our tetraquark operators make these operators seem very similar
to our meson-meson operators, the individual color-contracted quark-antiquark pairs in a tetraquark
are not each formed with a separate individual well-defined momentum nor does each pair transform
irreducibly under any symmetry except color. Only the full combination of the two quark-antiquark
pairs is formed with well defined momentum and spin/orbital transformation properties.

\subsection{Calculational details}
\label{sec:calcdetails}

This exploratory study was conducted using $N_{\rm cf}=412$ gauge field configurations generated by
the Hadron Spectrum collaboration~\cite{Edwards:2008ja, Lin:2008pr} on an anisotropic lattice
of size $32^3\times 256$ with a spatial length $L=3.74$ fm and a pion mass of approximately $230$ MeV,
using $N_f=2+1$ Wilson clover-improved fermions. Ensemble details are given in
Table~\ref{table:ensemble}, and the action is described in Ref.~\cite{Edwards:2008ja}.

The temporal correlators are evaluated using the Monte Carlo method as usual
in lattice QCD. For the LapH smearing of the quark fields, the number of eigenvectors
retained is $N_v=264$, and the three-dimensional covariant Laplacian operator
is constructed using stout-smeared spatial links with a staple weight of $\rho = 0.1$ and
$N_\rho = 10$ iterations.  Including multi-hadron operators in our correlation matrices
requires the use of time-slice to time-slice quark propagators.  To make the calculations
of the temporal correlators feasible, we resort to employing stochastic estimates of such
quark propagators. The stochastic LapH method\cite{Morningstar:2011ka} is used.  For quark
lines that connect the source and sink time slices, the noise dilution scheme (TF, SF, LI8)
is utilized (see Ref.~\cite{Morningstar:2011ka} for more details). This indicates full time
dilution (TF), full spin dilution (SF), and interlace-8 for the LapH eigenvector dilution
(LI8).  For correlators which involve quark lines that originate and terminate at the final
sink time, the dilution scheme (TI16, SF, LI8) was found to work well.

\begin{table}[t]
\caption{Details of the Monte Carlo ensemble used in this work and some stable meson masses.
 $N_{\rm cf}$ is the number of gauge-field configurations used. The temporal lattice spacing
 $a_t=0.033357(59)$~fm is taken from Ref.~\cite{Brett:2018jqw}. Results from our determinations
 of the masses $m_\pi, m_K, m_\eta, m_{\eta^\prime}$ of the pion, kaon, $\eta$, and
 $\eta^\prime$ mesons, respectively, are given in terms of $a_t$ and are described later in
 Sec.~\ref{sec:fvspec}.  The renormalized anisotropy $\xi=a_s/a_t=3.451(11)$ is taken from
 Ref.~\cite{Bulava:2016mks}, where $a_s$ is the spatial lattice spacing.  The lattice extent
 $(L/a_s)^3 \times (T/a_t)$ is $32^3\times 256$.
\label{table:ensemble}}
\begin{ruledtabular}
\begin{tabular}{ccccc}
  $N_{\rm{cf}}$ & $a_t m_\pi$ & $a_t m_K$ & $a_t m_\eta$ & $a_tm_{\eta^\prime}$ \\
  \hline
  $412$ & $0.03953(17)$ & $0.08348(14)$ & $0.0979(30)$ & $0.165(12)$
\end{tabular}
\end{ruledtabular}
\end{table}

\subsection{Operator selection}
\label{sec:opselect}

The quantum numbers associated with the $a_0(980)$ resonance are $I^G(J^{PC}) = 1^-(0^{++})$ and
strangeness $S = 0$, so the first channel of interest in this lattice QCD study is the isotriplet,
nonstrange, $A_{1g}^-$ symmetry sector having total zero momentum. In this nomenclature, the
subscript $g$ indicates positive parity and the superscript, -, indicates negative $G$-parity.
The quantum numbers associated with the $\kappa$ resonance are $I(J^{P})=\frac{1}{2}(0^+)$ and
strangeness $S=1$, so the second channel of interest in our lattice QCD study is the isodoublet,
strangeness 1, $A_{1g}$ symmetry sector with total zero momentum.  To probe these two resonances,
we need to include operators capable of creating all of the stationary states in finite volume
whose energies lie below $E_{\rm max}$, where $E_{\rm max}$ should lie somewhat above the expected
energies of these resonances.  We decided to choose  $E_{\rm max}\sim 3\, m_K$, where $m_K$ is the
mass of the kaon.

First, we choose a handful of single-meson operators in order to capture any states that are
predominantly quark-antiquark states.  The single-meson operators we selected in the two channels are
listed in Table~\ref{tab:single-meson-ops}.  More will be said about these single-meson operators
below.  From prior computations, the energies of a single pion, kaon, $\eta$ and $\phi$ having
particular momenta have been determined, so the non-interacting energies of any two of these particles
are known.  We determine all of the two particle combinations that produce non-interacting energies
below $E_{\rm max}$ in these channels, then include the two-meson operators expected to create these
states.  The two-meson operators we selected in the two channels are listed in
Table~\ref{tab:two-meson-ops}. The lowest non-interacting energy of three mesons in each channel lies
above $E_{\rm max}$, so no three-meson operators are needed.  The four-meson thresholds in the
$\kappa$ and $a_0(980)$ channels occur near $2.4\,m_K$ and $2.6\,m_K$, respectively.  Since
incorporating four-meson operators into this exploratory study would be very costly, we did not
include them.  In so doing, our results above these thresholds must be considered with caution.
Since the effects of the tetraquark operators occur well below the four-meson thresholds, the
main conclusions of this work are not affected by this decision to exclude four-meson operators.

\begin{table}
\caption{Single-meson operators used in the $a_0(980)$ and $\kappa$ channels. These operators
 annihilate states of total zero momentum. The particle names refer to flavor content only, as
 described in Table~\ref{tab:mesonflav}. The spatial and spin configuration of each operator is
 indicated by its superscript. The letters in the superscripts have been described in
 Sec.~\ref{sec:mesonops}, and the final number is an index that differentiates the different
 operators of the same spatial configuration.
\label{tab:single-meson-ops}}
\begin{ruledtabular}
\begin{tabular}{cc}
   $A_{1g}^-,\ I=1,\ S=0$ & $A_{1g},\ I=\frac{1}{2},\ S=1$\\
   \hline\\[-2mm]
   $\pi_{A_{1g}^-}^{SD2}$ & $K_{A_{1g}}^{DDL2}$ \\
   $\pi_{A_{1g}^-}^{TDO3}$ & $K_{A_{1g}}^{TDO3}$ \\
   & $K_{A_{1g}}^{TDU5}$
\end{tabular}
\end{ruledtabular}
\end{table}

In Tables~\ref{tab:single-meson-ops} and \ref{tab:two-meson-ops}, the spatial and spin configuration
of each single-meson operator is indicated by its superscript.  The letters in the superscripts have
been previously described in Sec.~\ref{sec:mesonops}, and the final integer in each superscript is
an index that differentiates the different single-meson operators of the same spatial/spin
configuration.  The complete details of our single-meson and two-meson operators are available upon
request.

The single-meson operators listed in Table~\ref{tab:single-meson-ops} are selected based on a
low-statistics Monte Carlo computation.  In each symmetry channel, the correlation matrix for a
fairly large number of single-meson operators is evaluated and the analysis procedure described in
Sec.~\ref{sec:temporal} is applied to obtain the $\vert Z_j^{(n)}\vert$ overlap factor magnitudes for all of the
operators. From this correlation matrix, a half dozen or so operators which produce a state with the
largest overlaps onto the ground state in each channel are identified and retained.  The correlation
matrix of the retained single-meson operators and all of the two-meson operators in
Table~\ref{tab:two-meson-ops} in each symmetry channel is then evaluated and examined.  In each
symmetry channel, some of the single-meson operators lead to an ill-conditioned correlator matrix,
suggesting a lack of linear independence, or introduce an excessive amount of statistical noise. 
These operators are removed to produce our final operator set. The remaining single-meson operators
are those listed in Table~\ref{tab:single-meson-ops}.  In the $a_0(980)$ channel, we are left with a
singly-displaced (SD) and a triply-displaced orthogonal (TDO) operator, and in the $\kappa$ channel,
a doubly-displaced-L (DDL), triply-displaced-orthogonal (TDO), and a triply-displaced-U (TDU)
operator survive the selection process.

\begin{table}
\caption{Two-meson operators used in the $a_0(980)$ and $\kappa$ channels. These operators
 annihilate states of total zero momentum. The particle names refer to flavor content only,
 as described in Table~\ref{tab:mesonflav}. The spatial and spin configuration of each operator
 is indicated by its superscript. The letters in the superscripts have been described in
 Sec.~\ref{sec:mesonops}, and the final number is an index that differentiates the different
 operators of the same spatial configuration. The number in parentheses following a particle
 name denotes $\bm{n_p}^2$, related to the square of its total momentum by
 $\bm{p}^2=(2\pi/L)^2 \bm{n_p}^2$ for spatial volume $L^3$, and the subscript denotes its
 little group irrep.
\label{tab:two-meson-ops}}
\begin{ruledtabular}
\begin{tabular}{cc}
  $A_{1g}^-,\ I=1,\ S=0$ & $A_{1g},\ I=\frac{1}{2},\ S=1$\\
  \hline\\[-2mm]
  $K(0)_{A_{1u}}^{SS0}\overline K(0)_{A_{1u}}^{SS0}$  & $K(0)_{A_{1u}}^{SS0}\pi(0)_{A_{1u}^-}^{SS0}$\\
  $K(1)_{A_2}^{SS1}\overline K(1)_{A_2}^{SS1}$        & $K(0)_{A_{1u}}^{SS0}\etaop(0)_{A_{1u}^+}^{SS0}$\\
  $K(2)_{A_2}^{SS0}\overline K(2)_{A_2}^{SS0}$        & $K(0)_{A_{1u}}^{SS0}\phiop(0)_{A_{1u}^+}^{SS0}$\\
  $K(2)_{A_2}^{SS1}\overline K(2)_{A_2}^{SS1}$        & $K(1)_{A_2}^{SS1}\pi(1)_{A_2^-}^{SS1}$\\
  $\etaop(0)_{A_{1u}^+}^{SS0}\pi(0)_{A_{1u}^-}^{SS0}$ & $K(1)_{A_2}^{SS1}\etaop(1)_{A_2^+}^{SS1}$\\
  $\etaop(1)_{A_2^+}^{SS0}\pi(1)_{A_2^-}^{SS0}$       & $K(1)_{A_2}^{SS1}\phiop(1)_{A_2^+}^{SS1}$\\
  $\etaop(2)_{A_2^+}^{SS1}\pi(2)_{A_2^-}^{SS1}$       & $K(2)_{A_2}^{SS0}\pi(2)_{A_2^-}^{SS0}$\\
  $\phiop(0)_{A_{1u}^+}^{SS0}\pi(0)_{A_{1u}^-}^{SS0}$ & $K(3)_{A_2}^{SS0}\pi(3)_{A_2^-}^{SS0}$ \\
  $\phiop(1)_{A_2^+}^{SS1}\pi(1)_{A_2^-}^{SS1}$       & \\
  $\phiop(2)_{A_2^+}^{SS0}\pi(2)_{A_2^-}^{SS0}$       &
\end{tabular}
\end{ruledtabular}
\end{table}

\begin{table}
\caption{The tetraquark operators selected in the $a_0(980)$ and $\kappa$ channels. These operators
 annihilate states of total zero momentum and are both single-site opeators, as indicated by the
 superscripts. The final number in the superscript is an index that differentiates the different
 operators of the same spatial configuration, and the $\pm$ in parentheses indicates the color
 structure.
\label{tab:tetraq}}
\begin{ruledtabular}
\begin{tabular}{cc}
  $A_{1g}^-,\ I=1,\ S=0$ & $A_{1g},\ I=\frac{1}{2},\ S=1$\\
  \hline\\[-2mm]
  $T[(\overline{u}u+\overline{d}d)\overline{d}u]_{A_{1g}^-}^{SS2(+)}$ 
  & $T[\overline{s}u\overline{s}s]_{A_{1g}}^{SS2(-)}$
\end{tabular}
\end{ruledtabular}
\end{table}

\begin{figure*}
\begin{center}
\includegraphics[width=0.32\textwidth]{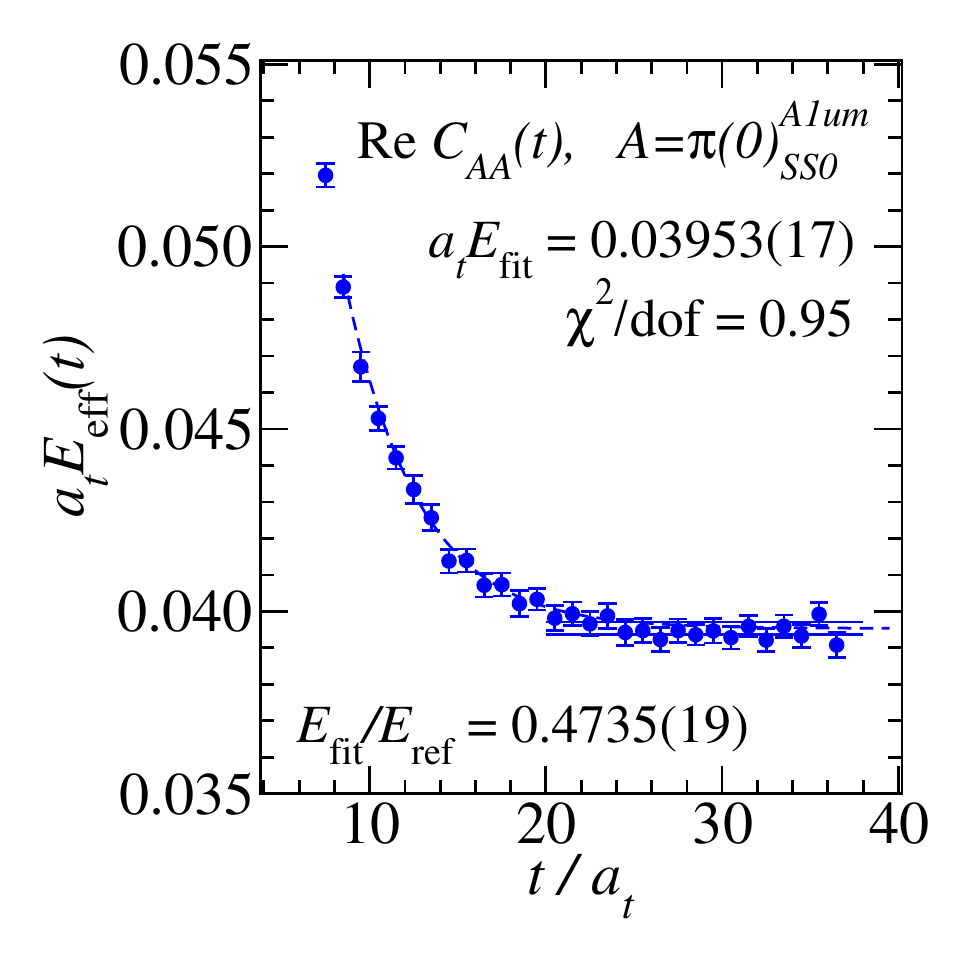}
\includegraphics[width=0.32\textwidth]{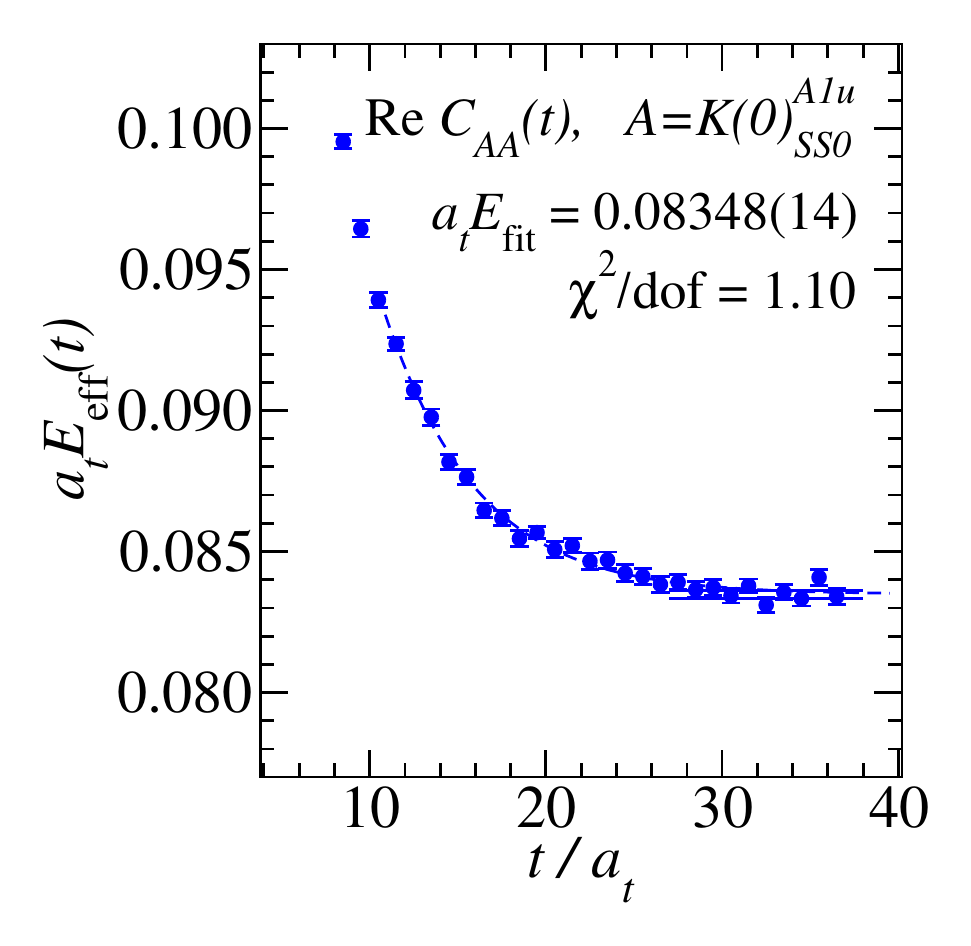}
\includegraphics[width=0.31\textwidth]{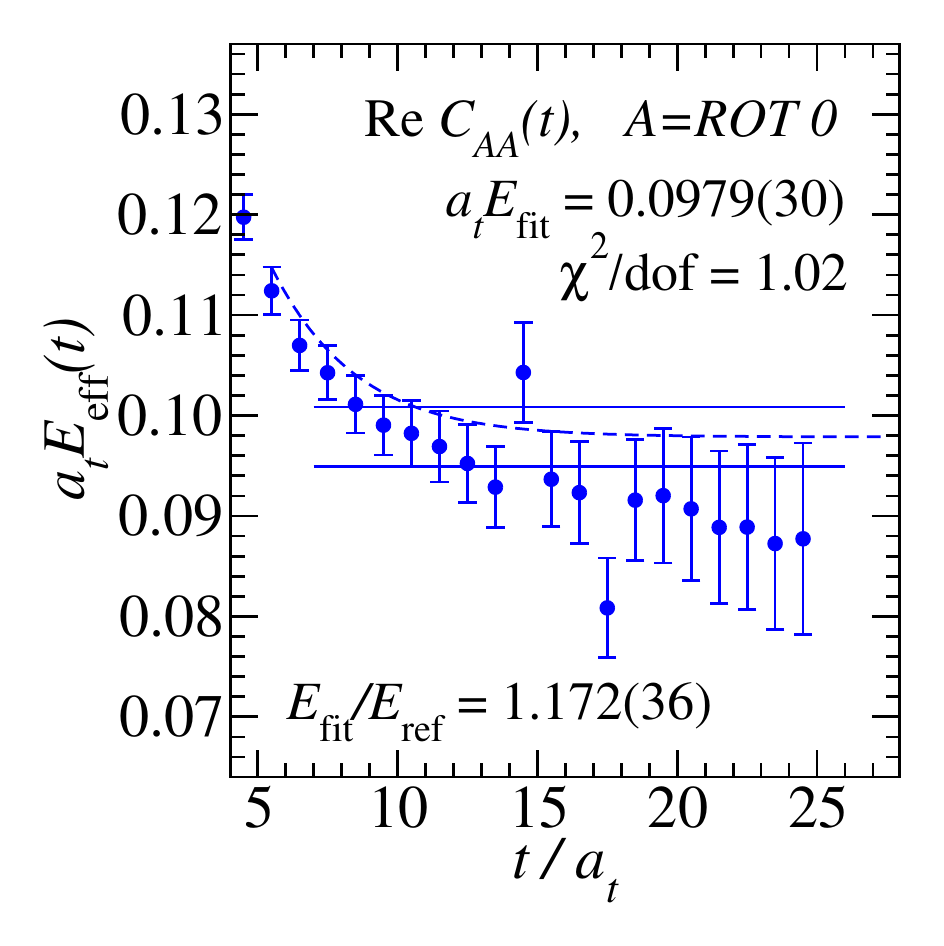}
\end{center}
\vspace*{-5mm}
\caption{Effective energies for the at-rest pion (left), kaon (center), and $\eta$ meson (right).
 The symbols with error bars indicate the Monte Carlo estimates of the effective energy functions,
 and fitting information is provided by the dashed and solid blue curves, which are explained in
 the text. Fit results for the dimensionless products of the temporal lattice spacing $a_t$ times
 the meson masses are shown, along with fit quality $\chi^2/N_{\rm dof}$.  Fits to the correlators
 are done using a sum of two exponentials as the fit model. The vertical displacements of the
 effective energy points for the $\eta$ meson at time separations 14.5~$a_t$ and 17.5~$a_t$ is
 caused by using an interlace-16 time dilution scheme for the disconnected contributions.  Energies
 were also obtained for these mesons at a variety of different momenta.  The operators used to
 obtain the single-meson masses are discussed in the text in Sec.~\ref{sec:singlemesons}. The
 reference energy $E_{\rm ref}$ is taken to be the mass of the kaon.
\label{fig:refmasses}}
\end{figure*}

With these reduced operator sets (Tables~\ref{tab:single-meson-ops} and \ref{tab:two-meson-ops}),
good quality fits to the diagonal elements of the rotated correlators are obtained in each channel
following the procedure described in Sec.~\ref{sec:temporal}.  The important question we address
in this work is: have any energy levels been missed which are important for revealing the
$a_0(980)$ and $\kappa$ resonances?

To obtain information about the $a_0(980)$ and $\kappa$ resonances in lattice QCD, the elements
of the infinite-volume scattering $K$ matrix must first be parametrized, then the best-fit values
of the parameters are obtained using a well-known quantization 
condition\cite{Luscher:1990ux,Rummukainen:1995vs,Kim:2005gf,Briceno:2014oea} involving the
finite-volume spectrum of states obtained in lattice QCD.  The masses and widths of the resonances
follow from the details of the $K$ matrix.  To apply this so-called L\"uscher method reliably, the
energies of \textit{all} of the finite-volume eigenstates with some overlap onto the two-particle
scattered states are required in the energy range near the energy of the resonance.  If an energy
is missed, the quantization condition could lead to an incorrect determination of the $K$ matrix.

Although tetraquark operators are similar to two-meson operators, they are different enough that
the possibility exists of missing an eigenstate by not including them. To test this, we designed
hundreds of tetraquark operators as described in Sec.~\ref{sec:tetrops} with differing flavor
structures, color structures, and displacements.  Of course, it is not feasible to include
hundreds of operators all at once in a correlation matrix.  Our approach for selecting the best
few tetraquark operators to include proceeds as follows. 

First, the diagonal correlators for each of the hundreds of tetraquark operators are evaluated 
in an initial low-statistics computation on 25 gauge configurations.  The effective energies 
corresponding to most of these diagonal correlators show very large statistical errors and/or are
very high-lying.  The tetraquark operators associated with these very noisy and/or high-lying
effective energies are discarded from consideration, leaving
60 operators in the $\kappa$ channel, and 32 in the $a_0(980)$ channel.

Next, we add, one by one, each of the remaining tetraquark operators
to the basis of single-meson and two-meson operators listed in
Tables~\ref{tab:single-meson-ops} and \ref{tab:two-meson-ops}, evaluate the correlation matrix in
each channel using 25 gauge configurations,
apply the analysis procedure described in Sec.~\ref{sec:temporal} using $\tau_N=3,\, \tau_0=4,\, \tau_D=7$,
and carry out appropriate fits to the diagonal elements of the rotated correlator matrix 
$\widetilde{C}(t)$ of Eq.~(\ref{eq:rotatedcorr}).  For the majority of the tetraquark 
operators, the spectrum determinations of low-lying states obtained by this procedure do not change 
significantly.  However, for a few similarly-structured tetraquark operators in each 
symmetry channel, some substantial changes in the low-lying energy
extractions are observed.  

In the $\kappa$ channel, the third and fourth lowest-lying energy extractions are 
substantially lowered for a few operators, while in the $a_0(980)$ channel, a lowering of the 
lowest and third lowest-lying energies is seen for a few operators, with some movements of other 
energies. In the $a_0(980)$ channel, we find that only tetraquark operators with the $(\overline u u
+\overline d d)\overline d u$ flavor structure yield a significant change in the spectrum
extraction of the low-lying finite-volume states, while in the $\kappa$ channel, we find that only
tetraquark operators with the $\overline s u \overline s s$ flavor structure yield substantial 
modifications of the energy determinations. The operators that produce the substantial changes
in each channel include both color structures and a variety of spatial configurations.  In each symmetry
channel, the operator yielding either the largest lowering of energy extractions or the effective 
energy with the smallest statistical errors and/or least amount of excited-state contamination is 
identified and selected as the best tetraquark operator to utilize.
The chosen best tetraquark operator in each symmetry channel is listed in Table~\ref{tab:tetraq}.
A color symmetric operator was chosen in the $a_0(980)$ sector, while a color antisymmetric
operator was selected in the $\kappa$ channel.  Further details about these operators, such as
the exact definition of the single-site operator designated by an SS2 superscript, are available
from the authors upon request. Note that the choice of the so-called best operator is
not likely to be crucial, as long as one of the few operators producing the substantial
changes is employed. Further details about our tetraquark operator selection procedure are
presented in the Supplemental Material\cite{suppl}.

This procedure is then repeated.  Each of the 59 $\kappa$-channel 
and 31 $a_0(980)$-channel tetraquark operators that are not
selected in the first step above are individually, one by one, added to the basis of operators
listed in Tables~\ref{tab:single-meson-ops}, \ref{tab:two-meson-ops}, and \ref{tab:tetraq},
the correlation matrix in each channel is evaluated using 25 gauge configurations, the
analysis procedure described in Sec.~\ref{sec:temporal} is again applied, and appropriate fits
to the diagonal elements of the rotated correlator matrix $\widetilde{C}(t)$ of
Eq.~(\ref{eq:rotatedcorr}) are carried out to see if any further changes occur to the
low-lying spectra.  In this second step, no new substantial
changes to any of the low-lying energy extractions
are found.  For this reason, no additional tetraquark operators are included in our basis
of interpolating operators.

Two-meson operators having larger relative momenta are also included in a low-statistics run
to see if such operators could have any effect on the low-lying spectrum determination, and no such effect
is observed.  Given this, we use the operators described in Tables~\ref{tab:single-meson-ops},
\ref{tab:two-meson-ops}, and \ref{tab:tetraq} for the higher-statistics computations.

\section{Finite-volume spectra}
\label{sec:fvspec}

Our results for the lowest-lying energies of the stationary states in finite volume are
presented and discussed in this section, with details of how our results are obtained. 
Energies are determined both as dimensionless products with the temporal lattice spacing
$a_t$ and as dimensionless ratios over a reference energy $E_{\rm ref}$, which we take to
be the kaon mass $E_{\rm ref}=m_K$. After discussing the determinations of the single-meson
masses and energies in Sec.~\ref{sec:singlemesons}, results for the $\kappa$ channel are
presented first in Sec.~\ref{sec:kappa_results} since statistical errors are smaller in
this channel, then results for the $a_0(980)$ channel are discussed in
Sec.~\ref{sec:a0_results}.

Below, so-called effective energy plots are used to illustrate some of our fits.  For a
given diagonal element $\widetilde{C}_{jj}(t)$ of a pivoted correlator matrix, the
effective energy is defined by
\beq
  a_t E_{\rm eff}^{(j)}(t) = \frac{1}{\Delta}\ln\left(\frac{\widetilde{C}_{jj}
  (t\!-\!\frac{1}{2}\Delta)}{\widetilde{C}_{jj}(t\!+\!\frac{1}{2}\Delta)}\right),
\label{eq:effenergy}
\eeq
where $\Delta=3a_t$ is used throughout this work, and the time separation
$t=\frac{5}{2}a_t,\frac{7}{2}a_t,\frac{9}{2}a_t,\dots$ takes half-integral multiples
of the temporal lattice spacing $a_t$.  The operator whose temporal correlations give
$\widetilde{C}_{jj}(t)$ typically creates many states when acting on the vacuum. 
Denote $E_{\rm lowest}^{(j)}$ as the lowest-lying energy of the states produced by
this operator, then as $t$ becomes large, the effective energy tends towards
$E_{\rm lowest}^{(j)}$.  Contributions from higher lying states are usually visible
for small values of $t$.

In Fig.~\ref{fig:refmasses}, as well as in all other effective energy plots shown
later in this work, the symbols with error bars indicate Monte Carlo estimates of
the effective energy function given by Eq.~(\ref{eq:effenergy}) associated with
pivoted temporal correlator $\widetilde{C}_{jj}(t)$.  To extract energies, fits
to these correlators using a model which is either a single exponential or a sum
of two exponentials are carried out. The effective energy reconstructed from the best
fit is shown as a dashed blue line in each effective energy plot, and this line is shown
only in the time separation range used in the fit.  Two horizontal solid blue lines
indicate the statistical uncertainty in the best fit value of the energy in each plot.

\subsection{Single-meson energies}
\label{sec:singlemesons}

Electromagnetic and weak nuclear effects are ignored in our study, and the masses of the
$u$ and $d$ quarks are taken to be the same so that isospin is a good symmetry.  With these
approximations, only decays by the strong force are permitted.  The pion and kaon are stable
since they are the lightest nonstrange and strange mesons.  Since the mass of the $\eta$
here is less than the energy of three pions at rest, the $\eta$ is a stable meson.  The
dominant decay mode of the $\eta^\prime$ meson is $\pi\pi\eta$, while its decay to three
pions is dramatically suppressed.  Given the suppression of the decay mode to three pions
and the fact that the $\eta^\prime$ mass here is less than sum of the masses of $\pi\pi\eta$,
the $\eta^\prime$ can approximated as stable.

Our results for the masses of these stable mesons are listed in Table~\ref{table:ensemble}.
Our determinations of the pion, kaon, and $\eta$ masses are illustrated in the effective
energy plots shown in Fig.~\ref{fig:refmasses}.  The pion and kaon energies are obtained
using single correlators involving only the simplest single-site pion and kaon operators,
respectively. Since our $\etaop$ and $\phiop$ operators actually have
$\overline{u}u+\overline{d}d$ and $\overline{s}s$ flavor structure, the $\eta$ and
$\eta^\prime$ energies are obtained using variationally-improved single-hadron operators,
which are linear combinations of the single-hadron operators determined by solving a GEVP
in the subspace created by the single-hadron $\etaop$ and $\phiop$ operators.  For each
momentum, two ``rotated" single-hadron operators are obtained in this way and are denoted
here by ROT~0 and ROT~1.  The operators named ROT~0 are used to obtain the $\eta$ energies,
and the ROT~1 operators are used to determine the $\eta^\prime$ energies.

\subsection{\boldmath Channel involving the $\kappa$ resonance}
\label{sec:kappa_results}

Results for the stationary-state energies in the isodoublet, strangeness 1,
zero-momentum $A_{1g}$ channel are presented in this section.  The results shown are
obtained using a normalization time, metric time, and diagonalization time in the GEVP
given by $\tau_N=3$, $\tau_0=6$, and $\tau_D=12$, respectively.

\begin{figure*}[pt]
  \raisebox{-.11cm}{\includegraphics[width=0.345\textwidth]{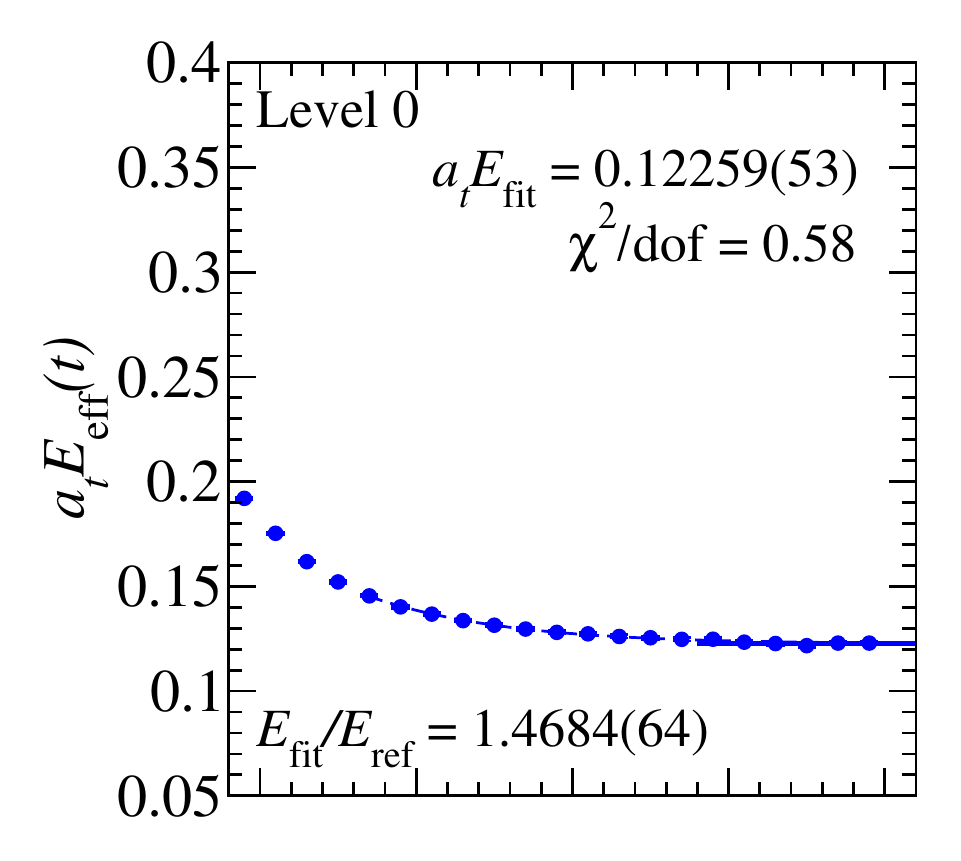}}
  \includegraphics[width=0.28\textwidth]{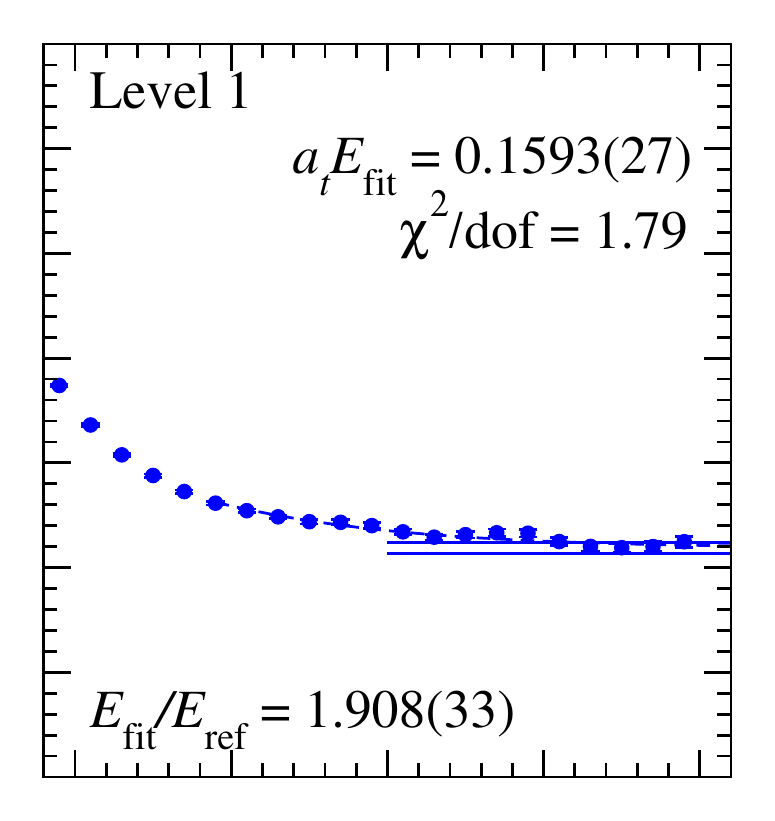}
  \includegraphics[width=0.28\textwidth]{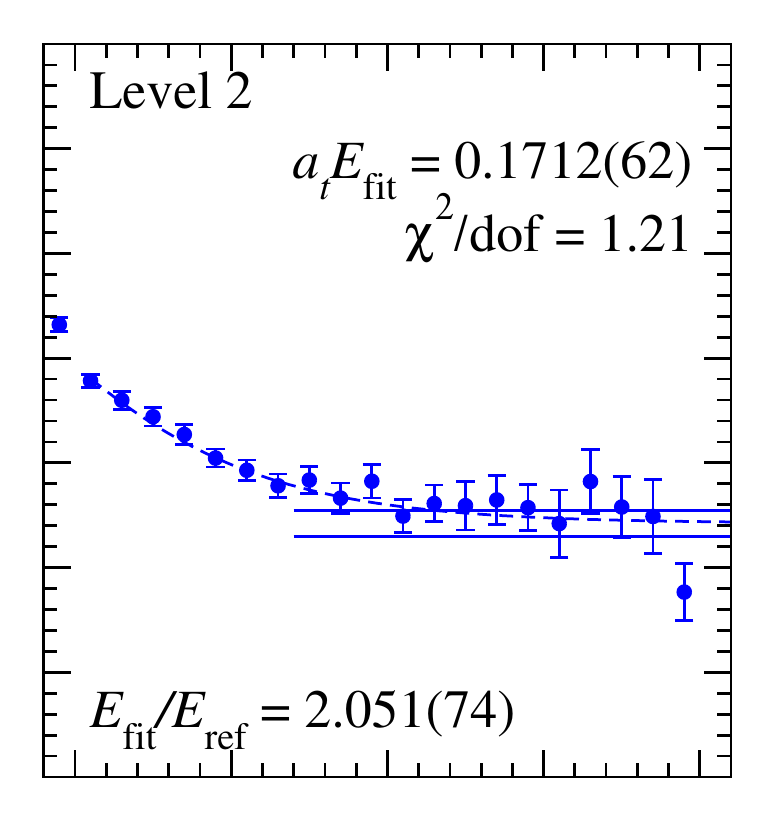}\\[-3mm]
  \raisebox{-.09cm}{\includegraphics[width=0.345\textwidth]{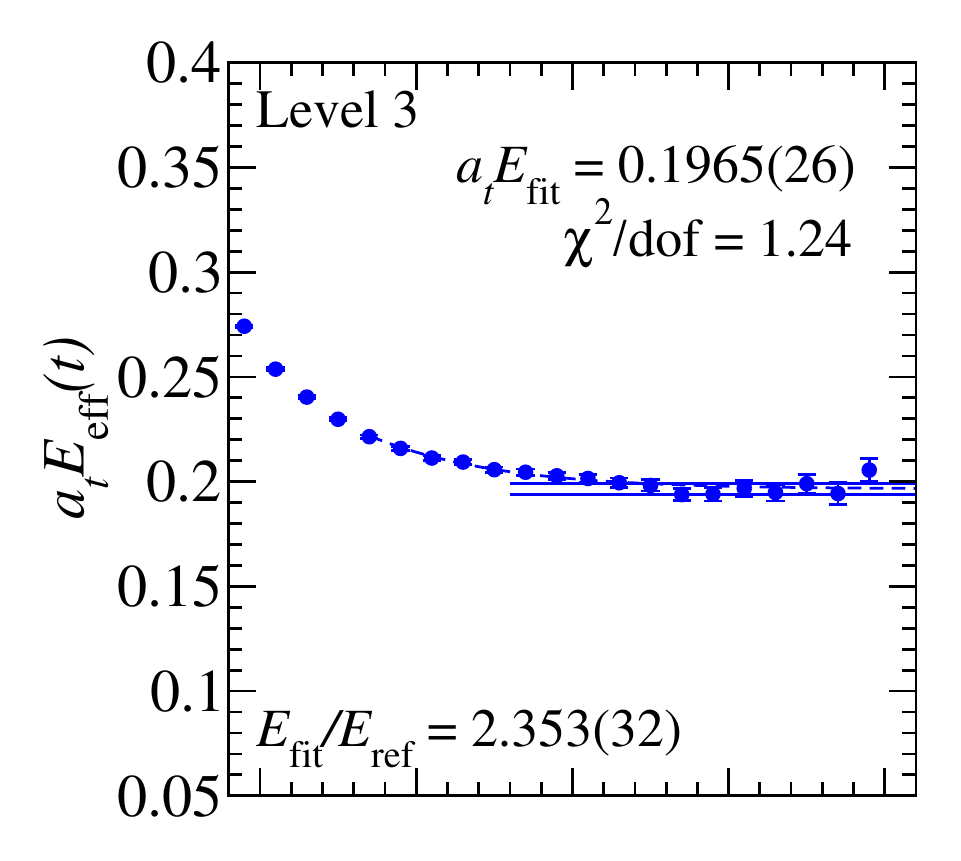}}
  \includegraphics[width=0.28\textwidth]{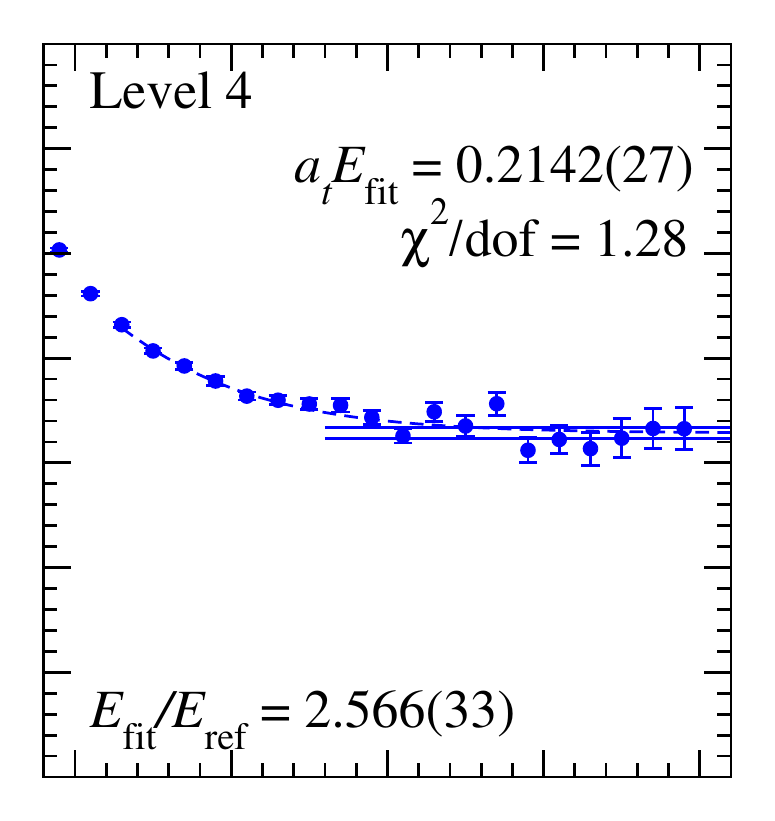}
  \includegraphics[width=0.28\textwidth]{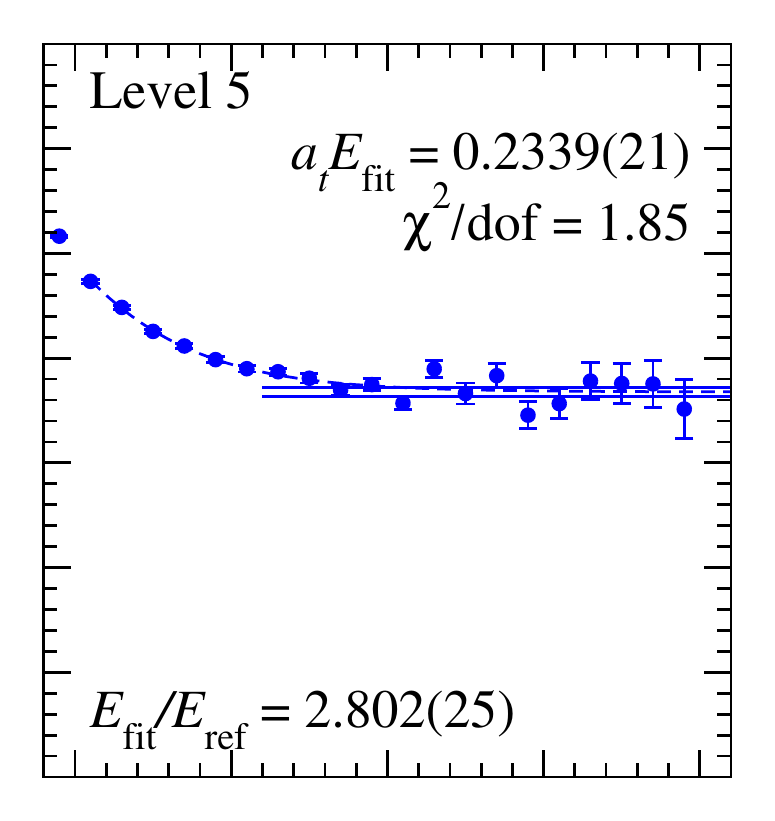}\\[-2mm]
  \par\noindent\rule{\textwidth}{0.5pt}
  \raisebox{-.11cm}{\includegraphics[width=0.345\textwidth]{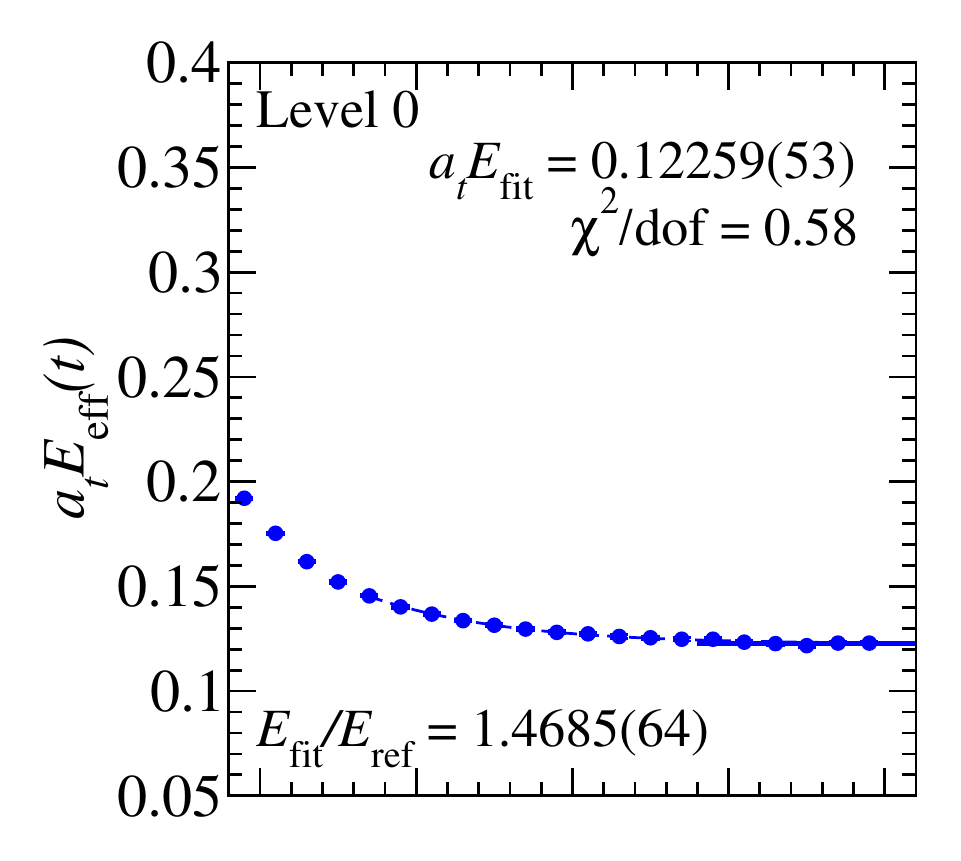}}
  \includegraphics[width=0.28\textwidth]{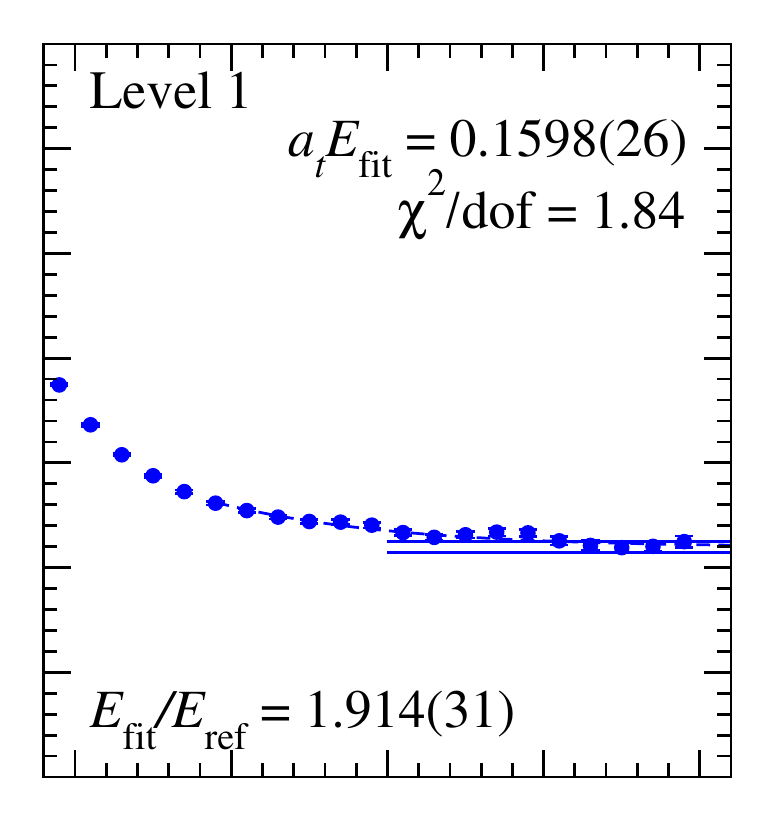}
  \includegraphics[width=0.28\textwidth]{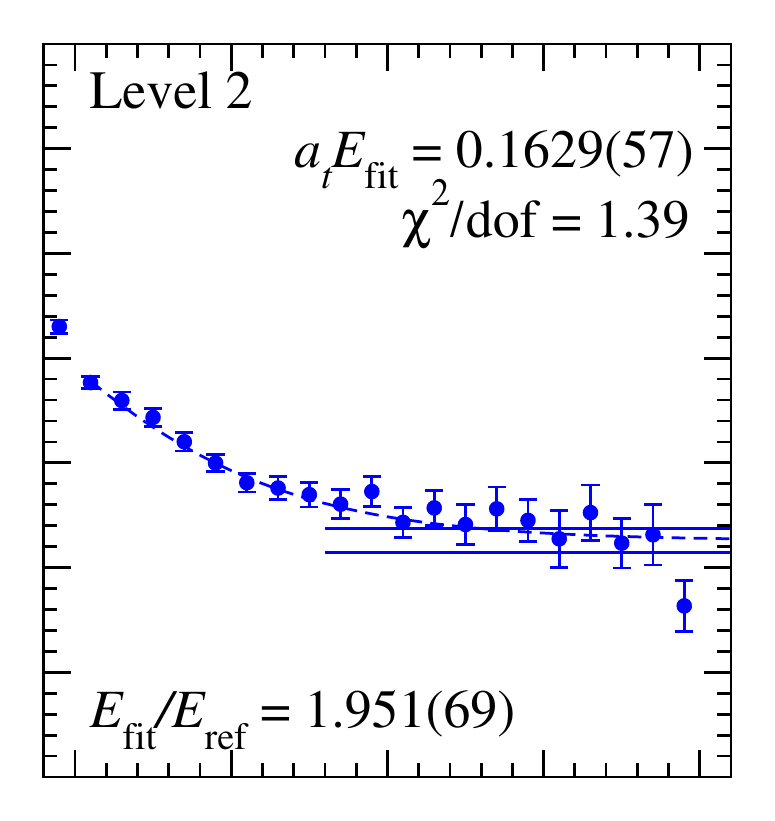}\\[-3mm]
  \raisebox{-0.00cm}{\includegraphics[width=0.345\textwidth]{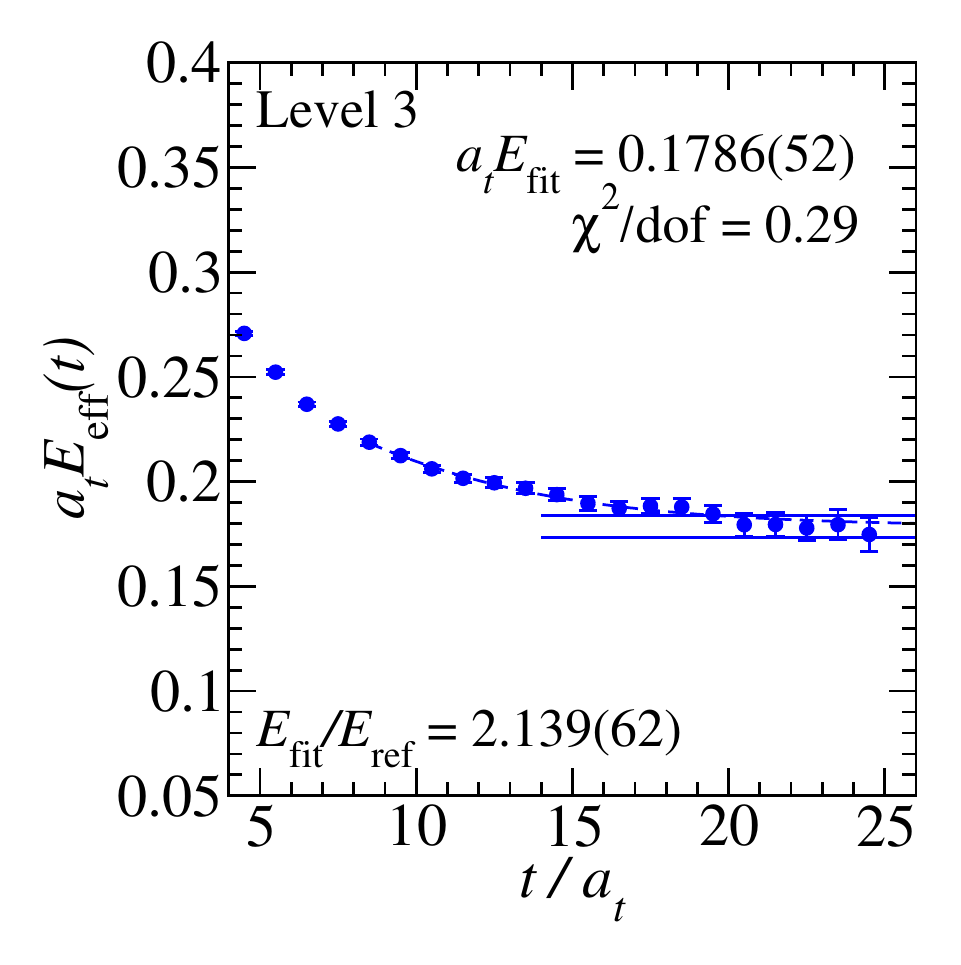}}
  \includegraphics[width=0.28\textwidth]{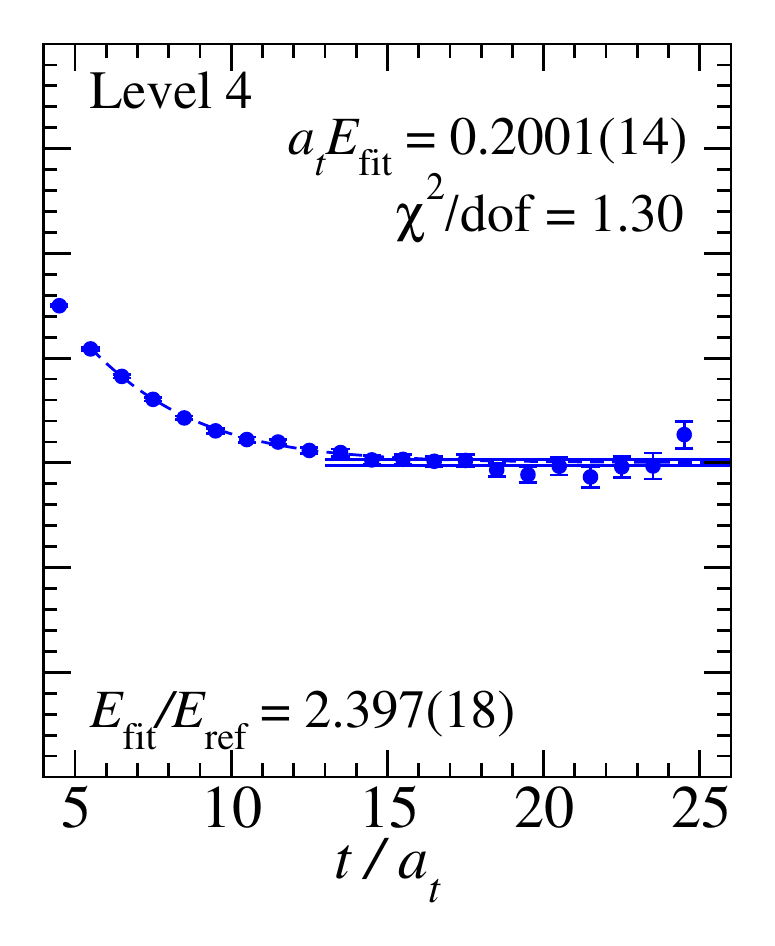}
  \includegraphics[width=0.28\textwidth]{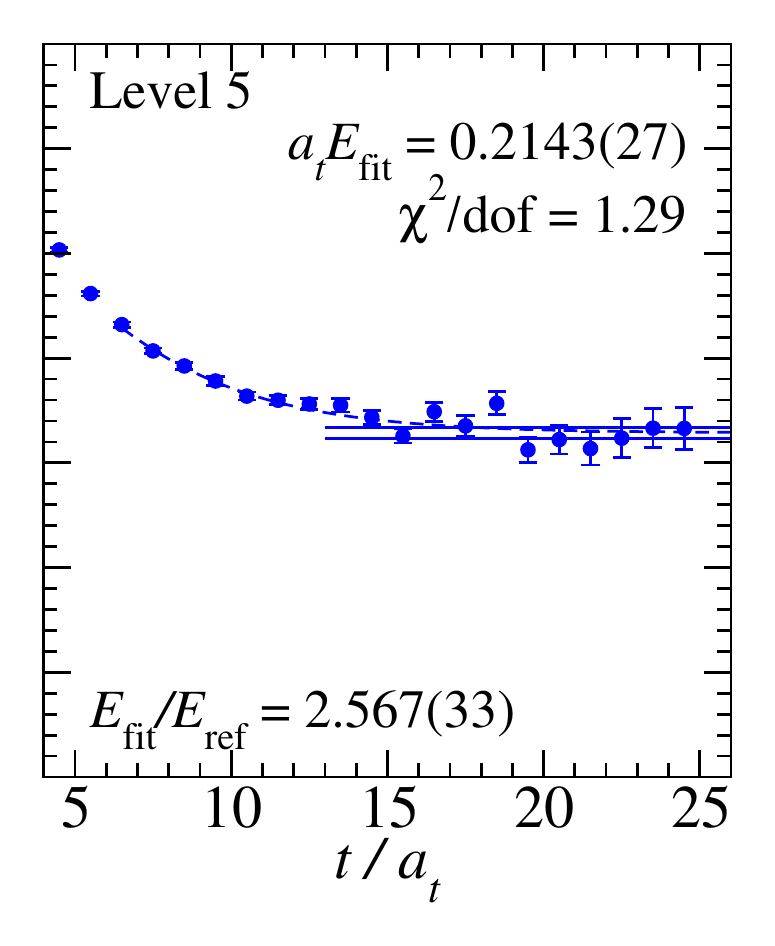}
\caption{Effective energies for the lowest six levels in the isodoublet, strangeness 1,
 zero-momentum, $A_{1g}$ channel. (Top two rows) Results obtained using only the operators
 in Tables~\ref{tab:single-meson-ops} and \ref{tab:two-meson-ops}, excluding the tetraquark
 operator.  (Bottom two rows) Results obtained using the operators in
 Tables~\ref{tab:single-meson-ops} and \ref{tab:two-meson-ops} and also including the
 tetraquark operator in Table~\ref{tab:tetraq}. Effective energy curves calculated from
 correlator fits are overlaid, and fit results are shown. Note that level 3 in the bottom
 row corresponds to the additional level found using the tetraquark operator.
\label{fig:kappa_fits}}
\end{figure*}

Fits to the diagonal elements of the correlator matrix $\widetilde{C}(t)$, obtained after
the single-pivot procedure described in Sec.~\ref{sec:temporal}, are shown as effective
energy plots in Fig.~\ref{fig:kappa_fits}.  In the top two rows are shown the fits which
yield the lowest six energies using only the single-meson and meson-meson operators (eleven
in total) listed in Tables~\ref{tab:single-meson-ops} and \ref{tab:two-meson-ops}, excluding
the tetraquark operator.  In the bottom two rows are presented the fits which yield the
lowest six energies using all twelve operators listed in Tables~\ref{tab:single-meson-ops}
and \ref{tab:two-meson-ops} and also including the tetraquark operator in Table~\ref{tab:tetraq}.
Only the fits for the six lowest-lying energies are shown, but higher energies were also
obtained. Fit details are given in Tables~\ref{tab:kappa_fit_details_no_tq} and
\ref{tab:kappa_fit_details_tq}.  Fits using a sum of two exponentials are mainly utilized,
but for a few of the higher-lying energies, the excited-state contamination at early times is
not sufficiently precise enough to allow the parameters of the second exponential to be
reliably determined, and so fits using only a single exponential are employed in these few
cases.  In each fit, only the correlator values for time separations in the range
$t_{\rm min}\leq t\leq t_{\rm max}$ are used. $t_{\rm{max}}$ is chosen to be the earliest
time at which the correlator value becomes statistically consistent with zero.
$t_{\rm{min}}$ is taken to be the earliest time at which the fit quality is reasonably
good and the best fit value of the energy is consistent with values obtained using larger
values of starting time.  Note that the fit qualities in
Tables~\ref{tab:kappa_fit_details_no_tq} and \ref{tab:kappa_fit_details_tq} are all
reasonably good.  The results obtained with only the meson and meson-meson operators
do not themselves give any hint of a missing energy level.

\begin{table}
\caption{Fit details for estimating the energies in the isodoublet strange zero-momentum
 $A_{1g}$ channel using the operator basis excluding the tetraquark operator. Energies $E$
 are listed in terms of the temporal lattice spacing $a_t$ and as ratios over the reference
 energy $E_{\rm ref}=m_K$.  Fits are done using either a two-exponential model (2-exp) or
 a single-exponential model (1-exp) to the diagonal elements of the $\widetilde{C}(t)$
 correlator matrix for times $t=t_{\rm min},\dots,t_{\rm max}$. Note that
 $\tau_{\rm min}=t_{\rm min}/a_t$ and $\tau_{\rm max}=t_{\rm max}/a_t$.
 The rightmost column gives the chi-square per degree of freedom of each best fit.
\label{tab:kappa_fit_details_no_tq}}
\begin{ruledtabular}
\begin{tabular}{llccl}
   $\ E / E_{\rm ref}$ & \quad $a_t E$ & Fit model & $(\tau_{\mathrm{min}},
   {\tau_\mathrm{max}})$ & $\!\!\!\!\chi^2 / N_{\rm dof}$\\
   \hline
   1.4684(64)&0.12259(53)&2{-}exp&$(7, 26)$&0.58\\
   1.908(33)&0.1593(27)&2{-}exp&$(8, 26)$&1.79\\
   2.051(74)&0.1712(62)&2{-}exp&$(4, 26)$&1.21\\
   2.353(32)&0.1965(26)&2{-}exp&$(7, 26)$&1.24\\
   2.566(33)&0.2142(27)&2{-}exp&$(5, 26)$&1.28\\
   2.802(25)&0.2339(21)&2{-}exp&$(4, 26)$&1.85\\
   2.924(92)&0.2441(76)&1{-}exp&$(11, 26)$&1.11\\
   2.95(18)&0.246(15)&1{-}exp&$(9, 19)$&0.97\\
   3.289(96)&0.2746(81)&2{-}exp&$(3, 26)$&1.49
\end{tabular}
\end{ruledtabular}
\end{table}

\begin{table}
\caption{Fit details for estimating the energies in the isodoublet strange zero-momentum
 $A_{1g}$ channel using the operator basis that includes the tetraquark operator.  See
 the caption of Table~\ref{tab:kappa_fit_details_no_tq} for further information about
 notation and headings.
\label{tab:kappa_fit_details_tq}}
\begin{ruledtabular}
\begin{tabular}{llccl}
   $\ E / E_{\rm ref}$ & \quad $a_t E$ & Fit model & $(\tau_{\mathrm{min}},
  {\tau_\mathrm{max}})$ & $\!\!\!\!\chi^2 / N_{\rm dof}$\\
   \hline
   1.4685(64)&0.12259(53)&2{-}exp&$(7, 26)$&0.58\\
   1.914(31)&0.1598(26)&2{-}exp&$(8, 26)$&1.84\\
   1.951(69)&0.1629(57)&2{-}exp&$(4, 26)$&1.39\\
   2.139(62)&0.1786(52)&2{-}exp&$(7, 26)$&0.29\\
   2.397(18)&0.2001(14)&2{-}exp&$(4, 26)$&1.3\\
   2.567(33)&0.2143(27)&2{-}exp&$(5, 26)$&1.29\\
   2.79(15)&0.233(13)&1{-}exp&$(11, 26)$&1.37\\
   2.797(24)&0.2335(20)&2{-}exp&$(4, 26)$&1.71\\
   2.82(16)&0.235(13)&1{-}exp&$(11, 26)$&1.04\\
   3.291(95)&0.2748(80)&2{-}exp&$(3, 26)$&1.5
\end{tabular}
\end{ruledtabular}
\end{table}

\begin{figure}
\centering
\includegraphics[width=0.95\columnwidth]{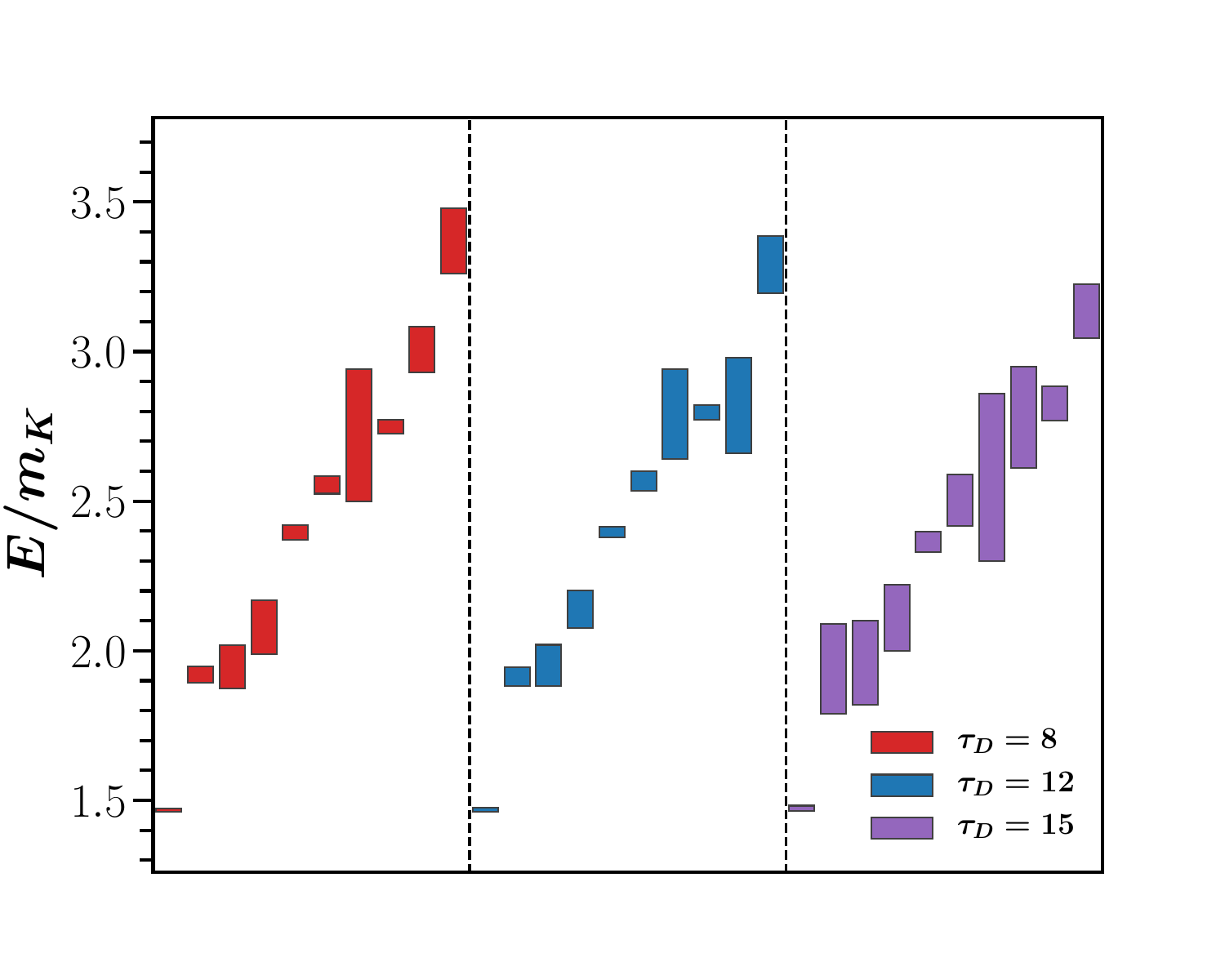}
\caption{Determinations of the spectrum in the isodoublet strange zero-momentum $A_{1g}$
 channel obtained using three different diagonalization times $\tau_D=8$ (left),
 $\tau_D=12$ (center), and $\tau_D=15$ (right) for the operator basis given in
 Tables~\ref{tab:single-meson-ops}, \ref{tab:two-meson-ops}, and \ref{tab:tetraq}, which
 includes the tetraquark operator. The metric times $\tau_0$ are chosen as $\tau_D/2$,
 rounded up.  Good agreement is found, but a significant increase
 in statistical uncertainties is evident for $\tau_D=15$.  
\label{fig:spectrum_td}}
\end{figure}

Our procedure for selecting the GEVP parameters $\tau_N, \tau_0,$ and $\tau_D$ is as follows. 
$\tau_N$ simply needs to be small.  The metric time $\tau_0$ is generally chosen to be
approximately $\tau_D/2$. In practice, we find results for the energies are very insensitive
to this parameter as long as it is not very small nor very large.  The diagonalization time
$\tau_D$ is the most important to choose well.  If too small a value is used, the
diagonalization may be contaminated by higher lying states and the $\widetilde{C}(t)$
correlator matrix does not remain diagonal within statistical errors for $t>t_D$, noting
that $t_D=\tau_D\,a_t$.  If too large a value is used, the diagonalization may be
contaminated by statistical noise, leading to larger uncertainties.  Our procedure is to
vary $\tau_D$ to find the intermediate window where errors are not overly large, all
off-diagonal elements of $\widetilde{C}(t)$ remain zero within statistical errors for all
$t>t_D$ that have reasonable uncertainties,  and the energy extractions are insensitive
to the choice of $\tau_D$.  We found that choosing $\tau_D=12$ yielded a good combination
of reduced noise and a sufficiently diagonal matrix $\widetilde{C}(t)$ for $t>t_D$.  To
ensure our results were insensitive to our choice of GEVP parameters within the window,
the spectrum was extracted for various choices and compared.  A comparison of the spectra
for three choices of $\tau_D$ is shown in Fig.~\ref{fig:spectrum_td}.

\begin{figure*}
\includegraphics[width=0.304\textwidth]{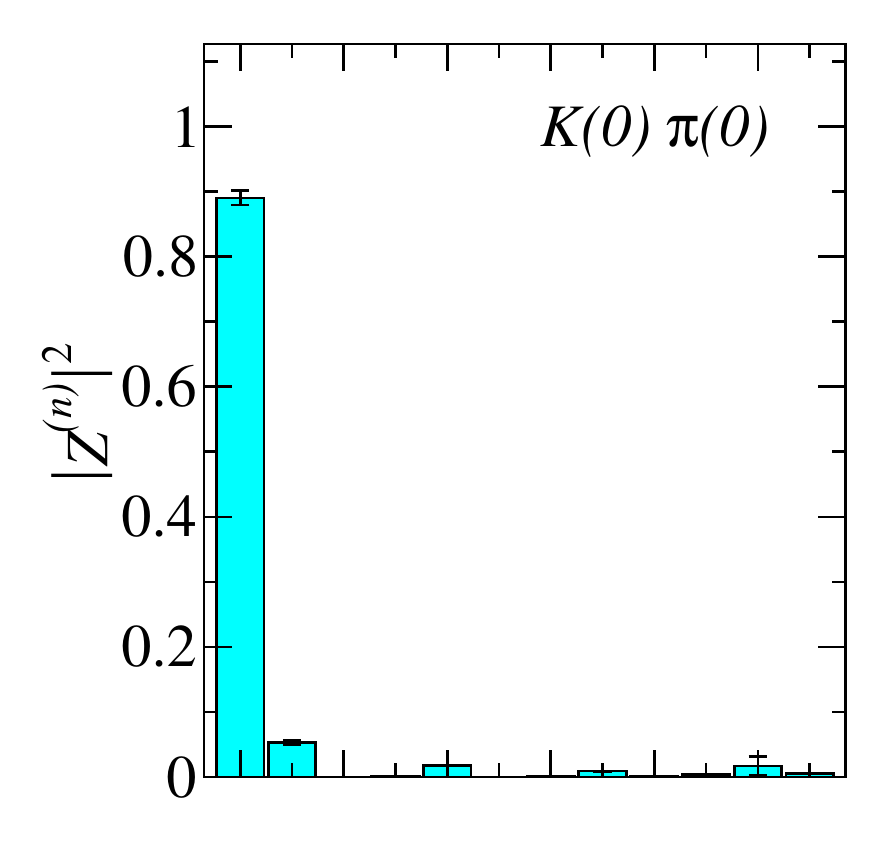}
\includegraphics[width=0.280\textwidth]{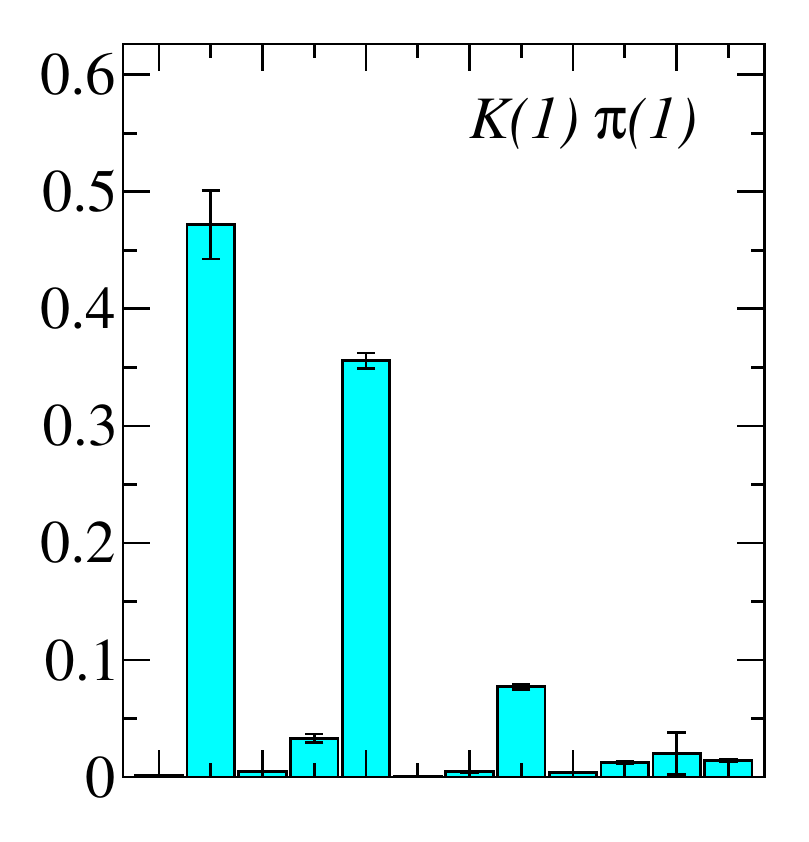}
\includegraphics[width=0.280\textwidth]{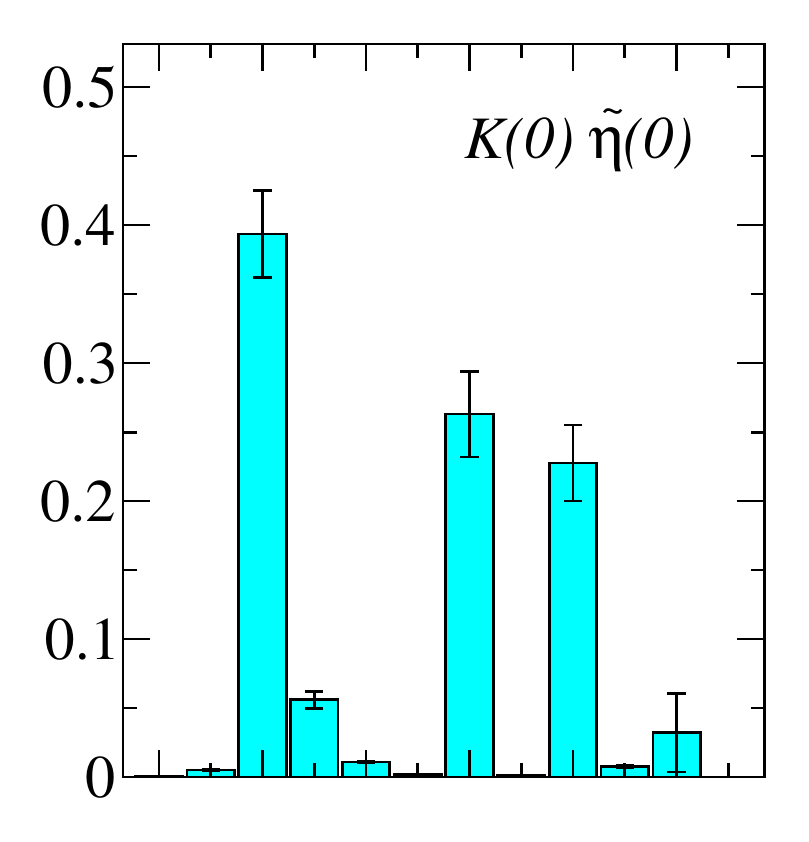}\\
\includegraphics[width=0.304\textwidth]{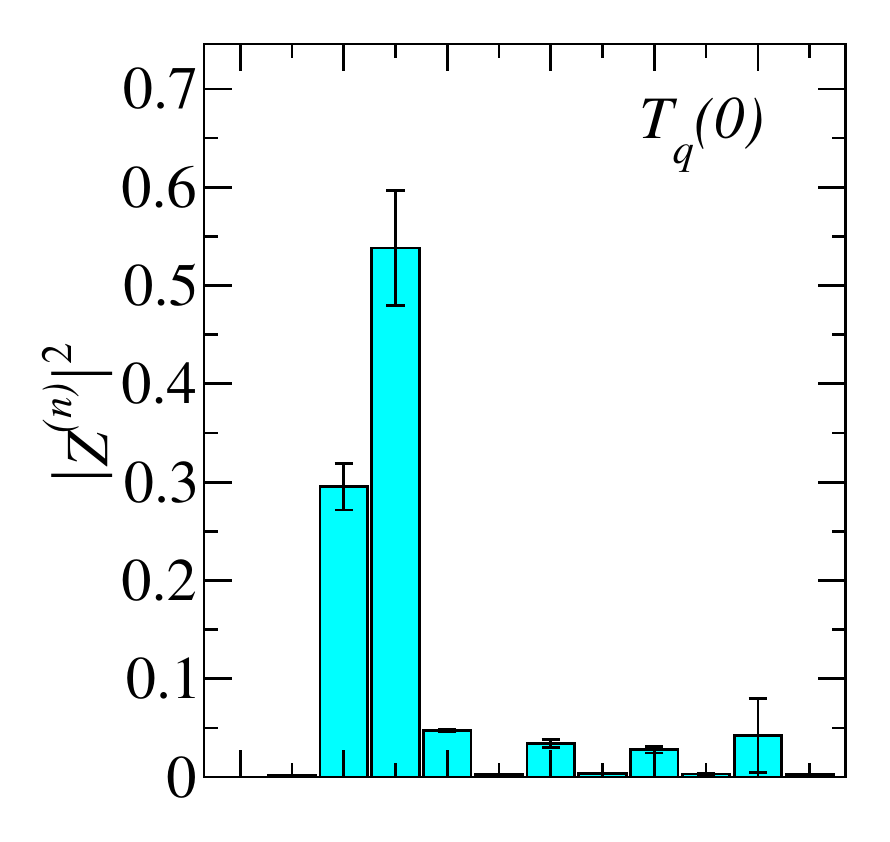}
\includegraphics[width=0.280\textwidth]{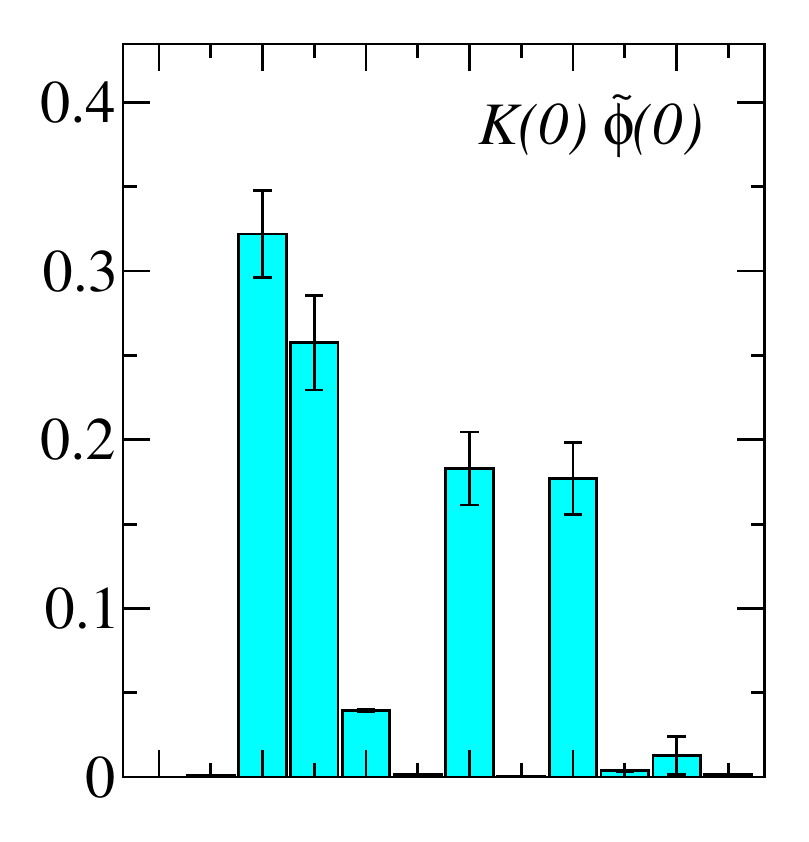}
\includegraphics[width=0.280\textwidth]{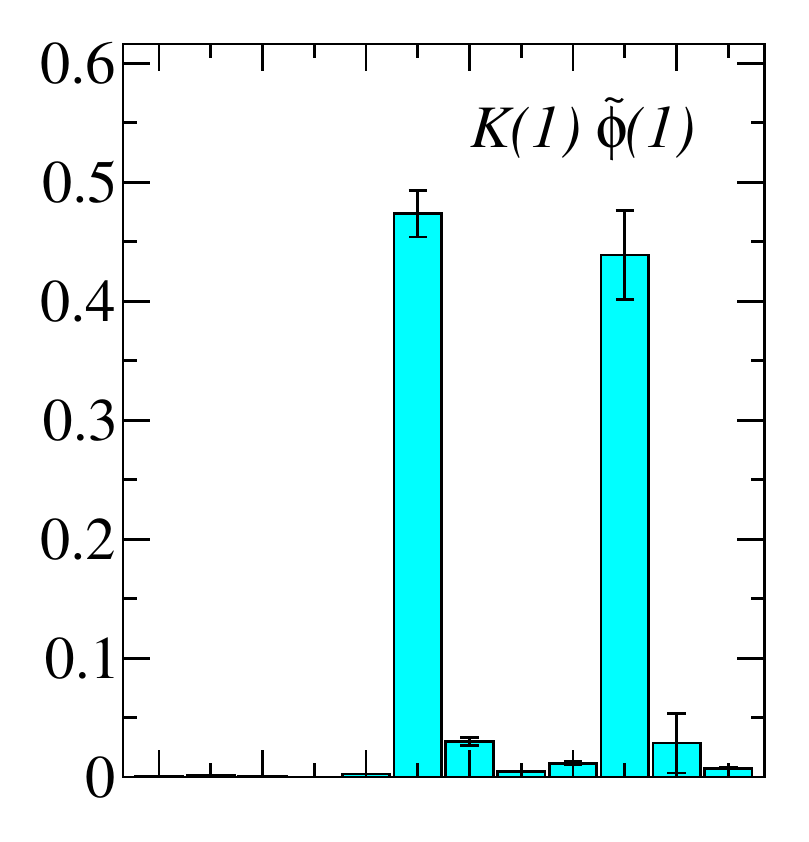}\\
\includegraphics[width=0.304\textwidth]{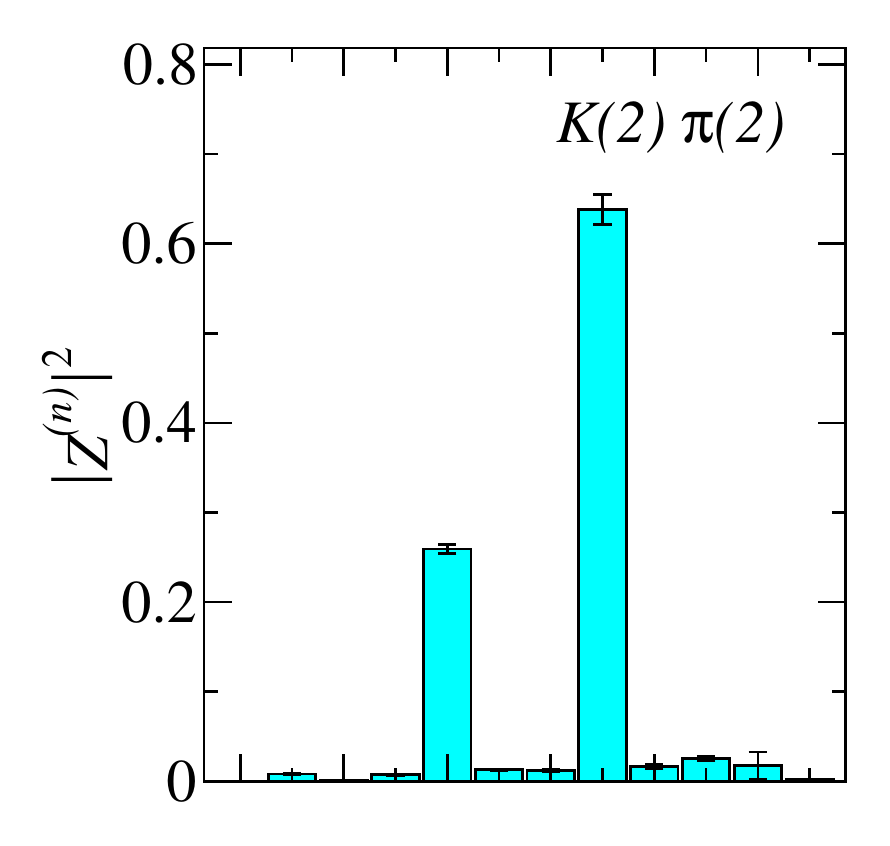}
\includegraphics[width=0.280\textwidth]{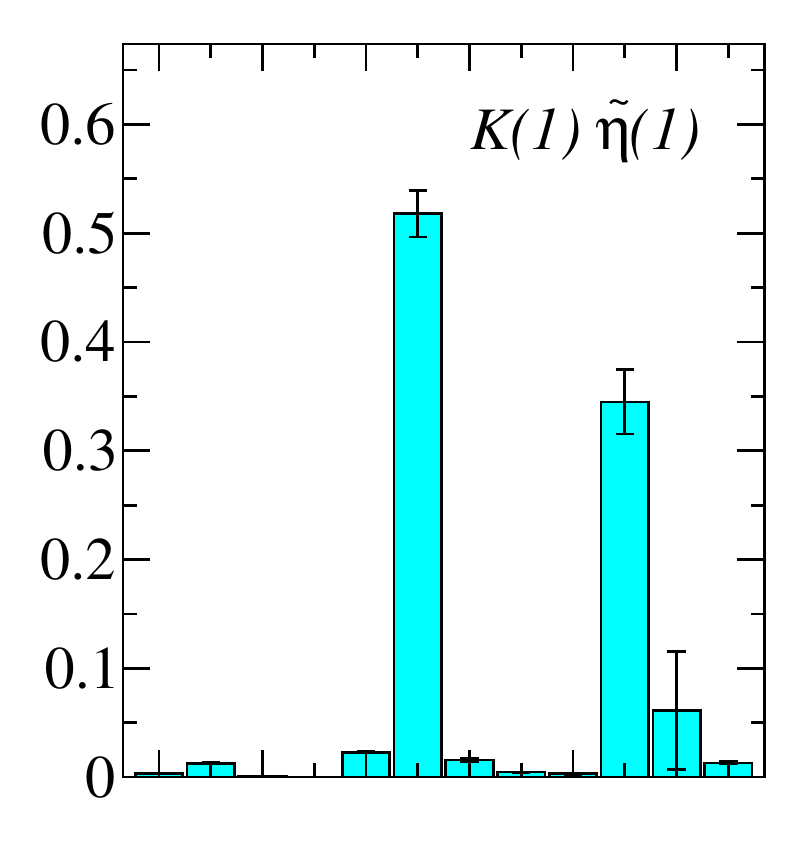}
\includegraphics[width=0.280\textwidth]{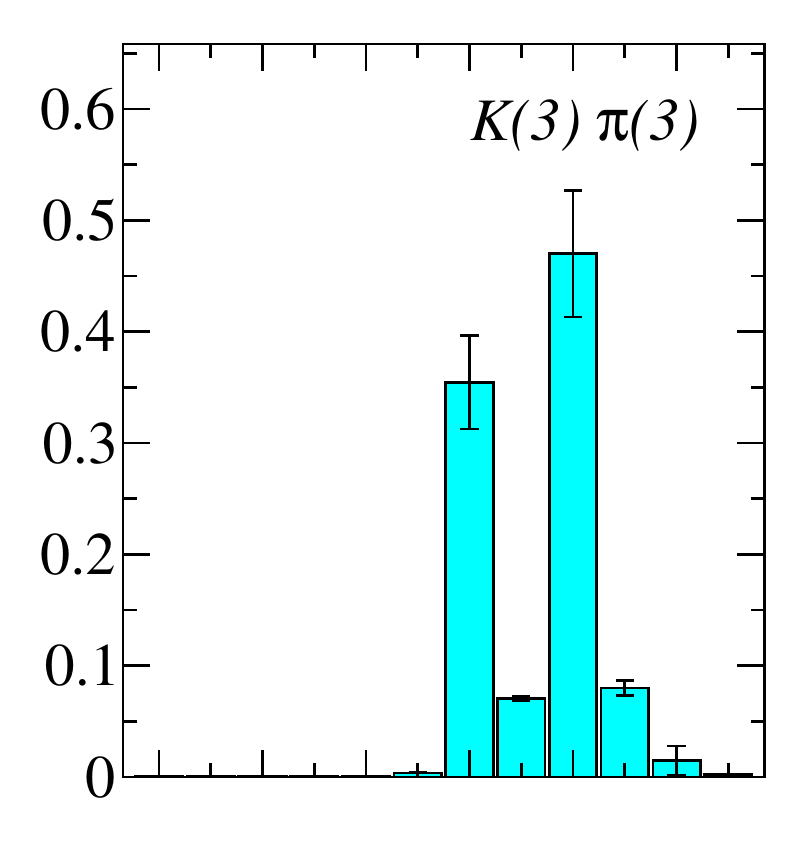}\\
\includegraphics[width=0.304\textwidth]{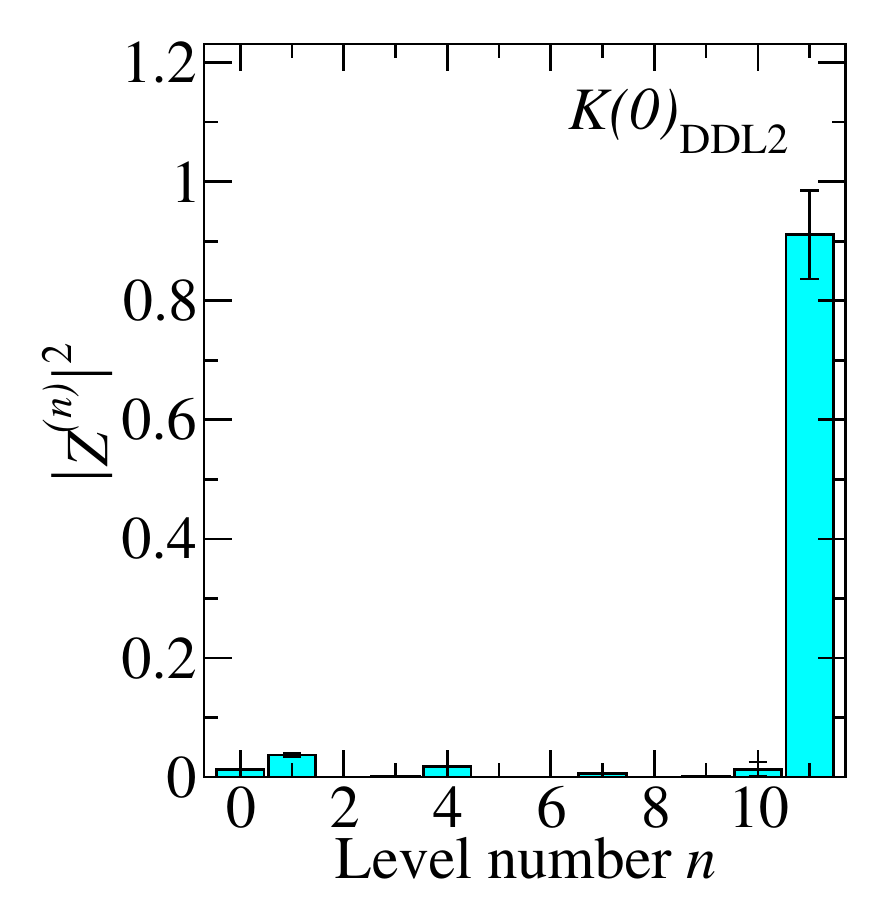}
\includegraphics[width=0.280\textwidth]{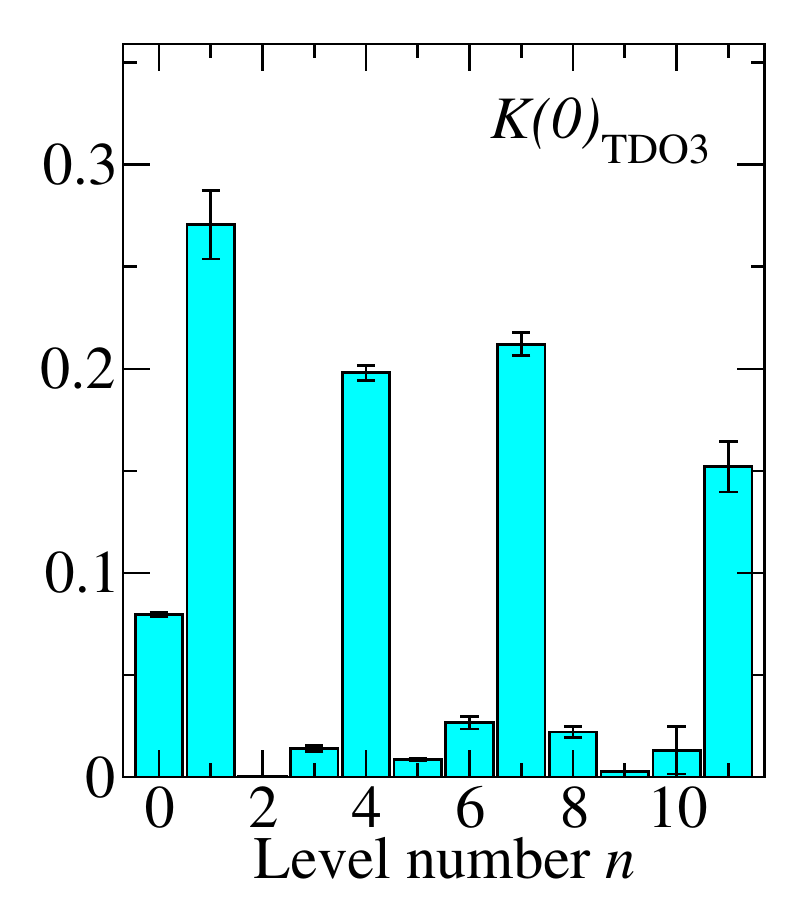}
\includegraphics[width=0.280\textwidth]{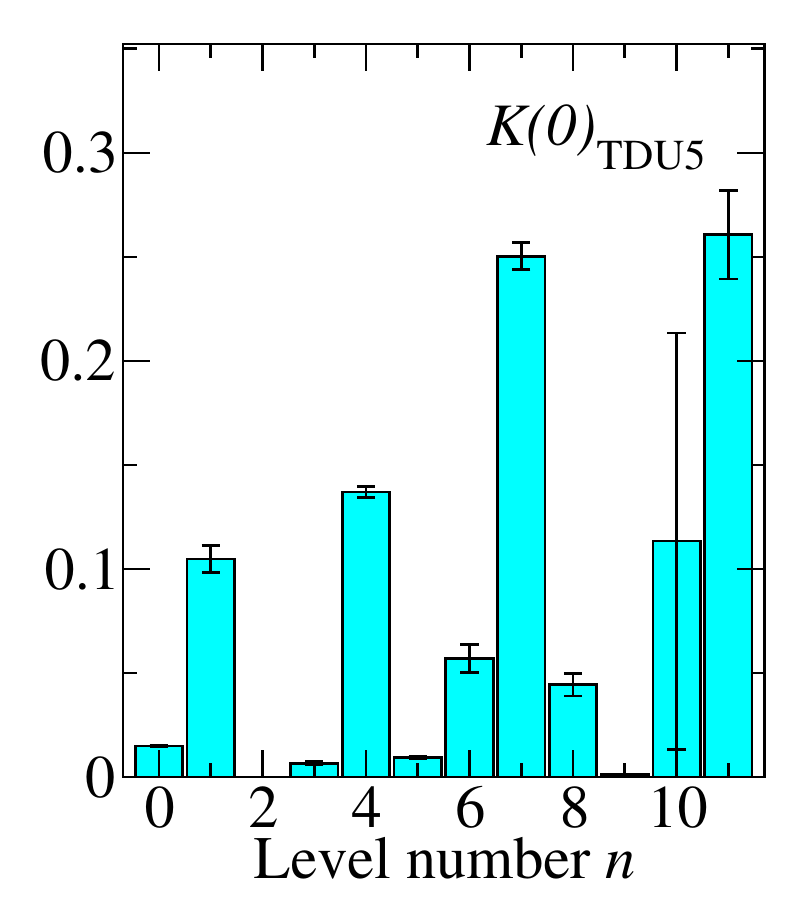}
\caption{Overlap factors $\vert Z^{(n)}_j\vert^2$ for the twelve operators used in the
 correlator matrix corresponding to the isodoublet strange zero-momentum $A_{1g}$ channel. 
 Each plot shows the factors for a single operator, with horizontal axis labelled by $n$,
 the index identifying the stationary states.  The factors for the tetraquark operator are
 shown in the left-most plot in the second row from the top.  Levels 3 and 2 have
 significant tetraquark, $K\etaop$, and $K\phiop$ content, where $\etaop$ means
 $\overline{u}u+\overline{d}d$ and $\phiop$ means $\overline{s}s$.
\label{fig:kappa_zfactors}}
\end{figure*}

In order to obtain some information about the structures of the stationary states, the
overlap factors $\vert Z^{(n)}_j\vert^2$ are determined for all twelve operators and are
shown in Fig.~\ref{fig:kappa_zfactors}.  Each plot shows the factors for a single operator,
with horizontal axis labelled by $n$, the index identifying the stationary states.  Within
each plot, the heights have been rescaled so that they add to unity, but keep in mind
that overlaps with levels for $n\geq 12$ have not been determined.

First, consider the lowest-lying level 0.  An inspection of the overlap plots in
Fig.~\ref{fig:kappa_zfactors} shows that this state is predominantly created by the $K(0)\pi(0)$
operator.  Hence, level 0 can be viewed as a state dominated by a pion at rest and a kaon
at rest with residual interactions.  Next, consider stationary state level 1.  Operators
$K(1)\pi(1)$, $K(0)_{\rm TDO3}$, and $K(0)_{\rm TDU5}$ produce states with non-negligible
overlaps with this level, with $K(1)\pi(1)$ producing the largest overlap factor.  Hence,
level 1 can be viewed as mainly a $K\pi$ with back-to-back minimal nonzero momenta and
residual interactions.  The $K(0)\etaop(0)$, $K(0)\phiop(0)$, and the tetraquark $T_q(0)$
operators create states with significant overlap onto level 2, and the $K(0)\phiop(0)$
and $T_q(0)$ operators create states with significant overlaps onto level 3.   Since the
state created by the tetraquark operator overlaps mainly with level 3, we qualitatively
identify level 2 as a mixed $K\etaop-K\phiop$ state (shown as an orange and blue box) and
level 3 predominantly as a tetraquark state with significant $K\phiop$ mixing (shown as
a hatched purple box). Again, keep in mind that for our operators,  $\etaop$ means
$\overline{u}u+\overline{d}d$ and $\phiop$ means $\overline{s}s$.

\begin{figure}
\centering
\includegraphics[width=\columnwidth]{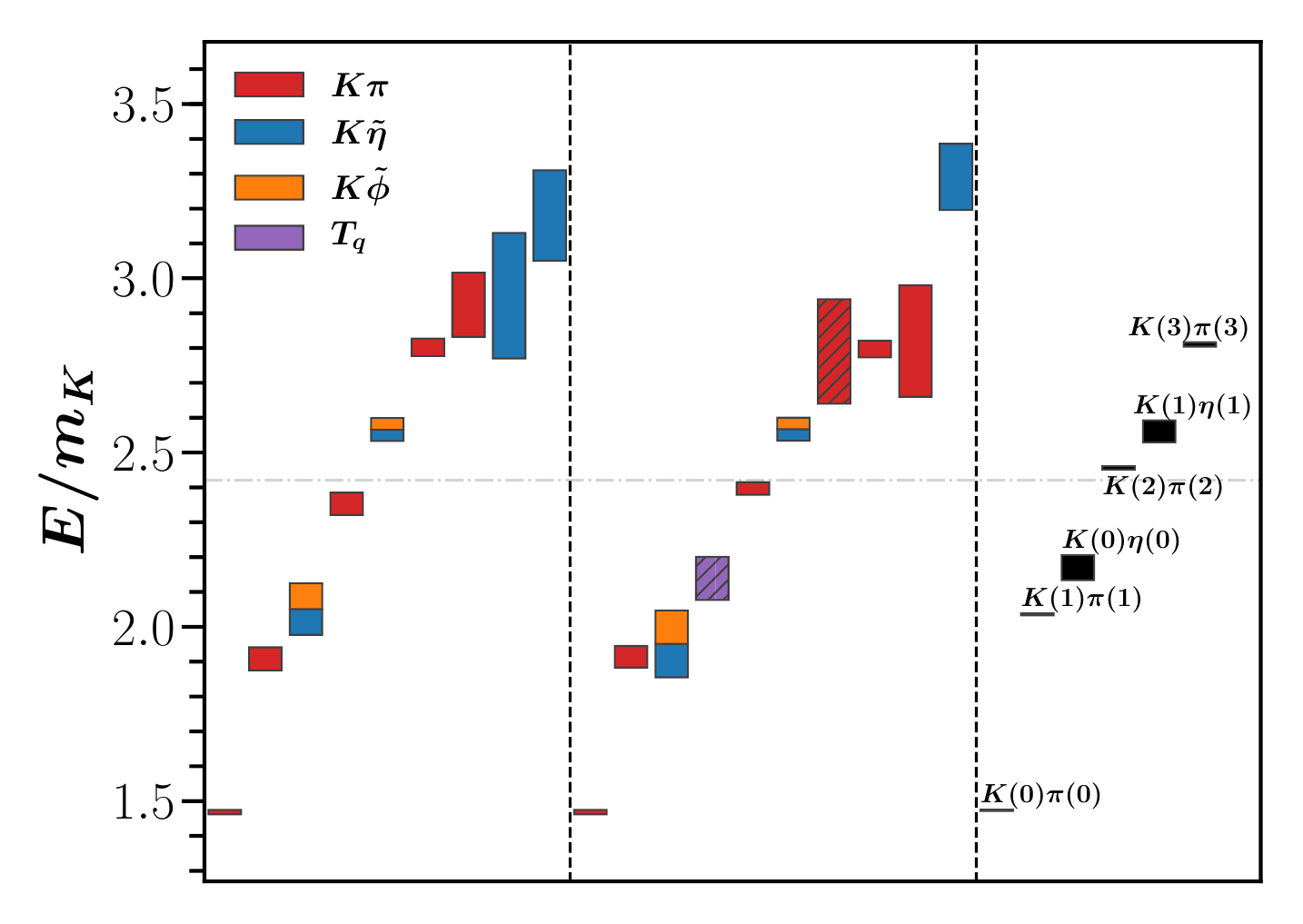}
\caption{(Left) Estimate of isodoublet, strangeness 1, zero-momentum, $A_{1g}$ spectrum
 using operators in Tables~\ref{tab:single-meson-ops} and \ref{tab:two-meson-ops},
 excluding the tetraquark operator.  (Center) Estimate of the spectrum obtained using
 operators in Tables~\ref{tab:single-meson-ops}, \ref{tab:two-meson-ops}, and
 \ref{tab:tetraq}, which includes the tetraquark operator. (Right) Two-particle
 non-interacting energies.  Statistical errors of the energy determinations are shown
 by the vertical extents of the boxes. The horizontal gray line indicates the
 four-particle $K\pi\pi\pi$ threshold (there are no three-particle states in this
 channel in this energy range).  The four flavor contents of the twelve basis operators
 are indicated by colors as shown in the legend.  Levels predominantly created by a
 single operator are indicated by a single color corresponding to the flavor of the
 operator. Dual colored levels are predominantly created by two of the basis operators
 with comparable overlap factors.  Levels created mostly by a single operator but with
 significant overlaps from one or more other operators are indicated by a hatched box.
\label{fig:kappa_spectrum}}
\end{figure}

A summary of the spectrum is shown in the center panel of Fig.~\ref{fig:kappa_spectrum}. 
The ten lowest-lying energy estimates are indicated by the colored boxes, horizontally
displaced for improved visibility. Statistical errors of the energy determinations are
shown by the vertical extents of the boxes. Using the overlap factors of
Fig.~\ref{fig:kappa_zfactors} as described above, the predominant content of each level
is determined and is indicated in Fig.~\ref{fig:kappa_spectrum} by the color of each
energy estimate. The four flavor contents of the twelve basis operators are indicated
by colors as shown in the legend.  Levels predominantly created by a single operator
are indicated by a single color corresponding to the flavor of the operator. Dual
colored levels are predominantly created by two of the basis operators with comparable
overlap factors.  Levels created mostly by a single operator but with significant
overlaps from one or more other operators are indicated by a hatched box.  The
horizontal gray line indicates the four-particle $K\pi\pi\pi$ threshold (there are no
three-particle states in this channel in this energy range).

The six lowest-lying two-particle non-interacting energies are shown in the right panel
of Fig.~\ref{fig:kappa_spectrum}.  For the non-interacting energies, $\eta$ and $\phi$
refer to the physical particles, in contrast to our operator nomenclature. The
lowest-lying level in the center panel corresponds to the $K(0)\pi(0)$ non-interacting
state with a small energy shift.  Level 4 corresponds with the $K(2)\pi(2)$
non-interacting level with a small shift, and level 5 appears to be associated with
the $K(1)\eta(1)$ non-interacting level.  The most substantial difference between the
center and right panels is in the energy range between $(1.8 - 2.3)~m_K$.  In this
energy range, there are only two non-interacting levels $K(1)\pi(1)$ and $K(0)\eta(0)$,
whereas three levels are observed in the center panel in the presence of interactions,
and the tetraquark operator clearly plays an important role in the emergence of this
additional level.

An estimate of the spectrum obtained using an operator set that excludes the
tetraquark operator is shown in the left panel in Fig.~\ref{fig:kappa_spectrum}.  In
comparing this spectrum with that shown in the center panel, one sees that there is
no change to the lowest-lying energy, and the levels at 2.35~$m_K$ and 2.6~$m_K$ are
changed very little.  In the left panel, the orange and blue box (level 2) near
2.05~$m_K$ becomes two distinct energies in the center panel (levels 2 and 3).  The
energy of level 3 (the hatched purple box) in the center panel is slightly above the
energy of level 2 in the left panel.  Level 2 in the center panel, indicated by the
orange and blue box, has an energy slightly below that of level 2 in the left panel.

This is the expected signature of a missed level in a lattice QCD correlator matrix
analysis.  In the left panel, there is only one operator, a linear superposition of
the $K(0)\etaop(0)$ and $K(0)\phiop(0)$ operators, with significant overlap onto the
true levels 2 and 3, as shown in the center panel.  With only one such operator, the
two distinct energies cannot be resolved, and a decay rate of the correlator is
extracted that lies between these two energies.  In the center panel, the addition of the
tetraquark operator yields two operators which produce states with significant
overlaps onto levels 2 and 3, and so, the two individual energies can be resolved.

Level 4 at an energy around 2.4~$m_K$ in the center panel, shown as a red box, can be
compared to level 3 in the left panel.  The energy of level 3 in the left panel is
slightly below that of the energy of level 4 in the center panel, which is consistent
with the signature of a missed level below these energies.  The operator which mainly
produces level 3 in the left panel creates a state which has a small overlap onto the
two lower-lying states which are levels 2 and 3 in the center panel.  This small
overlap occurs because the state produced by the tetraquark operator is needed to
fully remove this overlap and it is not available.  This small leakage to lower-lying
states leads to an energy extraction which is very slightly too low.  Although this
leakage causes this very slight downward shift in the estimate of the energy, its
effect is too small to spoil an observed plateau in the effective energy.

Levels 0, 1, and 4 in the center panel (red boxes) of Fig.~\ref{fig:kappa_spectrum}
are predominantly $K\pi$ states.  The overlap factors in Fig.~\ref{fig:kappa_zfactors}
associated with the tetraquark operator are essentially zero for levels 0 and 1, and
very small for level 4.  The tetraquark operator is observed to impact levels 2 and 3
mostly. An inspection of the overlap factors in Fig.~\ref{fig:kappa_zfactors} shows
that $K\pi$ operators produce states with very little overlaps on levels 2 and 3. 
These facts indicate that the tetraquark operator is having an effect mainly in the
$K\eta$ decay channel and not in the $K\pi$ channel.

We conclude that the tetraquark operator is very important for obtaining a reliable
estimate of the finite-volume spectrum in this channel for energies below 2.7~$m_K$. 
Without the tetraquark operator, one level is missed that could be important in a
study of resonances involving the $K\eta$ decay channel.  The $\kappa$ or
$K_0^\ast(700)$ resonance decays mainly to $K\pi$, so it is not clear if missing
the level created by the tetraquark operator would adversely affect a study of
the $\kappa$ resonance.  However, the $K_0^\ast(1430)$ decays to both $K\pi$ and
$K\eta$, so a missed low-lying state in the $K\eta$ channel could adversely impact
a study of this resonance.

Above an energy of 2.7~$m_K$, statistical uncertainties in the energy extractions
are very large, making comparisons between the left and center panels difficult. 
Fortunately, these levels are not relevant for an analysis of the $\kappa$ resonance.

It is interesting to observe that our results shown in the left panel of
Fig.~\ref{fig:kappa_spectrum}, determined excluding the tetraquark operator, qualitatively
agree with the results shown in Fig.~5 of Ref.~\cite{Wilson:2014cna} obtained also without
any tetraquark operators, once adjustments for the different allowed momenta magnitudes
are taken into account.  The results in that work were obtained using ensembles with the
same lattice action but with $m_\pi\approx400$~MeV and small spatial extents of 16, 20
and 24, as compared to 32 in this work.  The overlap factors in Fig.~5 of
Ref.~\cite{Wilson:2014cna} show significant mixings between the $K\pi$ and single
quark-antiquark operators.  In the absence of the tetraquark operator, our overlap factors
show a similar pattern, but this tendency disappears when the tetraquark operator is included.
Without the tetraquark operator, the energy determinations become plagued by false plateaux,
unfortunately leading to incorrect determinations of the overlap factors. It is also
interesting to note that the three red $K\pi$ energies in the left panel of
Fig.~\ref{fig:kappa_spectrum} are in reasonable agreement with results shown in the upper
right plot of Fig.~4 in Ref.~\cite{Rendon:2020rtw}.  Our $K\pi$ levels agree with those
shown (the two lowest) in Fig.~4 of Ref.~\cite{Brett:2018jqw}, which used the same ensemble.

\begin{figure*}[p]
\raisebox{0cm}{\includegraphics[width=0.329\textwidth]{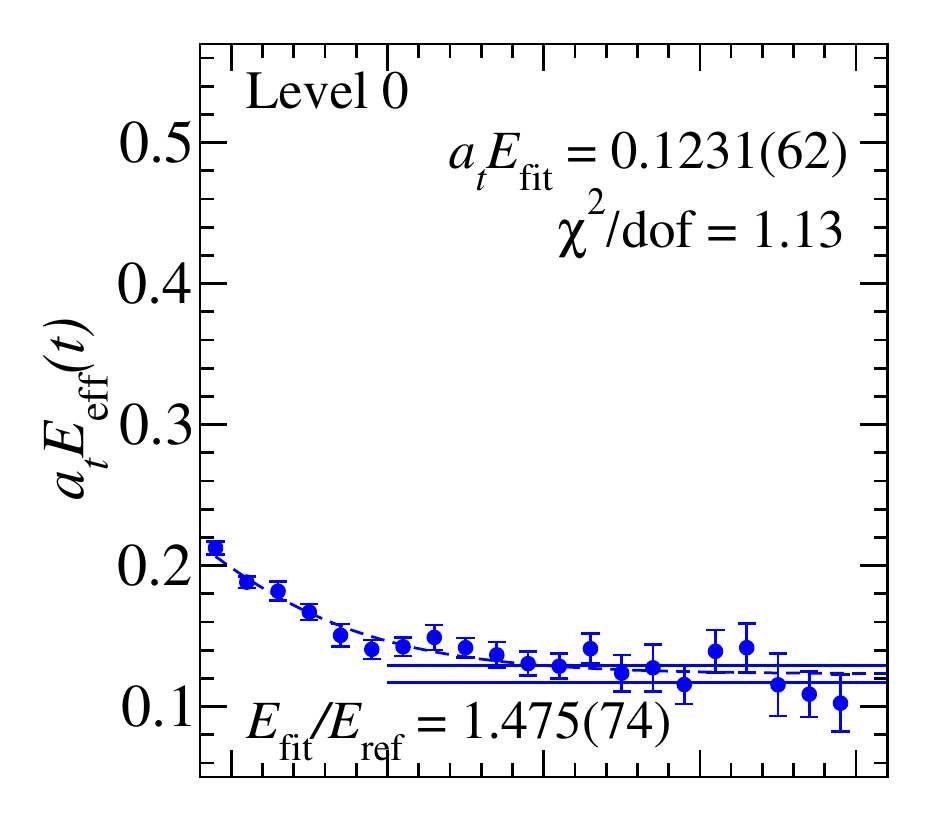}}
\includegraphics[width=0.28\textwidth]{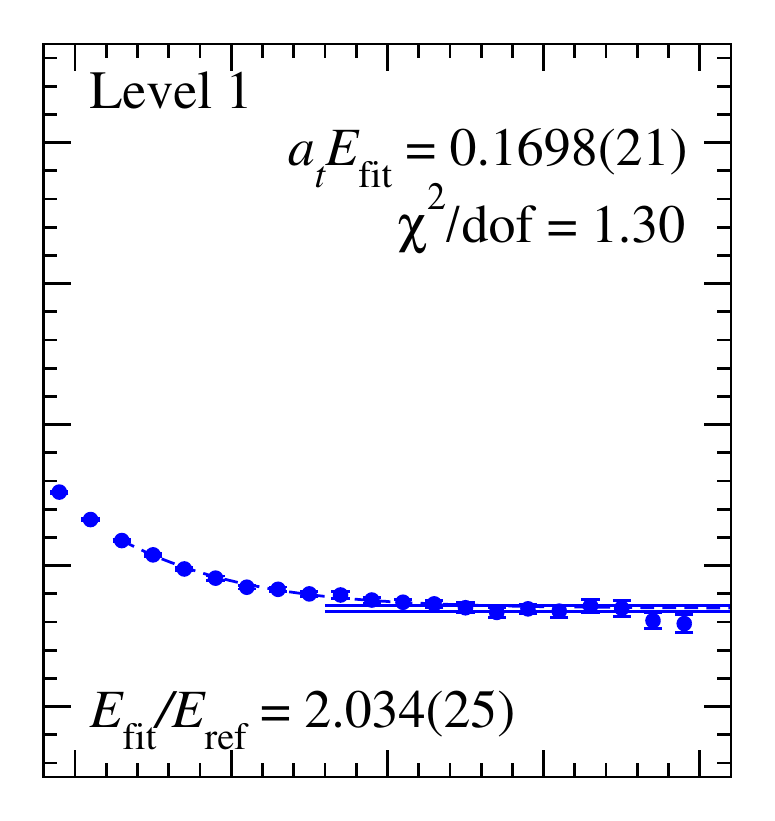}
\includegraphics[width=0.28\textwidth]{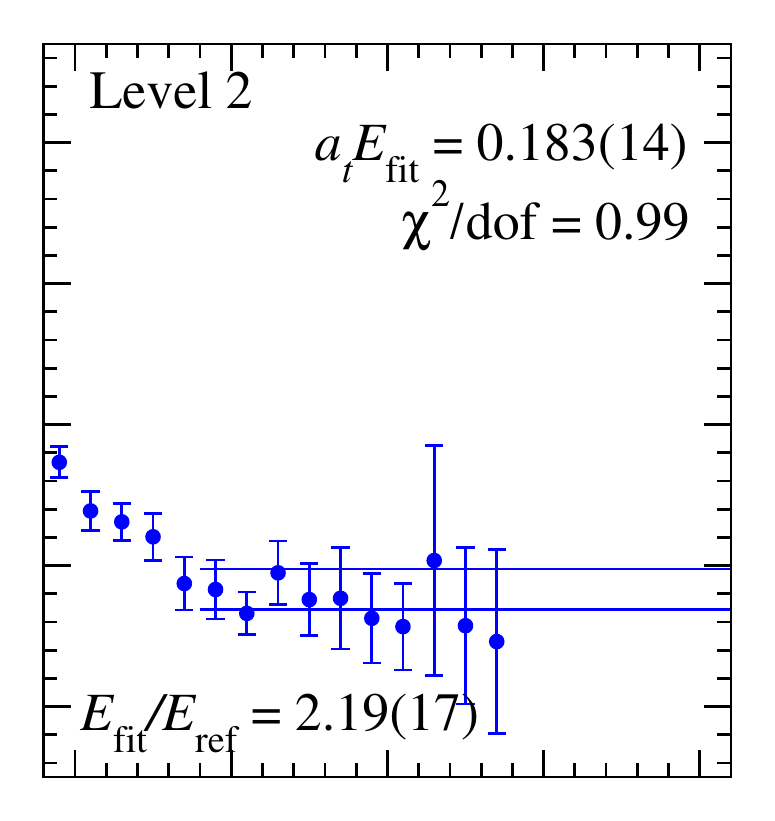}\\
\raisebox{-0.0in}{\includegraphics[width=0.329\textwidth]{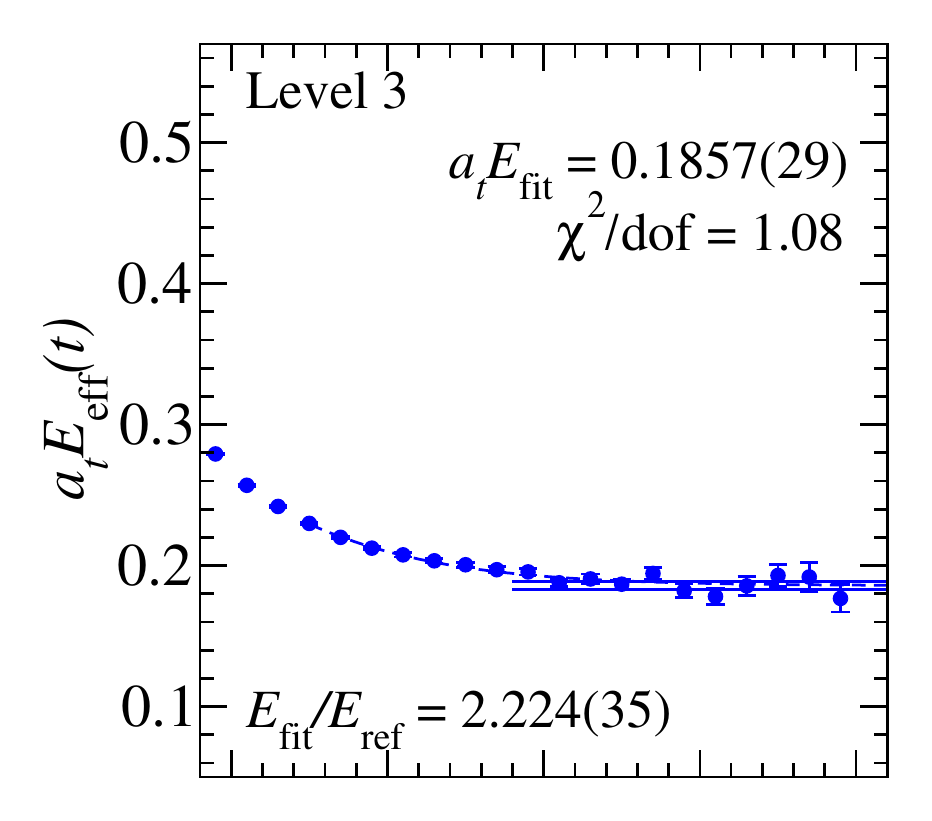}}
\includegraphics[width=0.28\textwidth]{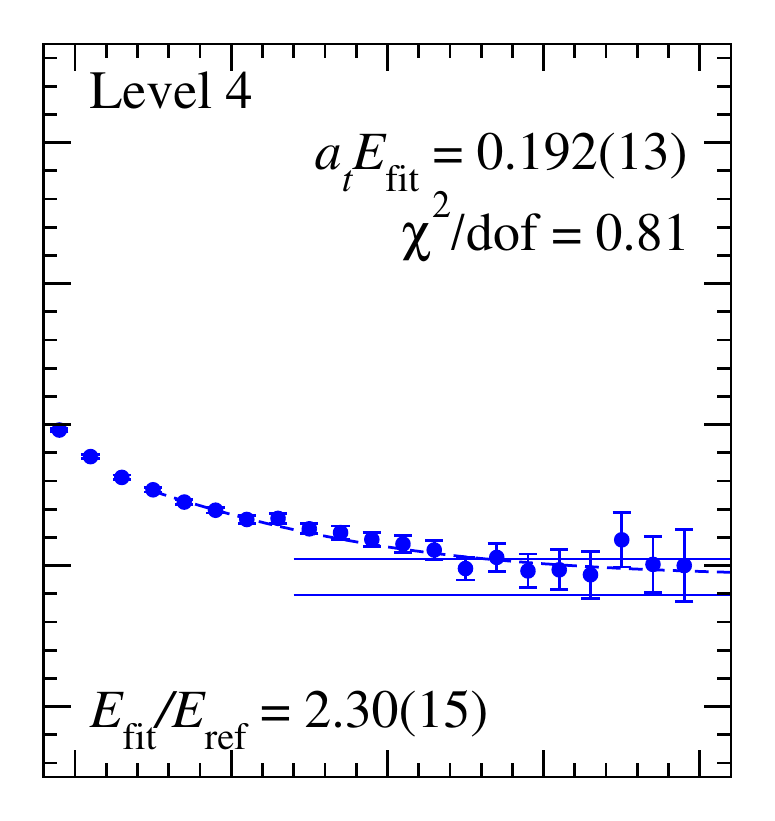}
\includegraphics[width=0.28\textwidth]{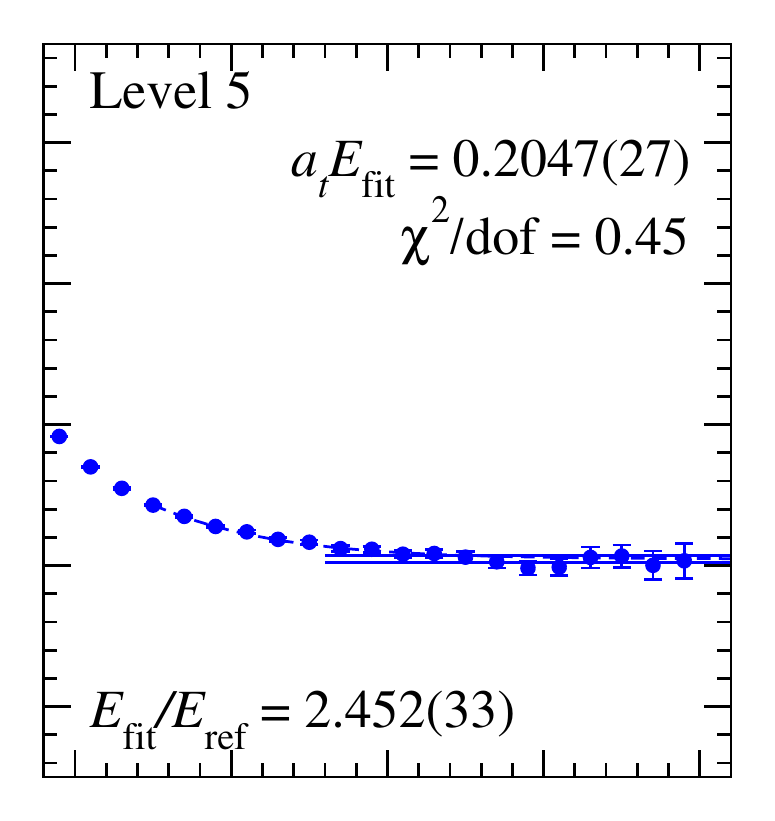}
\\[-2mm]
\par\noindent\rule{\textwidth}{0.5pt}
\includegraphics[width=0.329\textwidth]{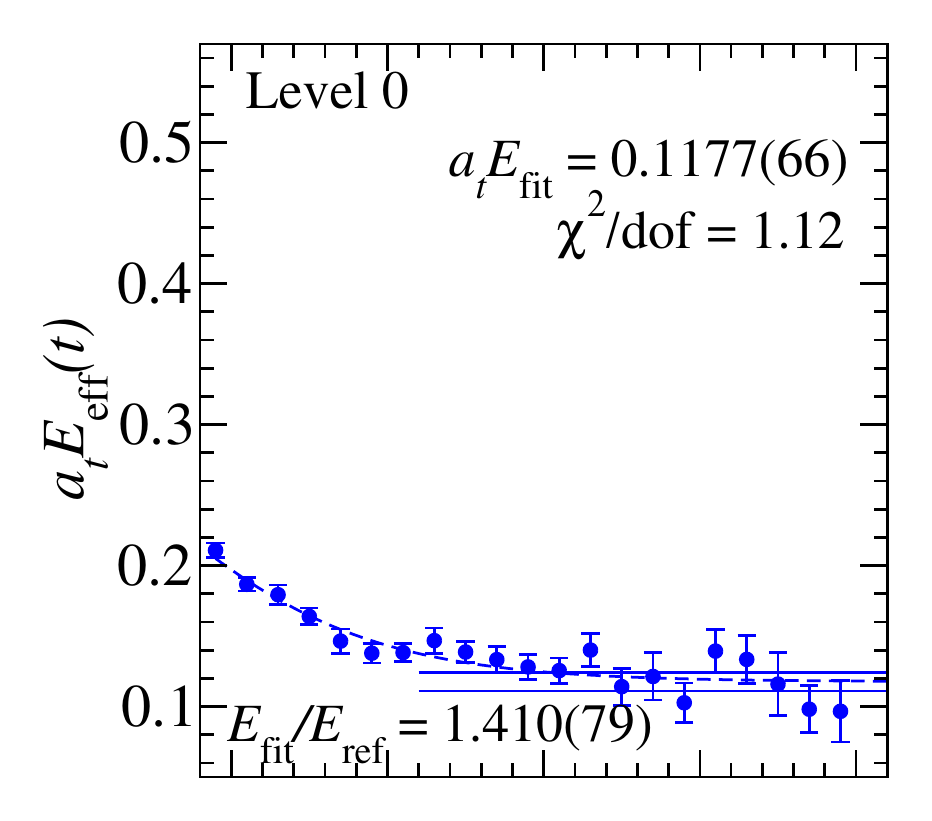}
\includegraphics[width=0.28\textwidth]{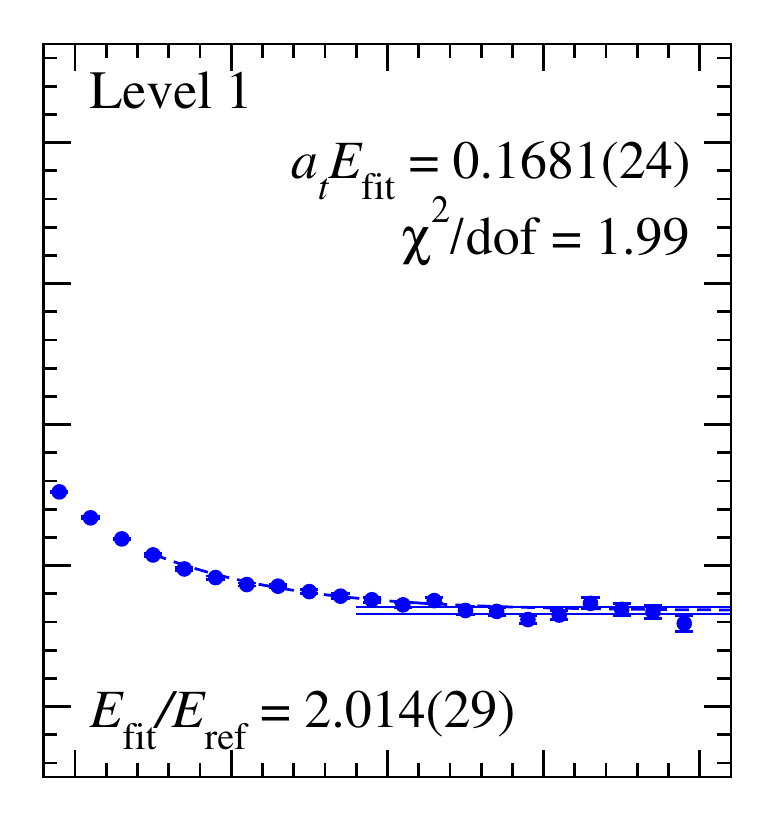}
\includegraphics[width=0.28\textwidth]{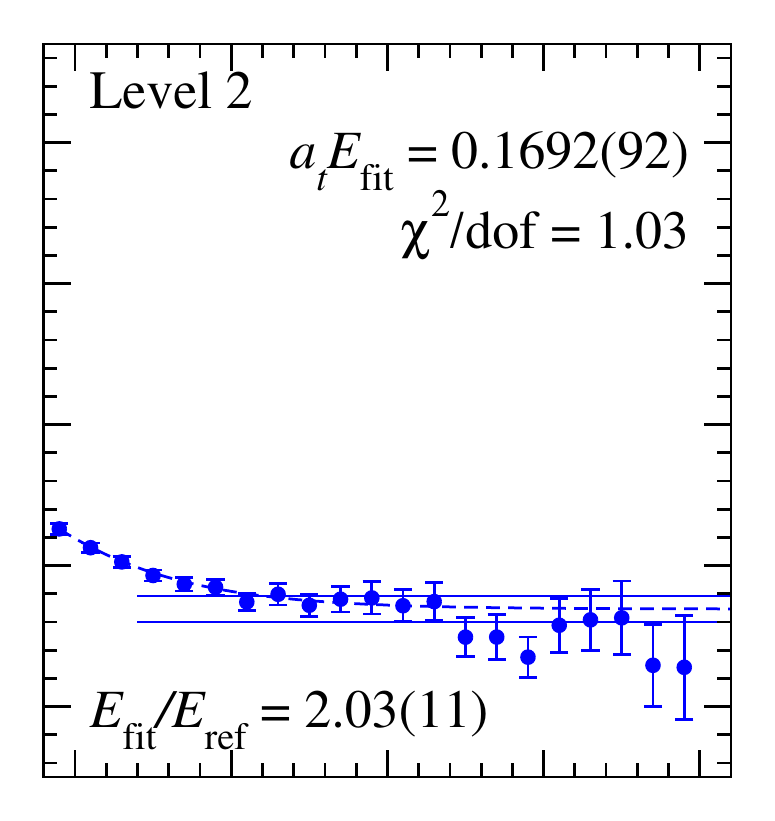}\\
\includegraphics[width=0.329\textwidth]{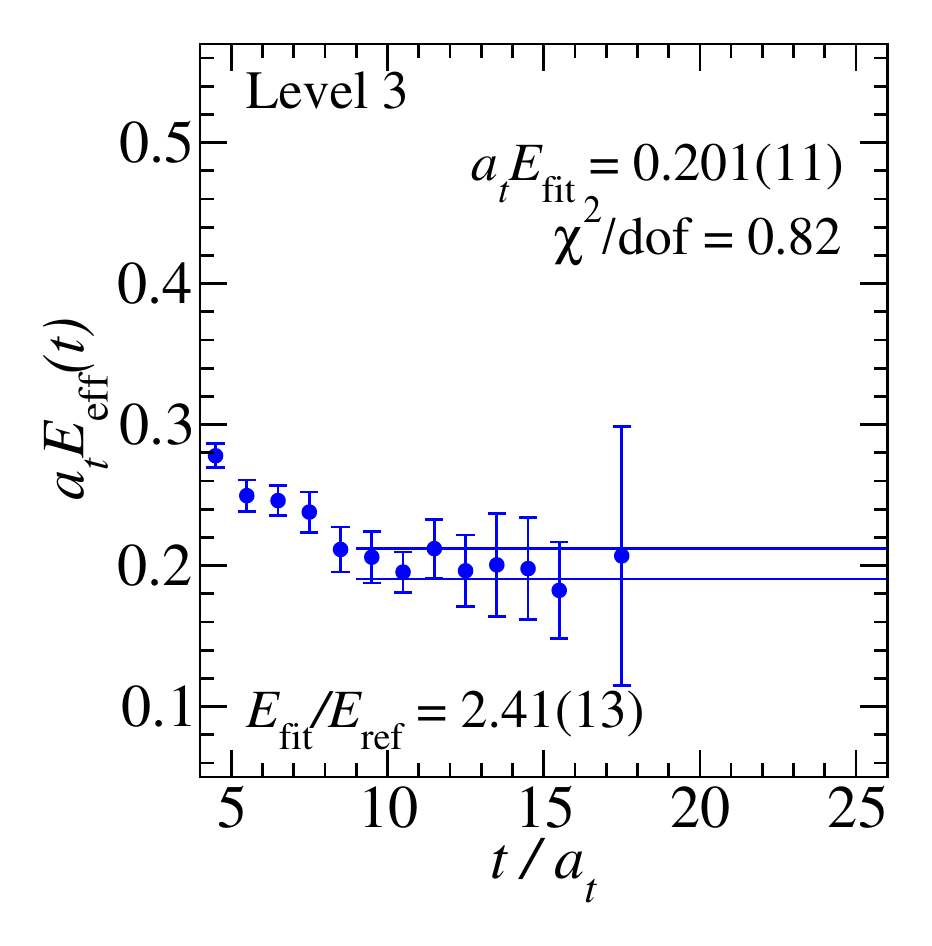}
\includegraphics[width=0.28\textwidth]{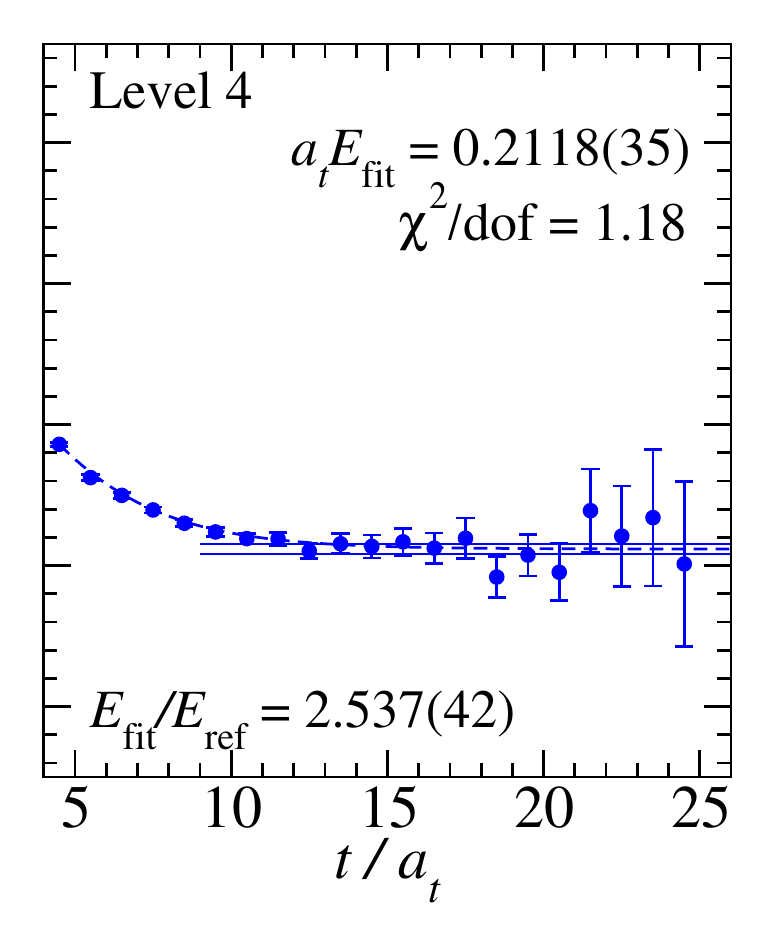}
\includegraphics[width=0.28\textwidth]{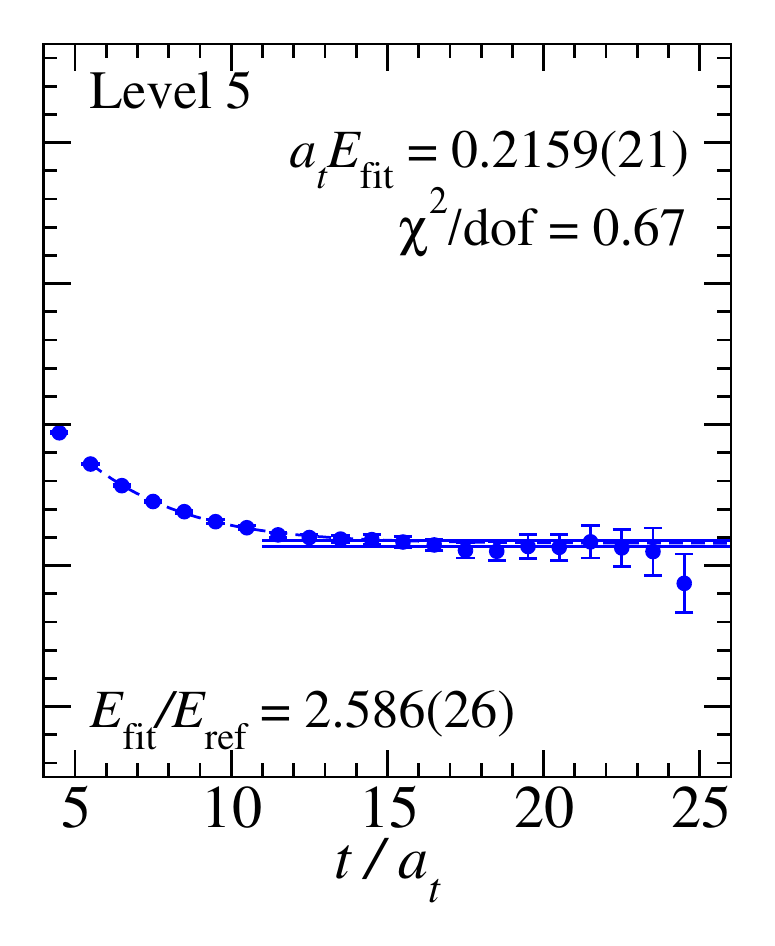}\\
\caption{Effective energies for the lowest six levels in the isotriplet, non-strange, zero-momentum
 $A_{1g}^-$ channel. (Top two rows) Results obtained using only the operators in
 Tables~\ref{tab:single-meson-ops} and \ref{tab:two-meson-ops}, excluding the tetraquark operator.
 (Bottom two rows) Results obtained using the operators in Tables~\ref{tab:single-meson-ops} and
 \ref{tab:two-meson-ops} and also including the tetraquark operator in Table~\ref{tab:tetraq}.
 Effective energy curves calculated from correlator fits are overlaid, and fit results are shown.
\label{fig:a0_fits}}
\end{figure*}

\begin{figure*}
\includegraphics[width=0.304\textwidth]{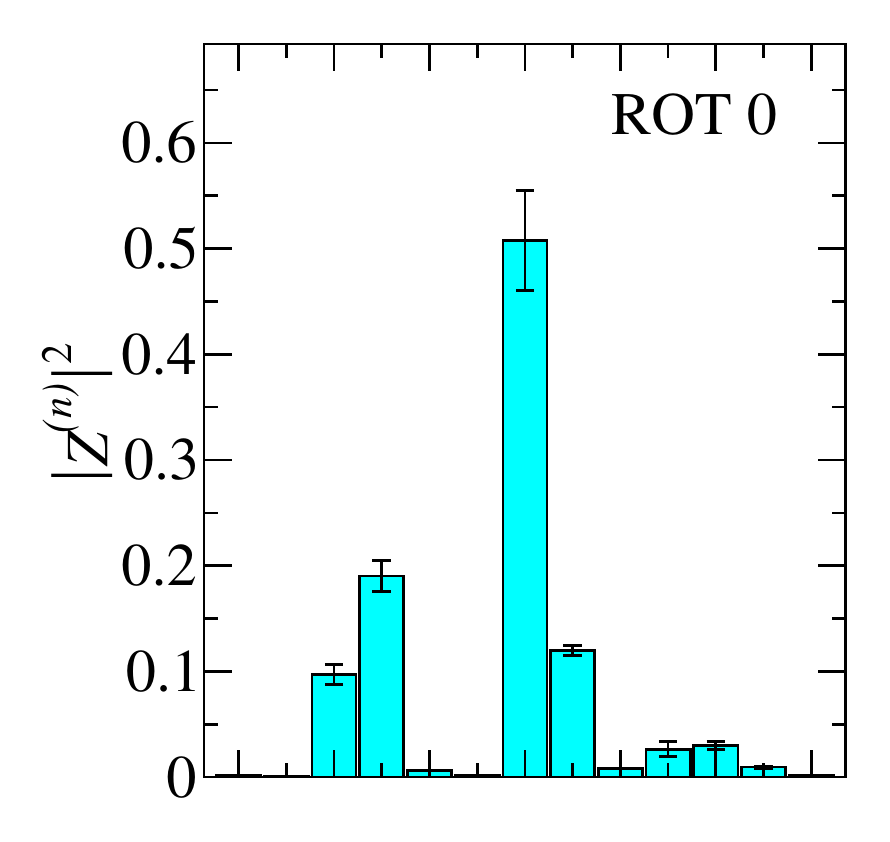}
\includegraphics[width=0.280\textwidth]{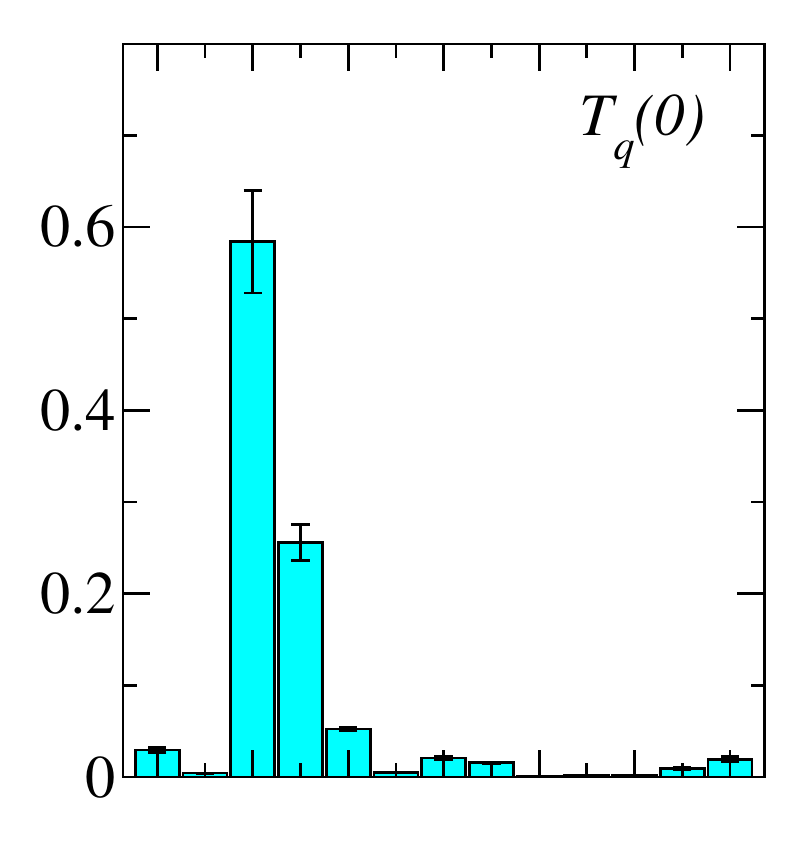}
\includegraphics[width=0.280\textwidth]{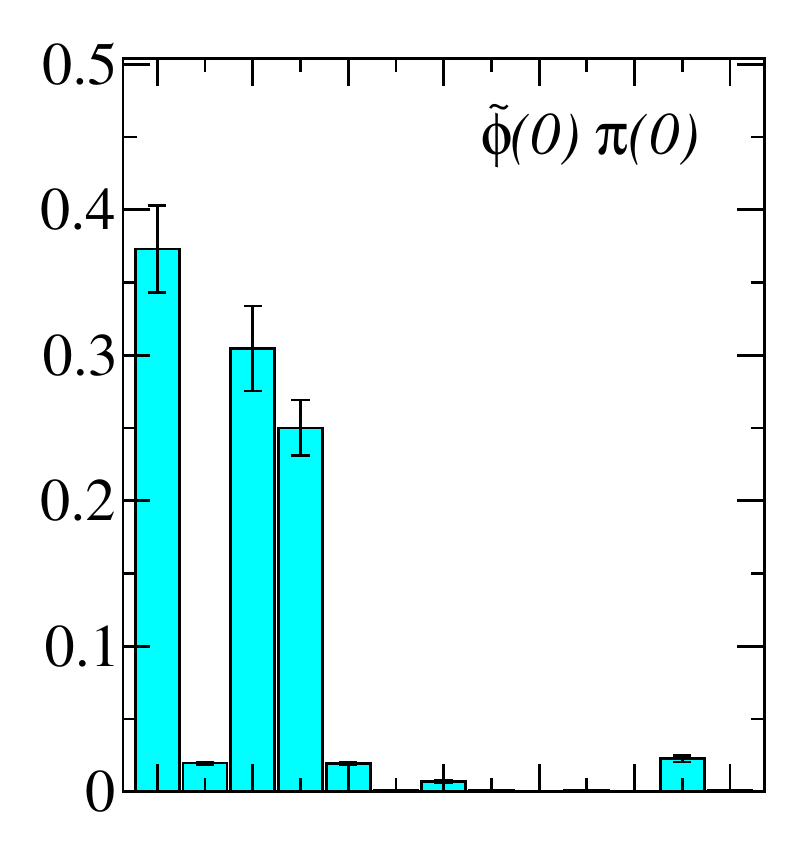}\\
\includegraphics[width=0.304\textwidth]{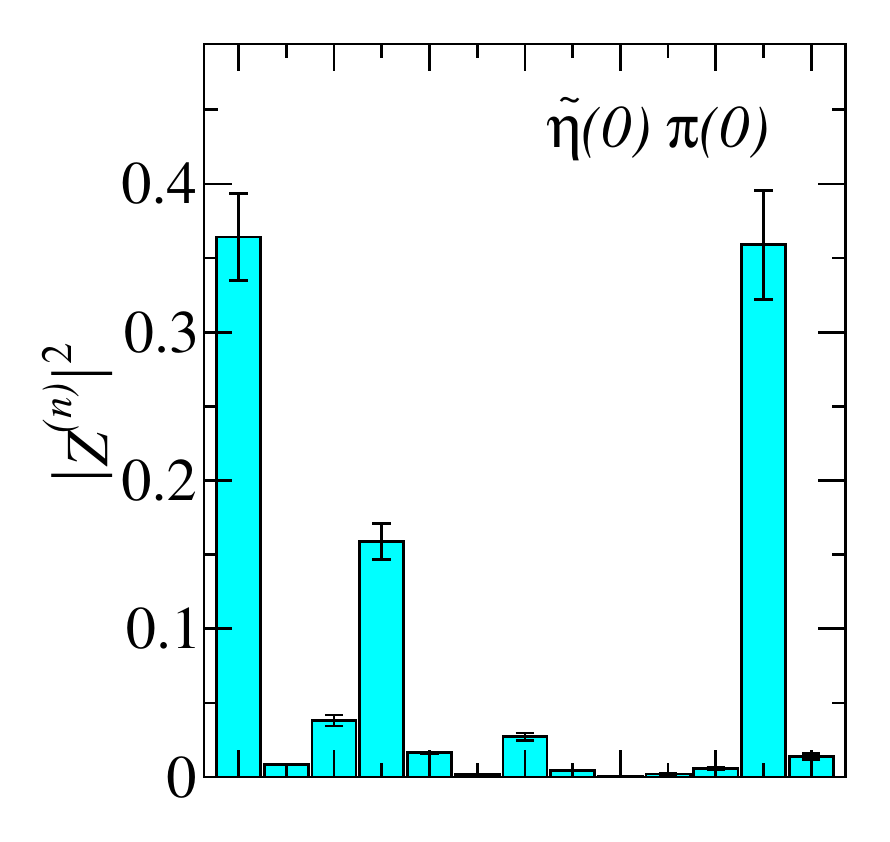}
\includegraphics[width=0.280\textwidth]{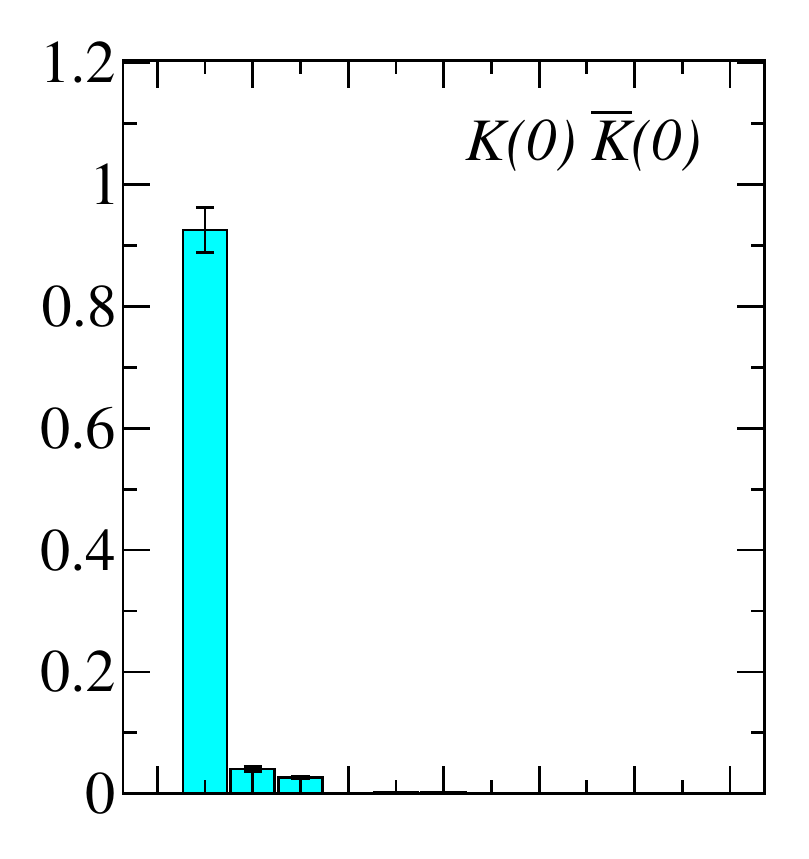}
\includegraphics[width=0.280\textwidth]{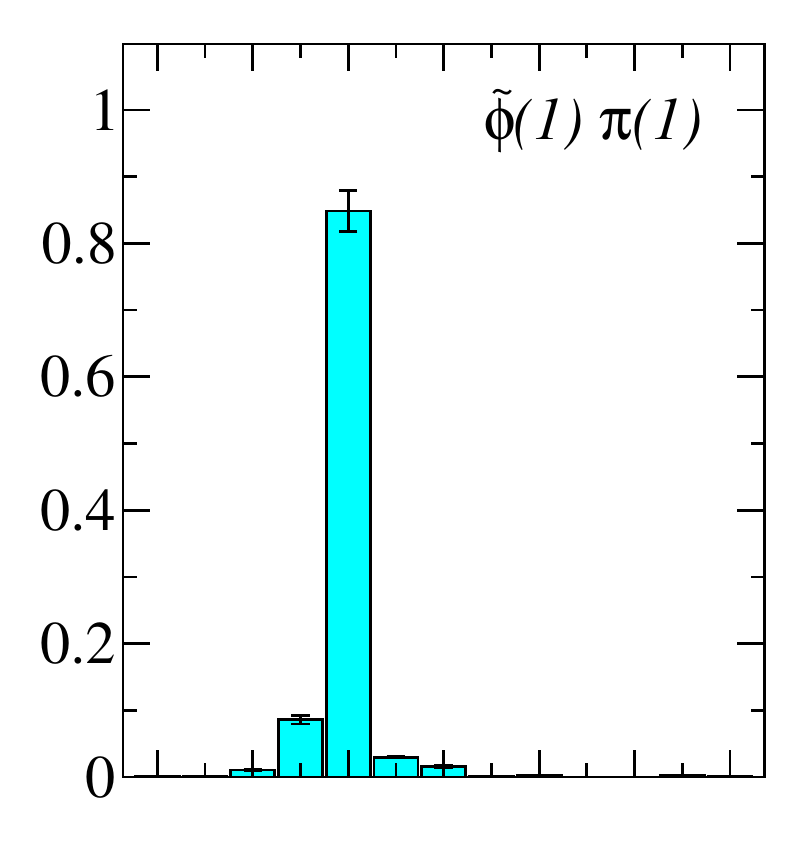}\\
\includegraphics[width=0.304\textwidth]{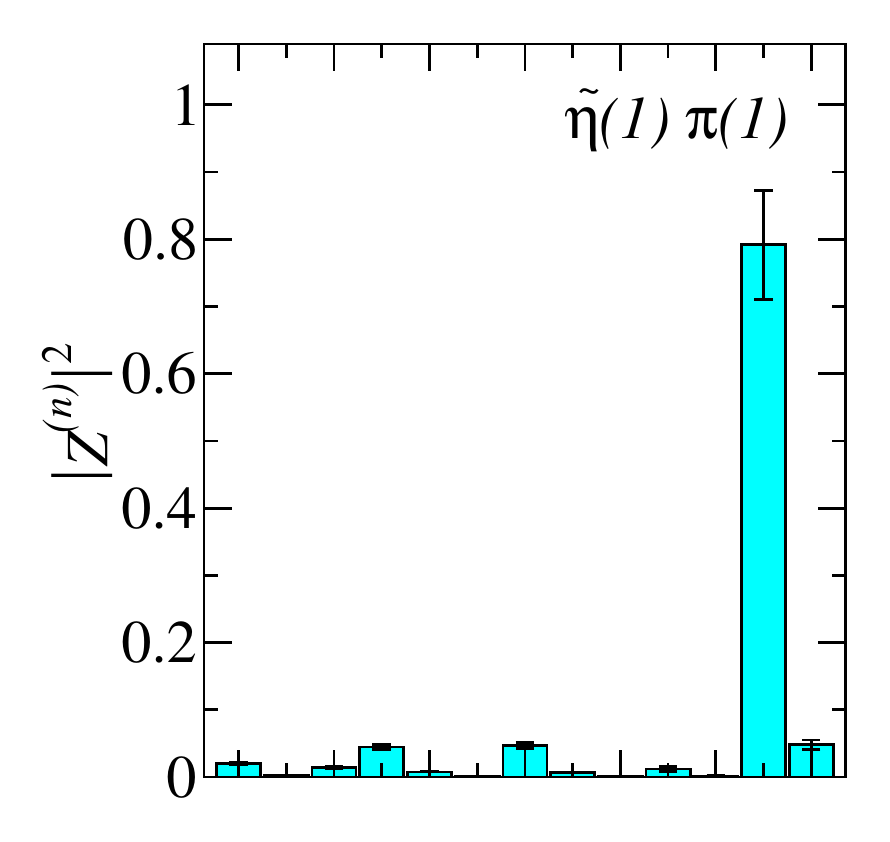}
\includegraphics[width=0.280\textwidth]{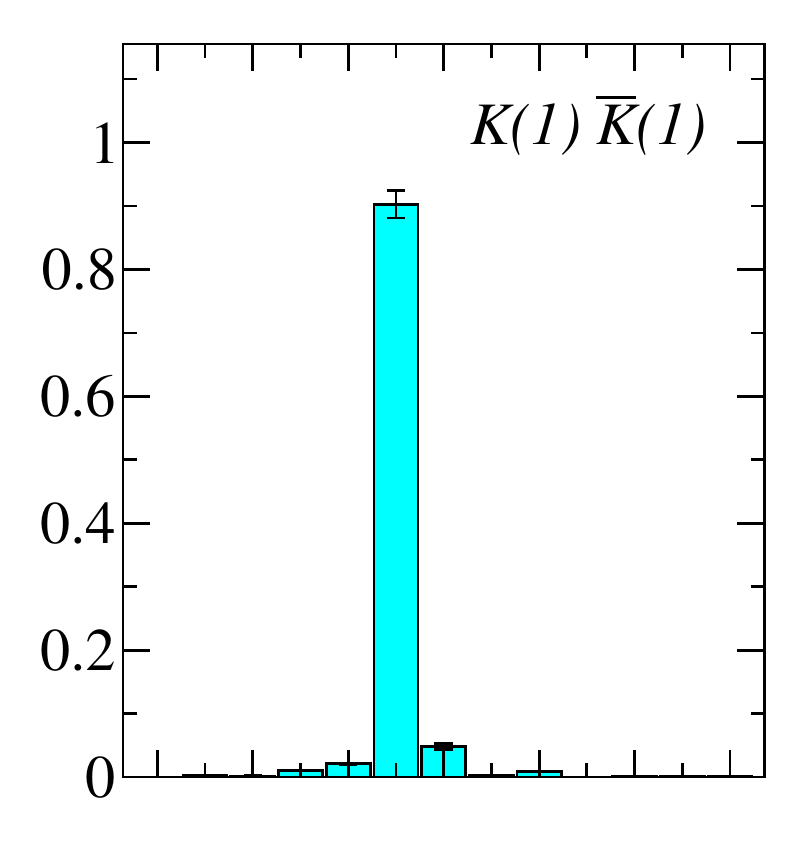}
\includegraphics[width=0.280\textwidth]{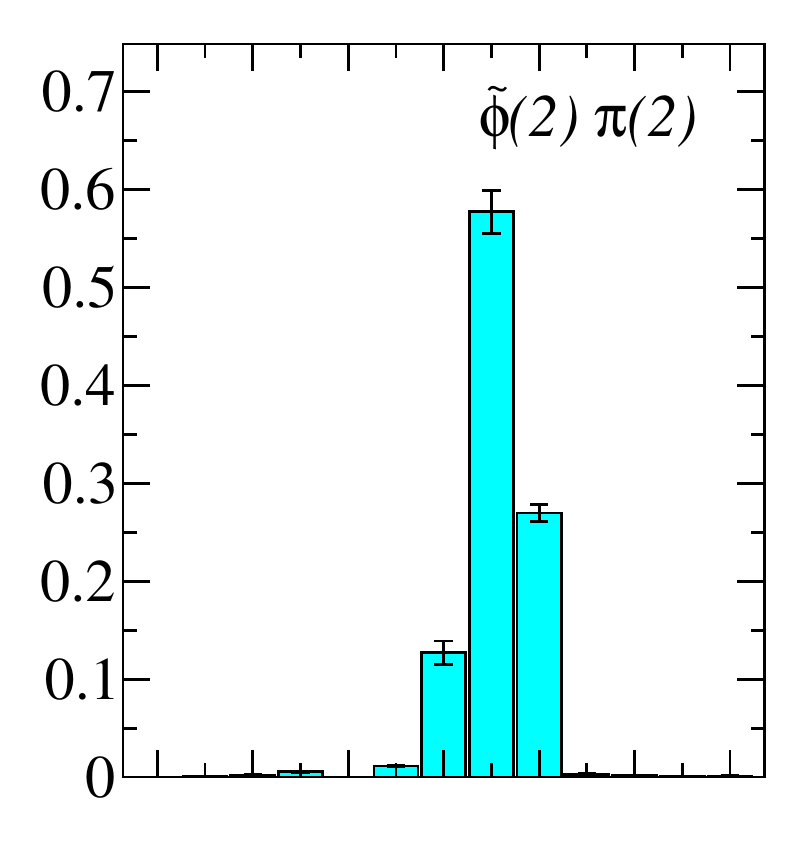}\\
\includegraphics[width=0.304\textwidth]{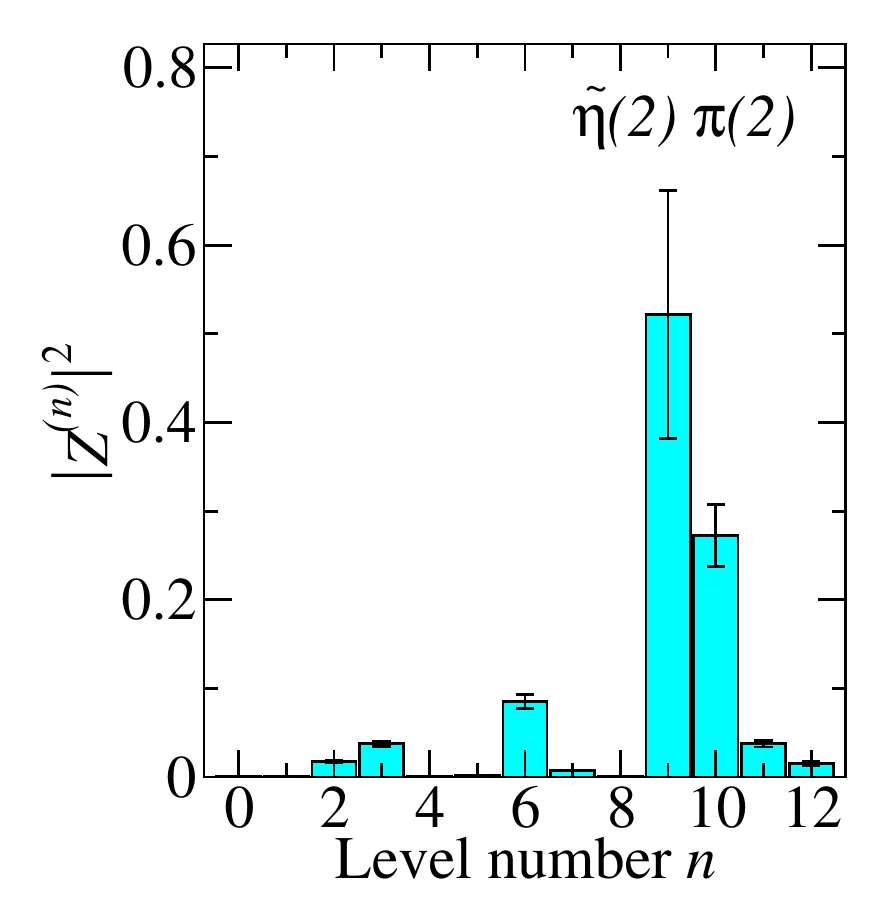}
\includegraphics[width=0.280\textwidth]{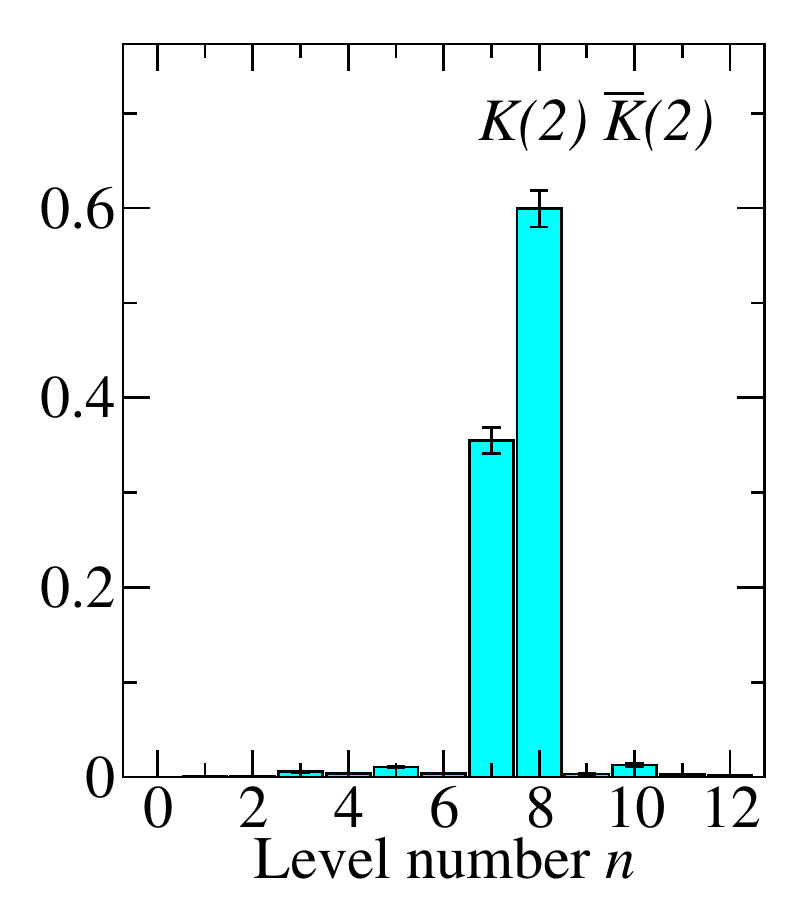}
\includegraphics[width=0.280\textwidth]{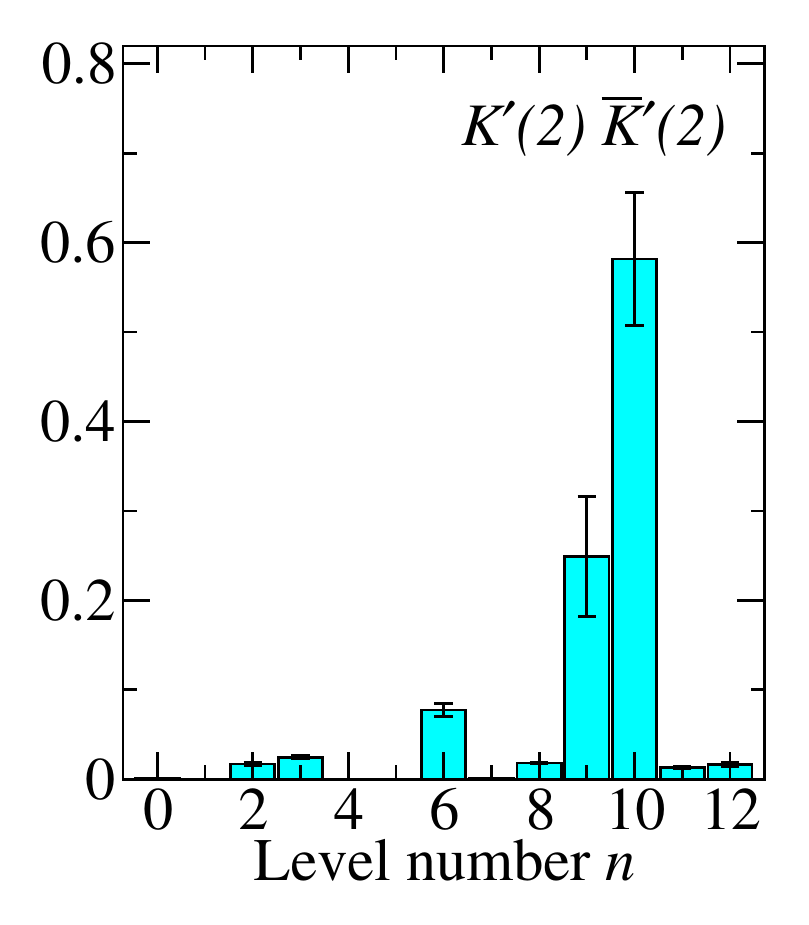}
\caption{Overlap factors $\vert Z^{(n)}_j\vert^2$ for twelve of the thirteen operators used in
 the correlator matrix corresponding to the isotriplet nonstrange zero-momentum $A_{1g}^-$
 channel.  Each plot shows the factors for a single operator, with horizontal axis labelled
 by $n$, the index identifying the stationary states.  The factors for the tetraquark operator
 are shown in the center plot in the top row. Operators labeled by \texttt{ROT N} denote our
 variationally improved single-hadron operators, as discussed in Sec.~\ref{sec:a0_results}.
 $K(2),K^\prime(2)$ refer to SS0, SS1 operators, respectively, and similarly for
 $\overline{K}(2),\,\overline{K}^\prime(2)$ (see Table~\ref{tab:two-meson-ops}). The factors for
 \texttt{ROT 1} (not shown) are all very small except for level 12.
\label{fig:a0_zfactors}}
\end{figure*}

\begin{table}
\caption{Fit details for estimating the energies in the isotriplet nonstrange zero-momentum
 $A_{1g}^-$ channel using the operator basis that excludes the tetraquark operator.  See
 the caption of Table~\ref{tab:kappa_fit_details_no_tq} for further information about
 notation and headings.
\label{tab:a0_fit_details_no_tq}}
\begin{ruledtabular}
\begin{tabular}{llccl}
   $\ E / E_K$ & \quad $a_t E$ & Fit model & $(\tau_{\mathrm{min}}, {\tau_\mathrm{max}})$
   & $\!\!\!\!\!\chi^2 / N_{\rm dof}$\\
   \hline
   1.475(74)&0.1231(62)&2{-}exp&$(3, 26)$&1.13\\
   2.034(25)&0.1698(21)&2{-}exp&$(5, 26)$&1.3\\
   2.19(17)&0.183(14)&1{-}exp&$(9, 26)$&0.99\\
   2.224(35)&0.1857(29)&2{-}exp&$(6, 26)$&1.08\\
   2.30(15)&0.192(13)&2{-}exp&$(6, 26)$&0.81\\
   2.452(33)&0.2047(27)&2{-}exp&$(6, 26)$&0.45\\
   2.828(45)&0.2361(37)&2{-}exp&$(6, 26)$&0.71\\
   2.971(32)&0.2480(26)&2{-}exp&$(4, 26)$&1.66\\
   5.00(39)&0.417(33)&1{-}exp&$(8, 18)$&1.23
\end{tabular}
\end{ruledtabular}
\end{table}

\subsection{\boldmath Channel involving the $a_0(980)$ resonance}
\label{sec:a0_results}

Results for the stationary-state energies in the isotriplet non-strange zero-momentum
$A_{1g}^-$ channel are presented in this section.  The results shown are obtained using
a normalization time, metric time, and diagonalization time of $\tau_N=3$, $\tau_0=4$,
and $\tau_D=7$, respectively.  The same procedure as described in
Sec.~\ref{sec:kappa_results} is used to arrive at these choices for the GEVP parameters
which yield a good combination
of reduced noise and a sufficiently diagonal matrix $\widetilde{C}(t)$ for $t > t_D$. 
To ensure our results were insensitive to our choice of GEVP parameters within the
window, the spectrum was extracted for various choices and compared.
The so-called disconnected contributions associated with the $\etaop$
operators in this channel lead to significantly larger statistical errors in the
determination of the correlation matrix, compared to the $\kappa$ channel.  This
is the main reason that we end up using $\tau_D=7$, which is substantially smaller
than the value of 12 used in the $\kappa$ channel.

\begin{table}
\caption{Fit details for estimating the energies in the isotriplet nonstrange
 zero-momentum $A_{1g}^-$ channel using the operator basis that includes the tetraquark
 operator.  See the caption of Table~\ref{tab:kappa_fit_details_no_tq} for further
 information about notation and headings.  \label{tab:a0_fit_details_tq}}
\begin{ruledtabular}
\begin{tabular}{llccl}
   $\ E / E_K$ & \quad $a_t E$ & Fit model & $(\tau_{\mathrm{min}}, {\tau_\mathrm{max}})$
   & $\!\!\!\!\!\chi^2 / N_{\rm dof}$\\
   \hline
   1.410(79)&0.1177(66)&2{-}exp&$(3, 26)$&1.12\\
   2.014(29)&0.1681(24)&2{-}exp&$(6, 26)$&1.99\\
   2.03(11)&0.1692(92)&2{-}exp&$(3, 26)$&1.03\\
   2.41(13)&0.201(11)&1{-}exp&$(9, 26)$&0.82\\
   2.537(42)&0.2118(35)&2{-}exp&$(3, 26)$&1.18\\
   2.586(26)&0.2159(21)&2{-}exp&$(4, 26)$&0.67\\
   2.84(12)&0.237(10)&2{-}exp&$(3, 24)$&0.99\\
   2.947(40)&0.2461(33)&2{-}exp&$(4, 26)$&0.9\\
   2.964(35)&0.2475(29)&2{-}exp&$(4, 26)$&1.64\\
   5.01(39)&0.418(33)&1{-}exp&$(8, 18)$&1.25
\end{tabular}
\end{ruledtabular}
\end{table}

Fits to the diagonal elements of the correlator matrix $\widetilde{C}(t)$, obtained
after the single-pivot procedure described in Sec.~\ref{sec:temporal}, are shown as
effective energy plots in Fig.~\ref{fig:a0_fits}.  In the top two rows are shown the
fits which yield the lowest six energies using only the single-meson and meson-meson
operators (twelve in total) listed in Tables~\ref{tab:single-meson-ops} and
\ref{tab:two-meson-ops}, excluding the tetraquark operator. In the bottom two rows
are presented the fits which yield the lowest six energies using all thirteen operators
listed in Tables~\ref{tab:single-meson-ops} and \ref{tab:two-meson-ops} and also
including the tetraquark operator in Table~\ref{tab:tetraq}. Only the fits for the
six lowest-lying energies are shown, but higher energies were also obtained. Fit
details are given in Tables~\ref{tab:a0_fit_details_no_tq} and \ref{tab:a0_fit_details_tq}.
As in the $\kappa$ channel, note that the fit qualities in
Tables~\ref{tab:a0_fit_details_no_tq} and \ref{tab:a0_fit_details_tq} are all
reasonably good.  The results obtained with only the meson and meson-meson operators
do not themselves give any hint of unreliable energy extractions.

In order to obtain some information about the structures of the stationary states,
the overlap factors $\vert Z^{(n)}_j\vert^2$ are computed.  In this channel, the
possibility exists that two states are present which are predominantly of
quark-antiquark content, corresponding to the $a_0(980)$ and the $a_0(1450)$ mesons. 
We do not expect a one-to-one correspondence between these two quark-antiquark states
and the two single-hadron operators in Table~\ref{tab:single-meson-ops}.  Instead, a
better one-to-one correspondence can be achieved using so-called
variationally-improved single-hadron operators, which are linear combinations of the
single-hadron operators determined by solving a GEVP in the subspace created by the
single-hadron operators.  The two single-hadron operators obtained in this way are
denoted here by ROT 0 and ROT 1.  The use of these operators in determining the
overlap factors greatly aids in the level identifications.  The overlap factors
$\vert Z^{(n)}_j\vert^2$ are determined for all thirteen operators and are shown
in Fig.~\ref{fig:a0_zfactors}.  Each plot shows the factors for a single operator,
with horizontal axis labelled by $n$, the index identifying the stationary states. 
Within each plot, the heights have been rescaled so that they add to unity, but
keep in mind that overlaps with levels for $n\geq 13$ have not been determined.

First, consider the lowest-lying level 0.  An inspection of the overlap plots in
Fig.~\ref{fig:a0_zfactors} shows that this state is predominantly created by the
$\etaop(0)\pi(0)$ and $\phiop(0)\pi(0)$ operators.  Given that our $\etaop$ operator
refers to $\overline{u}u+\overline{d}d$ and our $\phiop$ operator is $\overline{s}s$,
level 0 can be viewed as a state dominated by a pion at rest and a physical $\eta$ at
rest with residual interactions.  Next, consider stationary state level 1.  This
state is mainly created by the $\overline{K}(0)K(0)$ operator, so level 1 can be
interpreted as a kaon-antikaon state with residual interactions.  Levels 2 and 3
are produced mainly by the tetraquark and $\phiop(0)\pi(0)$ operators, with small
overlap factors from the $\etaop(0)\pi(0)$ and ROT 0 operators.  This leads to an
identification of level 2 as predominantly a tetraquark state and level 3 as a
$\pi\eta^\prime$ state.  Level 4 is mainly produced by the $\phiop(1)\pi(1)$
operator, so can be interpreted as a pion and a physical $\eta$ with back-to-back
minimal relative momenta. Level 5 is mainly produced by the $\overline{K}(1)K(1)$
operator, so can be interpreted as a kaon-antikaon state having back-to-back
minimal relative momenta.

\begin{figure}
\centering
\includegraphics[width=\columnwidth]{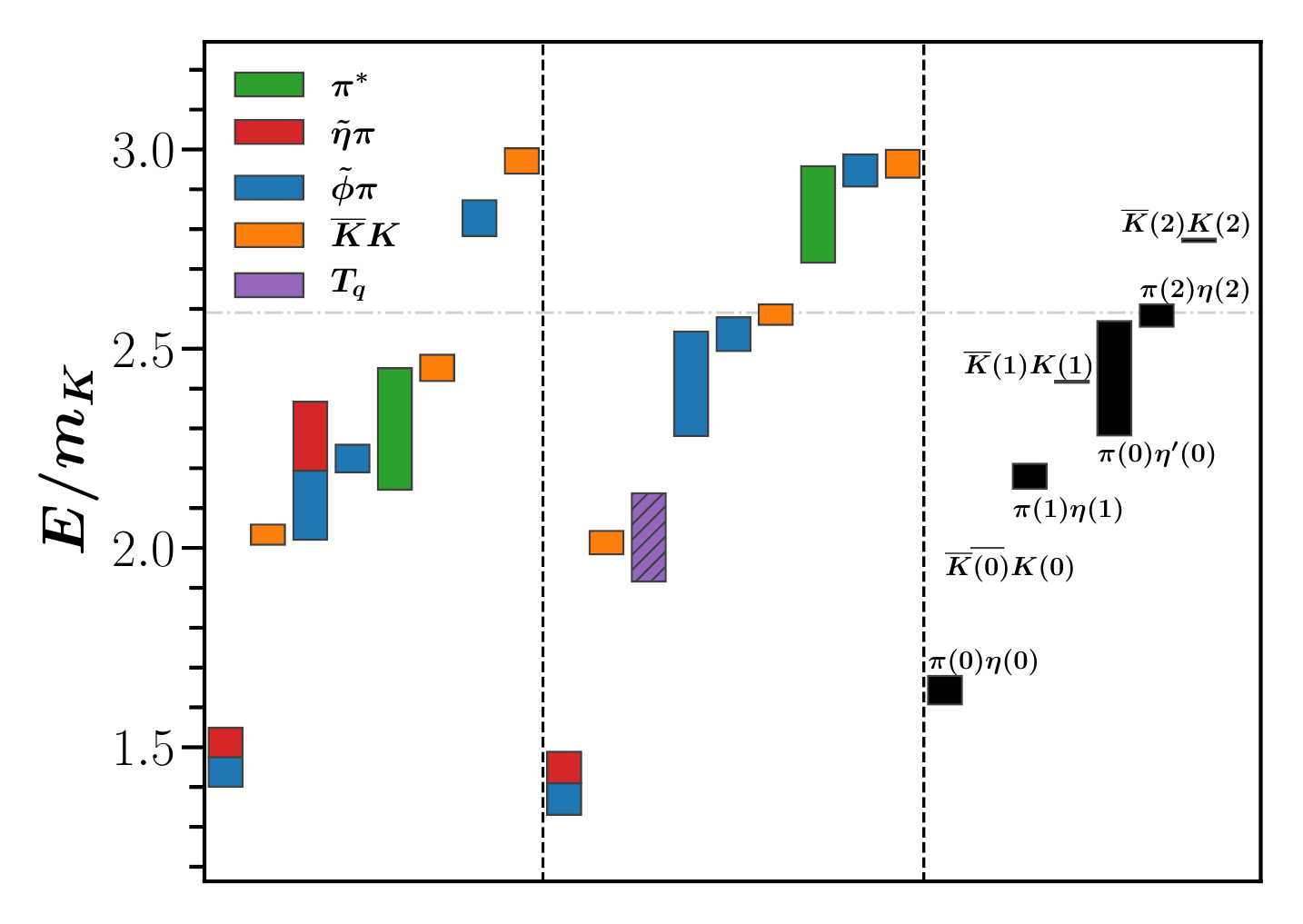}
\caption{
 (Left) Estimate of isotriplet, strangeness 0, zero-momentum, $A_{1g}^-$ spectrum
 using operators in Tables~\ref{tab:single-meson-ops} and \ref{tab:two-meson-ops},
 excluding the tetraquark operator.  (Center) Estimate of the spectrum obtained
 using operators in Tables~\ref{tab:single-meson-ops}, \ref{tab:two-meson-ops},
 and \ref{tab:tetraq}, which includes the tetraquark operator. (Right)
 Two-particle non-interacting energies.  Statistical errors of the energy
 determinations are shown by the vertical extents of the boxes. The horizontal gray
 line indicates the four-particle $\pi\pi\pi\eta$ threshold (there are no
 three-particle states in this channel in this energy range).  The flavor contents
 of the 13 basis operators are indicated by colors as shown in the legend. 
 Levels predominantly created by a single operator are indicated by a single color
 corresponding to the flavor of the operator. Dual colored levels are predominantly
 created by two of the basis operators with comparable overlap factors.  Levels
 created mostly by a single operator but with significant overlaps from one or more
 other operators are indicated by a hatched box.
\label{fig:a0_spectrum}}
\end{figure}

A summary of the spectrum is shown in the center panel of Fig.~\ref{fig:a0_spectrum},
which is the analog of Fig.~\ref{fig:kappa_spectrum} in this channel.  The nine
lowest-lying energy estimates are indicated by the colored boxes, horizontally
displaced for improved visibility. The horizontal gray line indicates the four-particle
$\pi\pi\pi\eta$ threshold (there are no three-particle states in this channel in
this energy range).

The seven lowest-lying two-particle non-interacting energies are shown in the right
panel of Fig.~\ref{fig:a0_spectrum}.  For the non-interacting energies, $\eta$ and
$\phi$ refer to the physical particles, in contrast to our operator nomenclature.
The lowest-lying level in the center panel corresponds to the $\pi(0)\eta(0)$
non-interacting state with a significant energy shift.  In fact, fairly large energy
shifts away from the non-interacting energies are observed for most of the levels.

The left panel in Fig.~\ref{fig:a0_spectrum} shows our best attempt to extract the
spectrum using the set of twelve operators in Tables~\ref{tab:single-meson-ops} and
\ref{tab:two-meson-ops} which excludes the tetraquark operator.  Note that including
more single-hadron and two-meson operators does \textit{not} change these extractions
of the lowest-lying energies; we explicitly checked this in low-statistics calculations. 
The pattern of energies in the left panel differs dramatically from that in the center
panel.  Levels 0 and 1 agree fairly well, but the pattern is very different above these
levels.  Without the presence of the tetraquark in the left panel, some of the other
operators which produce small overlap factors with level 2 must fill in the role of
producing this level and simply cannot do so very well by a temporal separation of
$t_{\rm max}$ in the correlators.  The ROT 1 operator especially results in a large
change to the extracted energies.  Without the tetraquark operator in the left panel,
one clearly sees that the remaining operators do a poor job of resolving the energy
spectrum obtained in the middle panel.  All of the single-meson and two-meson operators
have very small overlap factors with stationary state 2 of the center panel and most
have large overlap factors with the higher-lying states.  Because of this, several
extracted energies have very large errors and occur erroneously at values above that
of stationary state 2 and below that of the higher states.  This phenomenon, which
occurs when a state is essentially missed by the entire operator set used, has been
commonly known in lattice QCD since the first glueball calculations in the 1980s,
but an explicit explanation can be found in Ref.~\cite{Dudek:2012xn}.  In the center
panel, the presence of the tetraquark operator dramatically improves the energy
extractions.  The purple hatched level 2 is reliably determined using the tetraquark
operator, and the other operators can then focus on tending towards the levels
corresponding to their largest overlap factors.

We conclude that the tetraquark operator is very important for obtaining a reliable
estimate of the finite-volume spectrum in this channel for energies below 2.7~$m_K$. 
Without the tetraquark operator, the pattern of extracted energies above level 1 is
simply incorrect.  Exclusion of the tetraquark operator has a dramatic adverse impact
on the extraction of the energy spectrum.

It is interesting to observe that our results shown in the left panel of Fig.~\ref{fig:a0_spectrum},
determined excluding the tetraquark operator, qualitatively agree with the results
shown in Fig.~3 of Ref.~\cite{Dudek:2016cru} obtained also without any tetraquark operators,
once adjustments for the different allowed momenta magnitudes are taken into account. 
The results in that work were obtained using ensembles with the same lattice action but
with $m_\pi\approx400$~MeV and small spatial extents of 16, 20 and 24, as compared to 32
in this work. The overlap factors shown in Fig.~3 of Ref.~\cite{Dudek:2016cru} show
significant mixings among nearly all of the different operators.  As in the $\kappa$ case, in the
absence of the tetraquark operator, our overlap factors show a similar pattern, but this tendency
disappears when the tetraquark operator is included, providing further evidence that without the
tetraquark operator, the energy determinations become plagued by false plateaux which can lead to
incorrect determinations of the overlap factors.

\section{Scattering amplitudes}
\label{sec:Kamp}

The last step in studying a resonance in lattice QCD is to apply the so-called L\"uscher
quantization condition using the finite-volume spectrum to determine the scattering
$K$-matrix from which the resonance parameters can be deduced.  Essentially, the goal
is to apply a suitable parametrization of the $K$-matrix elements in terms of the
center-of-mass energy and determine the best-fit values of the parameters by matching
the energy spectra determined from the quantization condition in a finite volume to
those obtained from the lattice QCD computations.

In this exploratory study, our goal is only to study the role of tetraquark operators
in reliably computing the finite volume spectrum in lattice QCD in the channels
associated with the $\kappa$ and $a_0(980)$ resonances.  For this reason, we
simplify the computations by focusing only on the channels having zero total
momentum.  For a robust determination of the $K$-matrix, it is well known that
it is best to use several total momenta.  This is left for a future study.  However,
it is still interesting to examine the quantization condition with our results for
zero total momentum.

This study employs a lattice of spatial volume $L^3$ and periodic spatial boundary
conditions. The form of the quantization condition used here is given by
\begin{equation}
\label{eq:det}
  {\rm det}\Bigl(\widetilde{K}^{-1}(E_{\rm cm}) - \Bp(E_{\rm cm}) \Bigr) = 0 \ ,
\end{equation}
where $\Ecm$ denotes center-of-mass energy, $\widetilde{K}(\Ecm)$ is related to the
usual scattering $K$-matrix as described below, and $\Bp(E_{\rm cm})$ for a particular
total momentum $\boldsymbol{P} = (2\pi/L) \boldsymbol{d}$, where $\boldsymbol{d}$ is a
three-vector of integers, is the so-called ``box matrix'', using the notation of
Ref.~\cite{Morningstar:2017spu}. Eq.~(\ref{eq:det}) applies only for energies below
all thresholds of states containing three or more particles.  The determinant is taken
over all unit normalized two-hadron states $\vert Jm_J\ell Sa\rangle$ specified by total
angular momentum $J$, the projection of $J$ along the $z$-axis $m_J$, the orbital
angular momentum $\ell$, the total intrinsic spin $S$, and particle species $a$.

Inserting the partial waves into the cubic box so as to maintain the periodic boundary
conditions leads to the presence of the box matrix in Eq.~(\ref{eq:det}). The elements
$\langle J'm'_J\ell' S'a'\vert\ \Bp(E_{\rm cm})\ \vert Jm_J\ell Sa\rangle$ of this matrix
are given by known but complicated functions of some of the kinematic variables involved
in the scattering process. This matrix is diagonal in the indices corresponding to total
intrinsic spin and particle species, but not to any of the other indices.  In particular,
states of different total angular momentum can mix. It is straightforward to show that,
under any symmetry transformation $Q$ of the cubic box which is an element of the little
group of $\bm{P}$, the box matrix transforms as $Q\Bp Q^\dagger=\Bp$.  This reveals that
the box matrix can be block diagonalized by projecting onto the superpositions of states
that transform according to the irreps of the little group.  The $\widetilde{K}$ matrix similarly
block diagonalizes in such a basis, except for the total intrinsic spin and particle
species indices.  Hence, the determinant in Eq.~(\ref{eq:det}) can be dealt with separately
block by block.  A particular block can be denoted by the finite-volume irrep
$\Lambda(\boldsymbol{d}^2)$ and a row of this irrep $\lambda$.  Since the spectrum is independent
of the row $\lambda$, this index is henceforth omitted. For a particular block, the
block-diagonalized box-matrix is denoted $B^{\Lambda(\boldsymbol{d}^2)}_{J'\ell' n'; J\ell n}$,
where $n,n'$ are irrep occurrence numbers.  The expressions for all elements of
$B^{\Lambda(\bm{d}^2)}$ relevant for this work are given in Ref.~\cite{Morningstar:2017spu}. 
After transforming to the block diagonal matrix, the $\widetilde{K}$ matrix has the form
given by Eq.~(35) in Ref.~\cite{Morningstar:2017spu}.  A truncation
$\ell\leq\ell_{\rm max}$ in each block then makes the determinant condition manageable.

\begin{figure}[t]
\centering
\includegraphics[width=\columnwidth]{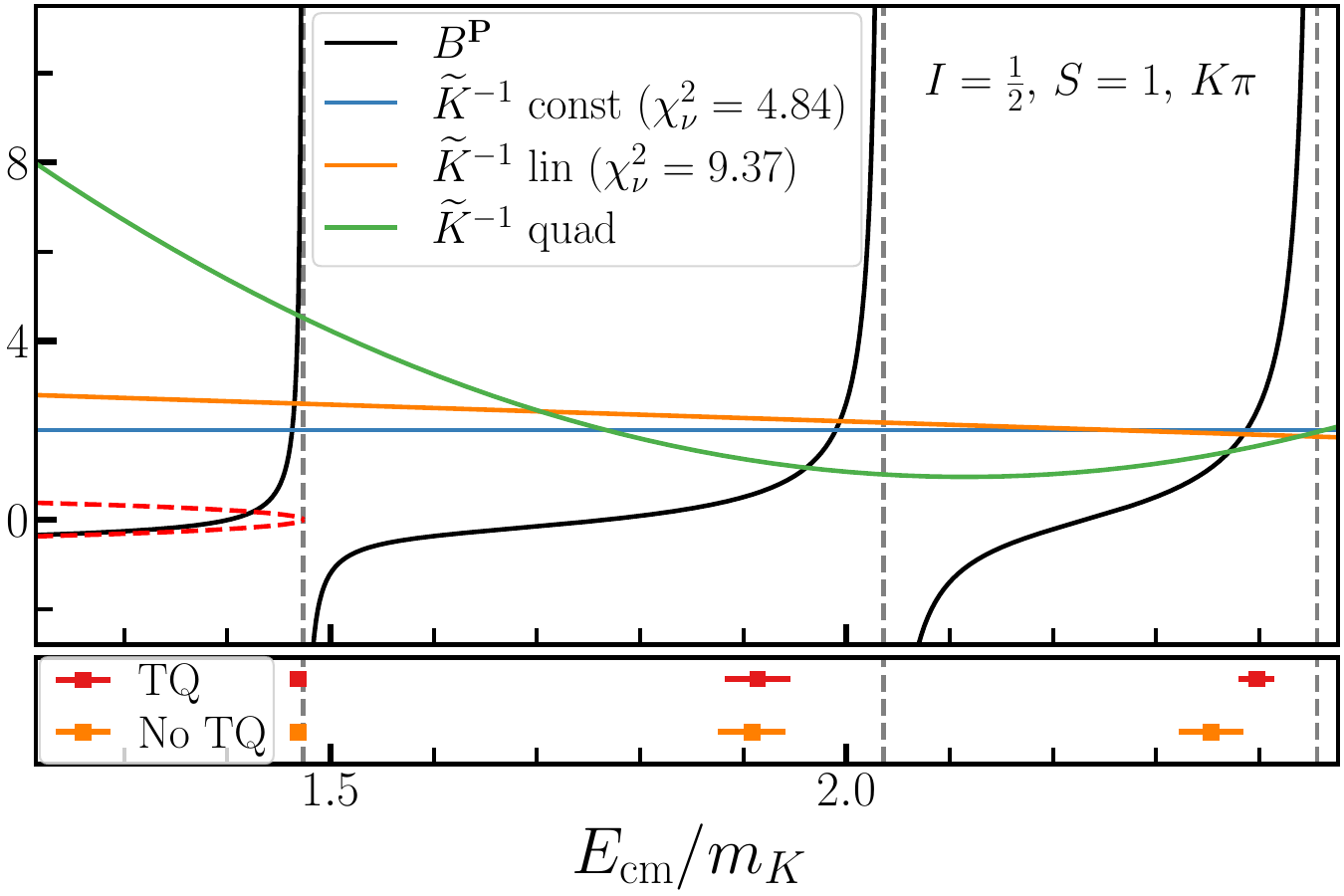}
\caption{Study of the L\"uscher quantization condition for the $I=\frac{1}{2}, S=1, A_{1g}$ channel
 with zero total momentum, restricted to relative orbital angular momentum $\ell=0$ and the $K\pi$
 decay channel.  (Upper panel) The black curve shows the box matrix $\Bp$ function against
 center-of-mass energy over kaon mass, with its singularities at the non-interacting energies
 indicated by vertical dashed lines.  Blue, orange, and green curves show the behavior of
 $\widetilde{K}^{-1}$ for constant, linear, and quadratic forms, respectively.  Intersections of
 these curves with the black curves indicate the energies that satisfy the quantization condition.
 Vertical dashed lines indicate the non-interacting energies, and the red dashed curve is
 $\pm\sqrt{-(k_{K\pi}/m_K)^2}$. In the legend, $\chi_\nu^2=\chi^2/N_{\rm dof}$. (Lower panel)
 Finite-volume energies for the predominantly $K\pi$ states from the lattice QCD computations
 which include (exclude) the tetraquark operator are shown by the red (orange) points.
\label{fig:kappa_amp_1}}
\end{figure}

We use the same definition of the $K$-matrix as described in Ref.~\cite{Morningstar:2017spu}.
This matrix is real and symmetric and diagonal in total angular momentum and its projection:
\begin{equation}
   \langle J'm'_J\ell' S'a'\vert K \vert Jm_J\ell Sa\rangle
   = \delta_{J'J}\delta_{m'_J m_J} K^{(J)}_{\ell'S'a'; \ell Sa}(E_{\rm cm}).
\end{equation}
The matrix $\widetilde{K}$ here is defined by
\begin{equation}
  \widetilde{K}^{(J)-1}_{\ell'S'a';\ \ell Sa}(E_{\rm cm})=k_{a'}^{\ell'+\frac{1}{2}}
  \ K^{(J)-1}_{\ell'S'a';\ \ell Sa}(E_{\rm cm})
  \ k_a^{\ell+\frac{1}{2}},
\end{equation}
where
\begin{equation}
   k_{a}^{2}=\frac{1}{4} \Ecm^2
   - \frac{1}{2}(m_{a_1}^2+m_{a_2}^2) + \frac{(m_{a_1}^2-m_{a_2}^2)^2}{4\Ecm^2},
\end{equation}
for a decay channel $a$ involving particles of mass $m_{a_1}$ and $m_{a_2}$.  Note that this
definition of $\widetilde{K}$ differs very slightly from that given in
Ref.~\cite{Morningstar:2017spu}, and hence, the box matrix here also is slightly different.
Simple factors related to the spatial extent $L$ have been removed here.

\subsection{\boldmath Scattering in the $\kappa$ channel}

\begin{figure}[t]
\centering
\includegraphics[width=\columnwidth]{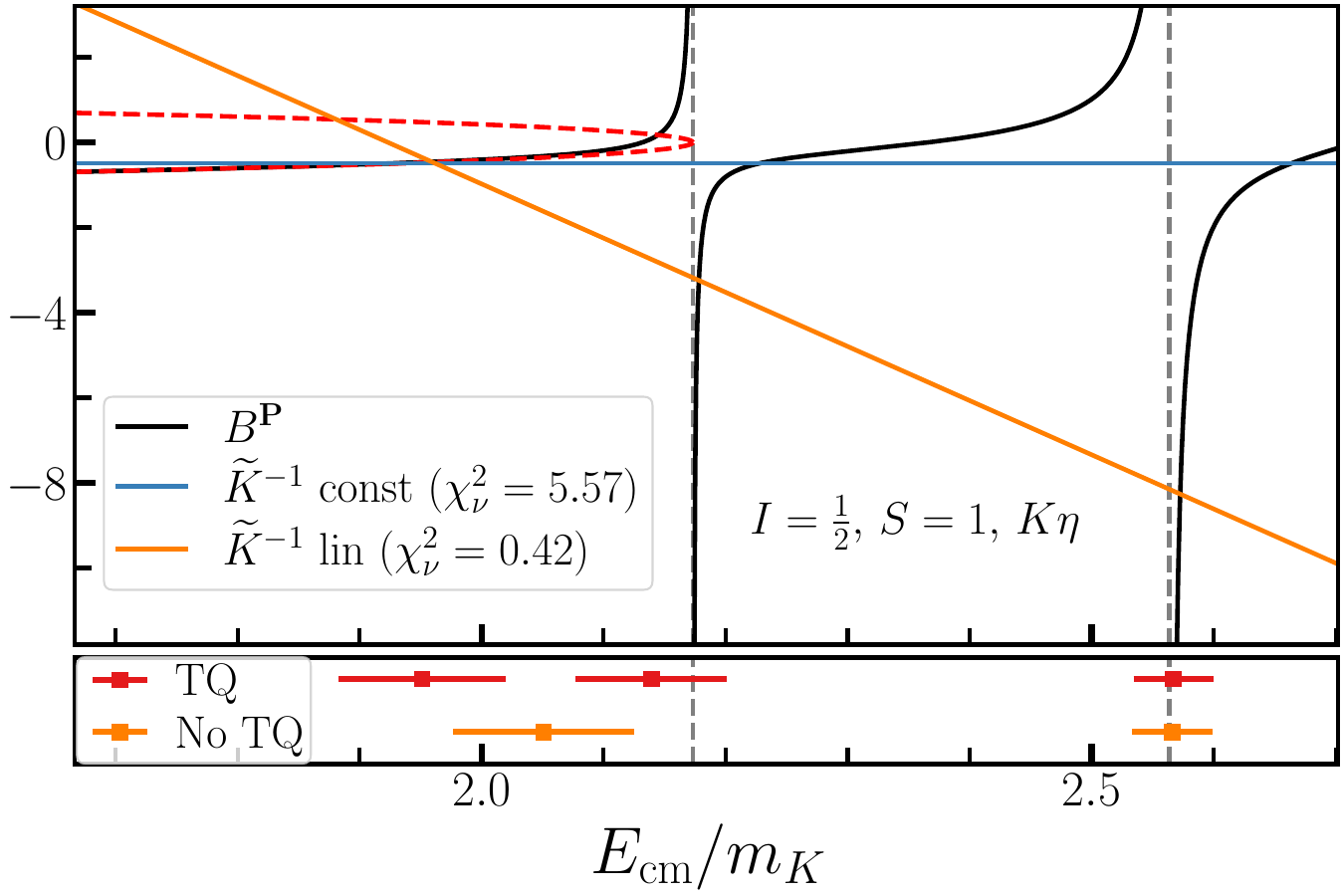}
\caption{Study of the L\"uscher quantization condition for the $I=\frac{1}{2}, S=1, A_{1g}$ channel
 with zero total momentum, restricted to relative orbital angular momentum $\ell=0$ and the $K\eta$
 decay channel.  (Upper panel) The black curve shows the box matrix $\Bp$ function against
 center-of-mass energy over kaon mass, with its singularities at the non-interacting energies
 indicated by vertical dashed lines.  Blue and orange curves show the behavior of $\widetilde{K}^{-1}$
 for constant and linear forms, respectively.  Intersections of these curves with the black curves
 indicate the energies that satisfy the quantization condition. Vertical dashed lines indicate
 the non-interacting energies, and the red dashed curve is $\pm\sqrt{-(k_{K\eta}/m_K)^2}$. In the
 legend, $\chi_\nu^2=\chi^2/N_{\rm dof}$. (Lower panel) Finite-volume energies for the predominantly
 $K\eta$ states from the lattice QCD computations which include (exclude) the tetraquark operator
 are shown by the red (orange) points.
\label{fig:kappa_amp_2}}
\end{figure}

In the isodoublet, strangeness 1, $A_{1g}$ channel, there are two decay channels for any low-lying
resonance: $K\pi$ and $K\eta$.  The expansion in $\ell$ is known to converge very quickly.  If we
focus on the dominant $\ell=0$ contribution, the quantization condition involves a $2\times 2$
matrix for the two channels.  The overlap factors show very little mixing between the $K\pi$ and
$K\eta$ channels, and this is also observed experimentally for the $K_0^\ast(700)$ resonance,
which decays overwhelmingly to $K\pi$ \cite{ParticleDataGroup:2024cfk}.  This means the
off-diagonal elements of the matrix whose determinant appears in the quantization condition in
Eq.~(\ref{eq:det}) can be taken to be negligible to a good approximation, and the quantization
condition simplifies to an equality with zero of a product of two separate scalar functions.

The quantization condition for the $K\pi$ channel is studied in Fig.~\ref{fig:kappa_amp_1}.  The black
curve in this figure shows the box matrix $\Bp$ as a function of $\Ecm/m_K$ using the mean values of
the kaon and pion masses, as well as the mean value determined for the lattice anisotropy, as
presented in Table~\ref{table:ensemble}.  The singularities of this function at the non-interacting
energies are indicated by the vertical dashed lines.  The intersections of the $\widetilde{K}^{-1}$
curve with the black curve satisfy the quantization condition.  The goal is to find a parametrization
of $\widetilde{K}^{-1}$ that produces intersections with the black curve at the energies shown by the red
points in the lower panel, which are the energies obtained in our lattice QCD computations that
correspond to the $K\pi$ channel (the red blocks of levels 0, 1, and 4 in the center panel of
Fig.~\ref{fig:kappa_spectrum}).  The blue and orange curves show the best fits when $\widetilde{K}^{-1}$
is taken to be a constant and linear function, respectively. In this energy range, the best-fit constant
and linear functions intersect with the black curve at three energies, but these energies do not
agree well with the energies in the lower panel, as indicated by the large values of chi-square per
degree of freedom shown in the legend. The $\chi^2$ used here to compute the fit qualities are
defined using the differences in the finite-volume $\Ecm/m_K$ values obtained from the
quantization condition and the lattice determinations as the residuals, taking all covariances into
account, except neglecting the small uncertainties in the single-meson masses. The green curve shows
that a quadratic function of $\Ecm/m_K$ for $\widetilde{K}^{-1}$ is able to intersect the black curve
at the energies shown in the lower panel reasonably well.  With only three energies in this channel of
zero total momentum and three parameters in the function to be fit, there are zero degrees of freedom
in such a fit, so additional total momenta are needed to reliably determine this function.  The main
point to be made here, however, is that the tetraquark operator appears to play no role in the $K\pi$
channel, so the omission of the tetraquark operator does not adversely affect any lattice QCD study of
the $\kappa$ resonance.

The quantization condition for the $K\eta$ channel is studied in Fig.~\ref{fig:kappa_amp_2}. The black
curve in this figure again shows the box matrix $\Bp$ as a function of $\Ecm/m_K$ using the mean values
of the kaon and $\eta$ masses and the lattice anisotropy.  In the lower panel, the red points show the
energies of levels 2, 3, and 5 in the center panel of Fig.~\ref{fig:kappa_spectrum}, obtained in
the lattice QCD computations including a tetraquark operator. The orange points in the lower panel
show the energies extracted when the tetraquark operator is excluded in the analysis. The blue and
orange curves show the best fits when $\widetilde{K}^{-1}$ is taken to be a constant and linear
function, respectively. In this energy range, the best-fit constant and linear functions intersect
with the black curve at three energies.  The intersections with the constant function do not agree
very well with the red points in the lower panel below, but the energies from the linear
function do.

Without the tetraquark operator, only two intersections with the black curve would be allowed in order
to agree with the orange points. Fig.~\ref{fig:kappa_amp_2} illustrates how difficult it would be to
find a suitable form for $\widetilde{K}^{-1}$ that could accomplish only two intersections which agree
with the orange points. This is further evidence that the analysis excluding the tetraquark operator
has led to missing an important energy level. Moreover, under the assumption of a decoupled $K\eta$
channel, the intersections of the $\tilde{K}^{-1}$ curves with the dashed red line in
Fig.~\ref{fig:kappa_amp_2} suggest the existence of a bound $K\eta$ state when the analysis includes
the tetraquark operator.  Recall that $k\cot\delta_0(k) = \varepsilon\sqrt{-k^2}$ is the $S$-wave
bound state condition for $\varepsilon=-1$ and the $S$-wave virtual bound state condition for
$\varepsilon=+1$. Without the tetraquark operator, such intersections seem unlikely to occur,
leading to an absence of any bound state.  We emphasize the qualitative nature of our analysis, noting
that a quantitative study of $\widetilde{K}^{-1}$ would require many more energy levels using a variety
of different total momenta.

Although the absence of the tetraquark operator does not adversely impact a study of the $\kappa$
resonance, the $K_0^\ast(1430)$ decays to both $K\pi$ and $K\eta$, so a missed low-lying state in the
$K\eta$ channel could adversely impact a study of this resonance.

\subsection{\boldmath Scattering in the $a_0(980)$ channel}

For the configuration ensemble used in this study, the $\pi,K,\eta,\eta^\prime$ are all stable against
strong decays.  Hence, in the isotriplet, strangeness 0, $A_{1g}^-$ channel, there are three decay
channels for any low-lying resonance: $\overline{K}K$, $\pi\eta$, and $\pi\eta^\prime$.  The expansion
in $\ell$ is known to converge very quickly.  If we focus on the dominant $\ell=0$ contribution, the
quantization condition involves a $3\times 3$ matrix for the three channels.  When the tetraquark
operator is included in the analysis, the overlap factors show very little mixing among the
$\overline{K}K$, $\pi\eta$, and $\pi\eta^\prime$  decay channels.  Also, in
Ref.~\cite{ParticleDataGroup:2024cfk}, the branching ratio of decays of the $a_0(980)$ to
$\overline{K}K$ over that of $\pi\eta$ is about 0.17, and the branching ratio to $\pi\eta^\prime$
decays over total decays is very small.  These facts suggest that the off-diagonal elements of the
matrix whose determinant appears in the quantization condition of Eq.~(\ref{eq:det}) should be
negligible to a good approximation, and so, the quantization condition simplifies to an equality with
zero of a product of three separate scalar functions.

\begin{figure}[t]
\centering
\includegraphics[width=\columnwidth]{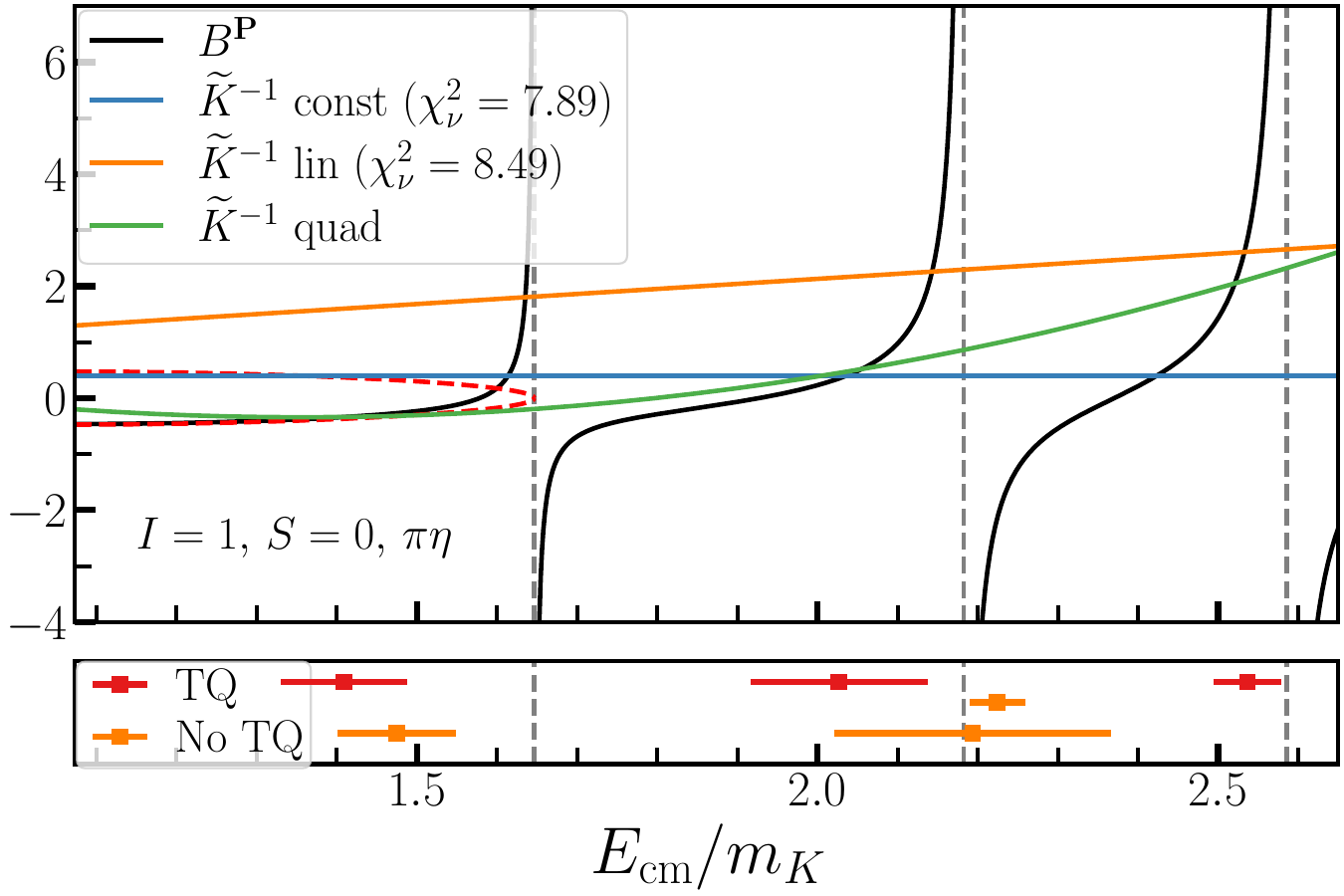}
\caption{Study of the L\"uscher quantization condition for the $I=1, S=0, A_{1g}^-$ channel with zero
 total momentum, restricted to relative orbital angular momentum $\ell=0$ and the $\pi\eta$ decay
 channel.  (Upper panel) The black curve shows the box matrix $\Bp$ function against center-of-mass
 energy over kaon mass, with its singularities at the non-interacting energies indicated by vertical
 dashed lines.  Blue, orange, and green curves show the behavior of $\widetilde{K}^{-1}$ for
 constant, linear, and quadratic forms, respectively.  Intersections of these curves with the black
 curves indicate the energies that satisfy the quantization condition. Vertical dashed lines
 indicate the non-interacting energies, and the red dashed curve is $\pm\sqrt{-(k_{\pi\eta}/m_K)^2}$.
 In the legend, $\chi_\nu^2=\chi^2/N_{\rm dof}$. (Lower panel)  Finite-volume energies for
 the predominantly $\pi\eta$ states (see text) from the lattice QCD computations which include
 (exclude) the tetraquark operator are shown by the red (orange) points.
\label{fig:a0_amp_1}}
\end{figure}

The quantization condition for the $\pi\eta$ channel is studied in Fig.~\ref{fig:a0_amp_1}. The black
curve in this figure shows the box matrix $\Bp$ function as a function of $\Ecm/m_K$ using the mean
values of the pion and $\eta$ masses and the lattice anisotropy.  In the lower panel, the red points
show the energies of levels 0, 2, and 4 in the center panel of Fig.~\ref{fig:a0_spectrum}, obtained in
the lattice QCD computations with the inclusion of a tetraquark operator. The $\overline{K}K$ energies
of levels 1 and 5, indicated by the orange boxes in Fig.~\ref{fig:a0_spectrum}, are excluded in the
lower panel of Fig.~\ref{fig:a0_amp_1}.  The energy of level 3 is also excluded since overlap factors
suggest this level is predominantly a $\pi\eta^\prime$ state.  The orange points in the lower panel
show the energies extracted when the tetraquark operator is excluded in the analysis.

Without the tetraquark operator, the overlap factors erroneously show much more mixing, making it
difficult to identify the state which is predominantly the $\pi\eta^\prime$ state.  In the lower
panel, we select one of two states which we believe have significant overlap with the
$\pi\eta^\prime$ so at least the number of states is correct. The blue and orange curves show the
best fits when $\widetilde{K}^{-1}$ is taken to be a constant and linear functions, respectively.
The green curve is a quadratic function engineered to intersect the black curve at energies within
the red error bars; with zero degrees of freedom, it is not a fit. In this energy range, the
best-fit constant and linear functions intersect with the black curve at three energies.  The
intersections with the constant and linear functions do not agree very well with the red points
in the lower panel below, as indicated by the large $\chi^2/N_{\rm dof}$ values of the fits.  To
achieve a reasonably good matching of the three intersections with the red points in the lower
panel, it appears that a function with some curvature, such as a quadratic function, is needed. 
Without the tetraquark operator, Fig.~\ref{fig:a0_amp_1} shows that matching the intersections
with the orange points would lead to a dramatically different $\widetilde{K}^{-1}$, requiring
significant downward curvature on the right to intersect the right most section of the black
curve at a significantly lower energy.

\begin{figure}[t]
\centering
\includegraphics[width=\columnwidth]{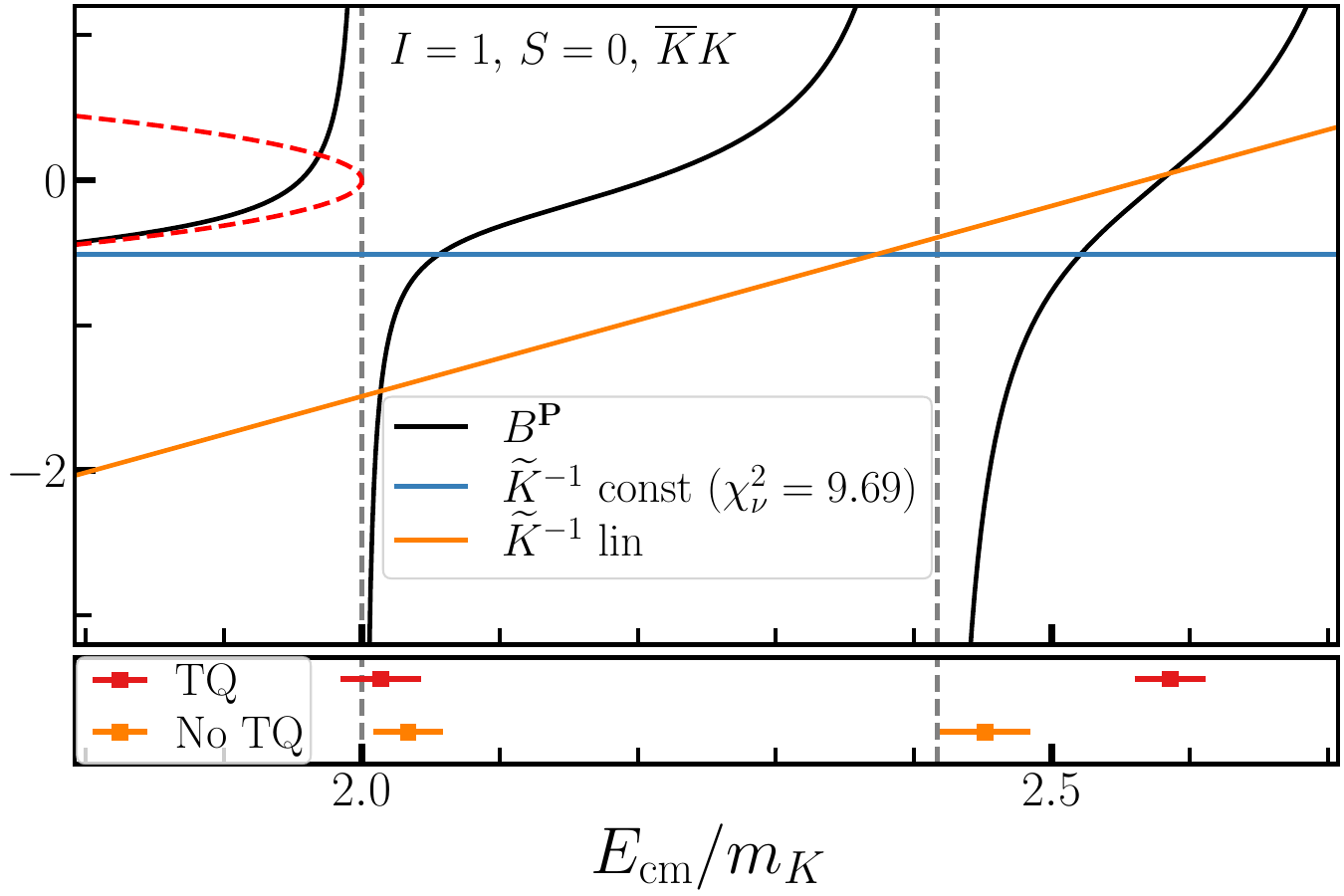}
\caption{Study of the L\"uscher quantization condition for the $I=1, S=0, A_{1g}^-$ channel with
 zero total momentum, restricted to relative orbital angular momentum $\ell=0$ and the $\overline{K}K$
 decay channel.  (Upper panel) The black curve shows the box matrix $\Bp$ function against
 center-of-mass energy over kaon mass, with its singularities at the non-interacting energies
 indicated by vertical dashed lines.  Blue and orange curves show the behavior of
 $\widetilde{K}^{-1}$ for constant and linear forms, respectively.  Intersections of these curves
 with the black curves indicate the energies that satisfy the quantization condition. Vertical
 dashed lines indicate the non-interacting energies, and the red dashed curve is
 $\pm\sqrt{-(k_{\overline{K}K}/m_K)^2}$. In the legend, $\chi_\nu^2=\chi^2/N_{\rm dof}$.
 (Lower panel) Finite-volume energies for the predominantly $\overline{K}K$ states from the lattice
 QCD computations which include (exclude) the tetraquark operator are shown by the red (orange) points.
\label{fig:a0_amp_2}}
\end{figure}

The quantization condition for the $\overline{K} K$ channel is studied in Fig.~\ref{fig:a0_amp_2}.
The black curve in this figure shows the box matrix $\Bp$ function as a function of $\Ecm/m_K$
using the mean values of the kaon and $\eta$ masses and the lattice anisotropy.  In the lower
panel, the red points show the energies of levels 1 and 5 in the center panel of
Fig.~\ref{fig:a0_spectrum}, obtained in the lattice QCD computations with the inclusion of
tetraquark operators. The orange points in the lower panel show the energies for the
$\overline{K} K$ dominated states extracted when the tetraquark operator is excluded in the
analysis. The blue curve shows the best fit when $\widetilde{K}^{-1}$ is taken to be a constant
function. The orange curve is a linear function engineered to intersect the black curve
at energies within the red error bars;  with zero degrees of freedom, it is not a fit. In this
energy range, the best-fit constant and the linear functions intersect with the black curve at
two energies.  The intersections with the constant function do not agree very well with the
red points in the lower panel below, as indicated by the large $\chi^2/N_{\rm dof}$ values of
the fit.  To achieve a reasonably good matching of the two intersections with the red points
in the lower panel, the linear function is needed.  Without the tetraquark operator,
Fig.~\ref{fig:a0_amp_2} shows that matching the intersections with the orange points would lead
to a linear function for $\widetilde{K}^{-1}$ with a dramatically different slope.

\section{Conclusion}
\label{sec:conclude}

Due to the manner in which stationary-state energies are extracted from matrices of temporal
correlations of single- and multi-hadron operators in lattice QCD, the unfortunate possibility
exists of flawed spectrum extractions whenever an operator set is insufficient to provide
overlaps onto all relevant finite-volume stationary states.  In this work, the isodoublet,
strangeness 1, zero-momentum, $A_{1g}$ channel relevant for the $\kappa$ resonance and the
isotriplet non-strange zero-momentum $A_{1g}^-$ channel relevant for the $a_0(980)$ channel were
studied.  We found that in both cases, neglecting to include a particular type of tetraquark
operator in the correlator matrices led to a flawed energy spectrum extraction.  The impact of
these erroneous energies on determinations of scattering amplitudes and resonance parameters
using the L\"uscher quantization condition was studied in both case.

In the $I=\frac{1}{2},\ S=1$ case, the energies of states dominated by $K\eta$ pairs were
significantly impacted, with one level completely missed when the tetraquark was excluded from
the analysis, whereas the $K\pi$ energies were only slightly affected.  Since the $\kappa$
resonance decays primarily to $K\pi$ states, omission of a tetraquark operator does not appear
to adversely impact studies of the $\kappa$ resonance.  However, this work suggests that the
use of a tetraquark operator will be important for any lattice QCD study of the
$K_0^\ast(1430)$ resonance.

In the $I=1,\, S=0$ case, omission of a tetraquark operator led to substantial changes in the
energies.  This strongly suggests that lattice QCD studies of the $a_0(980)$ resonance can
obtain reliable results only if a tetraquark operator of the type described in this work
is included in the analysis.

The results of this work underscore the importance of including tetraquark operators in studying
low-lying mesonic resonances, such as the $a_0(980)$.  This exploratory investigation focused
only on correlator matrices involving states of total zero momentum.  Future work will include
increased statistics, channels involving several nonzero total momenta, as well as a variety of
different lattice volumes and lattice spacings.  The importance of tetraquark operators
in lattice QCD studies of other light meson resonances, such as the isoscalar $\sigma$, should
also be investigated.

\begin{acknowledgments}
We acknowledge helpful discussions with Ben~H\"orz. Computations were carried out at the Oak
Ridge Leadership Computing facility at Oak Ridge National Laboratory and on
Frontera~\cite{frontera} at the Texas Advanced Computing Center (TACC).  Our software to
evaluate the correlators made use of the CHROMA~\cite{Edwards:2004sx} library.
C.M., J.M., and S.S.~acknowledge support from the U.S.~National Science Foundation (NSF)
under awards PHY-2209167 and PHY-2514831.  NSF awards PHY-1306805 and PHY-1613449
supported A.D.H., J.F., R.B, and D.D.~during early stages of this investigation.  Any
opinions, findings, and conclusions or recommendations expressed in this material are those
of the author(s) and do not necessarily reflect the views of the National Science Foundation.
This work was also supported by the U.S. Department of Energy (DOE), Office of Science,
Office of Nuclear Physics, under grant contract number DE-AC02-05CH11231 (A.W.L.). The work
of F.R.L. was supported in part by the Swiss National Science Foundation (SNSF) through
grant No. 200021-236432. J.B. is supported by the European Research Council (ERC)
consolidator grant StrangeScatt-101088506.
\end{acknowledgments}

\bibliography{thebib}
\end{document}